\documentclass[twocolumn,useAMS,usenatbib]{mn2e}
\usepackage{natbib}
\usepackage{amsmath,latexsym,amssymb}
\usepackage{epsfig,graphicx}
\def\reff@jnl#1{{\rm#1\/}}

\def\aj{\reff@jnl{AJ}}                  % Astronomical Journal
\def\araa{\reff@jnl{ARA\&A}}            % Annual Review of Astron and Astrophys
\def\apj{\reff@jnl{ApJ}}                        % Astrophysical Journal
\def\apjl{\reff@jnl{ApJ}}               % Astrophysical Journal, Letters
\def\apjs{\reff@jnl{ApJS}}              % Astrophysical Journal, Supplement
\def\apss{\reff@jnl{Ap\&SS}}            % Astrophysics and Space Science
\def\aap{\reff@jnl{A\&A}}               % Astronomy and Astrophysics
\def\aapr{\reff@jnl{A\&A~Rev.}}         % Astronomy and Astrophysics Reviews
\def\aaps{\reff@jnl{A\&AS}}             % Astronomy and Astrophysics, Supplement
\def\baas{\reff@jnl{BAAS}}              % Bulletin of the AAS
\def\jrasc{\reff@jnl{JRASC}}            % Journal of the RAS of Canada
\def\memras{\reff@jnl{MmRAS}}           % Memoirs of the RAS
\def\mnras{\reff@jnl{MNRAS}}            % Monthly Notices of the RAS
\def\physrep{\reff@jnl{Phys.Rep.}}
\def\pra{\reff@jnl{Phys.Rev.A}}         % Physical Review A: General Physics
\def\prb{\reff@jnl{Phys.Rev.B}}         % Physical Review B: Solid State
\def\prc{\reff@jnl{Phys.Rev.C}}         % Physical Review C
\def\prd{\reff@jnl{Phys.Rev.D}}         % Physical Review D
\def\prl{\reff@jnl{Phys.Rev.Lett}}      % Physical Review Letters
\def\pasp{\reff@jnl{PASP}}              % Publications of the ASP
\def\pasj{\reff@jnl{PASJ}}              % Publications of the ASJ
\def\skytel{\reff@jnl{S\&T}}            % Sky and Telescope
\def\solphys{\reff@jnl{Solar~Phys.}}    % Solar Physics
\def\sovast{\reff@jnl{Soviet~Ast.}}     % Soviet Astronomy
\def\ssr{\reff@jnl{Space~Sci.Rev.}}     % Space Science Reviews
\def\nat{\reff@jnl{Nature}}             % Nature

\newcommand{\beq}{\begin{equation}}
\newcommand{\eeq}{\end{equation}}
\newcommand{\beqa}{\begin{eqnarray}}
\newcommand{\eeqa}{\end{eqnarray}}

\newcommand{\rmd}{\mathrm{d}}

\newcommand{\ic}{\ensuremath{I_\mathrm{C}}}
\newcommand{\pc}{\ensuremath{G_\mathrm{C}}}
\newcommand{\ps}{\ensuremath{G_\mathrm{T}}}
\newcommand{\tic}{\ensuremath{\tilde{I}_\mathrm{C}}}
\newcommand{\tpc}{\ensuremath{\tilde{G}_\mathrm{C}}}
\newcommand{\tps}{\ensuremath{\tilde{G}_\mathrm{T}}}
\newcommand{\tkpd}{\ensuremath{\tilde{T}}}
\newcommand{\ticpd}{\ensuremath{\tilde{P}}}
\newcommand{\ticpdr}{\ensuremath{\tilde{P}^\mathrm{(rot)}}}
\newcommand{\ticpds}{\ensuremath{\tilde{P}^{(\gamma)}}}
\newcommand{\ticpdrs}{\ensuremath{\tilde{P}^{(\mathrm{rot},\gamma)}}}
\newcommand{\tks}{\ensuremath{\tilde{K}_\mathrm{T}}}
\newcommand{\tksr}{\ensuremath{\tilde{K}_\mathrm{T}^\mathrm{(rot)}}}
\newcommand{\tis}{\ensuremath{\tilde{I}^{(\gamma)}_\mathrm{T}}}
\newcommand{\tisr}{\ensuremath{\tilde{I}^{(\mathrm{rot},\gamma)}_\mathrm{T}}}

\newcommand{\rpix}{\ensuremath{R_\mathrm{pix}}}

\newcommand{\nc}{\ensuremath{N_\mathrm{C}}}

\newcommand{\nt}{\ensuremath{N_\mathrm{T}}}
\newcommand{\erms}{\ensuremath{e_\mathrm{rms}}}
\newcommand{\etot}{\ensuremath{e_\mathrm{tot}}}
\newcommand{\newtext}{}
\newcommand{\reftext}[1]{#1}

\title[Ground-based image simulation]{Precision simulation of
  ground-based lensing data using observations from space}

\author[Mandelbaum et al.]
{Rachel Mandelbaum$^1$\thanks{\tt rmandelb@astro.princeton.edu}, 
Christopher~M. Hirata$^2$,
Alexie Leauthaud$^3$, 
\newauthor
Richard J. Massey$^4$, Jason Rhodes$^5$
\\$^1$Department of Astrophysical Sciences, Princeton University,
Peyton Hall, Princeton, NJ 08544, USA
\\$^2$Department of Astronomy, Caltech M/C 350-17, Pasadena, CA 91125, USA
\\$^3$Lawrence Berkeley National Laboratory, Berkeley, CA 94720, USA
\\$^4$Institute for Astronomy, Royal Observatory, Blackford Hill, Edinburgh EH9 3HJ, UK
\\$^5$Jet Propulsion Laboratory, California Institute of Technology,
Pasadena, CA 91109
}

\date{\today}

\begin{document}
\maketitle

\begin{abstract}
  Current and upcoming wide-field, ground-based, broad-band imaging
  surveys promise to address a wide range of outstanding problems in
  galaxy formation and cosmology.  Several such uses of ground-based
  data, especially weak gravitational lensing, require highly precise
  measurements of galaxy image statistics with careful correction for
  the effects of the point-spread function (PSF).  In this paper, we
  introduce the {\sc shera} (SHEar Reconvolution Analysis) software to
  simulate ground-based imaging data with realistic galaxy
  morphologies and observing conditions, starting from space-based
  data (from COSMOS, the Cosmological Evolution Survey) and accounting
  for the effects of the space-based PSF.  This code simulates
  ground-based data, optionally with a weak lensing shear applied, in
  a model-independent way using a general Fourier space formalism.
  The utility of this pipeline is that it allows for a precise,
  realistic assessment of systematic errors due to the method of data
  processing, for example in extracting weak lensing galaxy shape
  measurements or galaxy radial profiles, given user-supplied
  observational conditions and real galaxy morphologies.  Moreover, the simulations
  allow for the empirical test of error estimates and determination of
  parameter degeneracies, via generation of many noise maps. The 
  public release of this software, along with a large sample of
  cleaned COSMOS galaxy images (corrected for charge transfer
  inefficiency), should enable upcoming ground-based imaging surveys
  to achieve their potential in the areas of precision weak lensing
  analysis, galaxy profile measurement, and other applications involving
  detailed image analysis.
\end{abstract}

\begin{keywords}
  methods: data analysis -- techniques: image processing --
  gravitational lensing: weak -- galaxies: structure
\end{keywords}

\section{Introduction}
\label{S:intro}

A tremendous variety of measurements are carried out on astronomical
images from ground-based telescopes.  A generic problem that often
arises is the question of how the intrinsically limited resolution of
ground-based images (both due to convolution with the atmospheric
point-spread function, or PSF, and due to the finite pixel size)
affects our ability to measure quantities such as the radial profiles
of galaxies $I(r)$, or statistics of the profile such as its slope,
half-light radius, and ellipticity.  Moreover the error distributions
of these parameters, which are often estimated via a highly non-linear
fitting procedure, are also unclear in many circumstances.

One application that particularly suffers from such uncertainties is
the field of weak gravitational lensing (for a review, see
\citealt{2001PhR...340..291B}, \citealt{2003ARA&A..41..645R},
\citealt{2008ARNPS..58...99H}, or \citealt{2010RPPh...73h6901M}).  In
the past decade, weak lensing has been used extensively for
measurements that can constrain cosmology and galaxy formation.
Cosmic shear measurements have constrained cosmological parameters
(e.g., most recently,
\citealt{2008A&A...479....9F,2010A&A...516A..63S}); cluster lensing
analyses (e.g., \citealt{2007MNRAS.379..317H,2010PASJ...62..811O})
have been used to understand the most massive structures in the
universe, the abundance of which can constrain cosmology through the
mass function \citep[e.g., ][]{2007ApJ...657..183R,2008MNRAS.387.1179M,2009ApJ...692.1060V,2010ApJ...708..645R};
and galaxy-galaxy lensing measurements have probed the connection
between galaxies, their dark matter halos, and their larger scale
environments \citep{2010MNRAS.408.1463S,2011arXiv1104.0928L}, as well
as constraining the theory of gravity on cosmological scales when
combined with other observational methods \citep{2010Natur.464..256R}.
The next decade promises a larger volume of weak lensing data that can
be used for more precise constraints on cosmology and galaxy
formation, from surveys such as Hyper Suprime-Cam 
(HSC, \citealt{2006SPIE.6269E...9M}),
Dark Energy Survey (DES\footnote{\texttt{https://www.darkenergysurvey.org/}}, \citealt{2005astro.ph.10346T}),
the KIlo-Degree Survey (KIDS\footnote{\texttt{http://www.astro-wise.org/projects/KIDS/}}),
the Panoramic Survey Telescope and Rapid Response System
(Pan-STARRS\footnote{\texttt{http://pan-starrs.ifa.hawaii.edu/public/}}, 
\citealt{2010SPIE.7733E..12K}); and even more ambitious programmes are planned such as
the Large Synoptic Survey Telescope
(LSST\footnote{\texttt{http://www.lsst.org/lsst}}, \citealt{2009arXiv0912.0201L}),
Euclid\footnote{\texttt{http://sci.esa.int/science-e/www/area/index.cfm?fareaid\
    =102}}, and the Wide-Field Infrared Survey Telescope 
(WFIRST\footnote{\texttt{http://wfirst.gsfc.nasa.gov/}}).

Weak lensing measurements depend on precise measurements of the shapes
of galaxies, in an attempt to infer the apparent 
shape distortions in distant `source' galaxies due to the mass in
nearby `lenses' (galaxies, clusters, or other mass distributions).
Weak lensing is a statistical measurement, with averages over
large numbers of sources in order to detect the $\sim 0.1$--$1$ per cent level
distortions within the noise of the intrinsic galaxy ellipticities, which
are typically a factor of $\sim 100$ larger.  Unfortunately, coherent
systematic distortions of galaxy
shapes due to the PSF caused by the atmosphere
(for a ground-based telescope), telescope optics, and detector are
significantly larger than typical weak lensing distortions, which
means that accurate PSF-correction is critical for current lensing
studies, and all the more so for future lensing surveys that aim for
$<1$ per cent statistical errors.

The weak lensing community has had several challenges, using blind
simulations, to identify the most promising methods of PSF correction.
The first of these, the Shear TEsting Programme-1
\citep[STEP1;][]{2006MNRAS.368.1323H}, included mock galaxies with
idealized radial profiles and several PSFs meant to mimic specific
observational issues (e.g., astigmatism).  The second, STEP2
\citep{2007MNRAS.376...13M}, used shapelets
\citep{2003MNRAS.338...35R,2003MNRAS.338...48R} decompositions of
COSMOS galaxies (including the COSMOS PSF, for which no correction was
imposed) as inputs, and then convolved them with a variety of
ground-based PSFs from Subaru Suprime-Cam \citep{2002PASJ...54..833M}.  Finally, the GRavitational
lEnsing Accuracy Testing-08
\citep[GREAT08;][]{2009AnApS...3....6B,2010MNRAS.405.2044B} and
GREAT10 \citep{2010arXiv1009.0779K} challenges reverted to composite
model galaxies (based on S\'ersic profiles;
\citealt{1968adga.book.....S}) with very specific sets of parameters,
and tested the recovery of the shear as a function of image $S/N$, PSF
size, and galaxy profile type.  All of these challenges were useful to
the lensing community in different ways, and in some cases led to
changes in attitudes towards (or a greater understanding of) common
methods of PSF correction.  However, their ability to identify, in a
broad sense, the most promising methods of PSF correction does not
mean that they can be used to constrain, to high precision, the shear
calibration in \newtext{all lensing observations} using those
PSF-correction methods\footnote{\newtext{The validity of this
    statement depends on the data for which the shear calibration is
    desired.  For example, the STEP2 simulations are likely to give a
    more realistic estimate of the calibration for the Subaru
    Suprime-Cam data that it was designed to mimic than for data from
    some other telescope (assuming that image combination and other
    steps in a realistic data analysis, which were not simulated, do
    not introduce additional biases).}}.  There are numerous reasons
why this is the case, such as galaxy models and observing conditions
(PSF and depth) that are not representative of a particular
survey. STEP2 \citep{2007MNRAS.376...13M} showed that the shear
systematics for nearly every method of PSF correction depend on the
observing conditions (the galaxy $S/N$ and resolution compared to the
PSF); GREAT08 \citep{2010MNRAS.405.2044B} showed a dependence on
galaxy type as well.

\newtext{More recent work has shown that the dependence of shear calibration factors on the galaxy population is generic. In particular, 
\citet{2007MNRAS.380..229M} and \citet{2011MNRAS.414.1047Z} showed that there is {\em no} stable shape measurement algorithm on finite-resolution data whose shear calibration factor 
is independent of the galaxy population\footnote{The argument hinges on the existence of a finite number of well-measured moments $M_{ij}$ of the 
galaxy, and the fact that the dependence of the $M_{ij}$ on shear
($\partial M_{ij}/\partial\gamma_k$) is determined in part by higher,
unmeasured moments. \reftext{A related issue occurs in Fourier space:
  when attempting to define a roundness test for a sheared galaxy,
  Bernstein (2010, last paragraph of section 4.2) finds that the finite
  extent of the modulation transfer function (MTF) prevents such a test from being shear-covariant,
  and argues that the issue is generic.}}.
%A related argument based on 
%Fourier modes can be found in \citet{2010MNRAS.406.2793B}.}. 
Instead, \citet{2010MNRAS.406.2793B} and \citet{2011MNRAS.414.1047Z} argue that the same 
lensing survey that measures shear could also determine the range of
galaxy models presented to us by the real Universe. \reftext{While we
  will explore this point in more detail in Sec.~\ref{S:psfcorrection},} we conclude that to precisely 
constrain the shear calibration or understand observational selection biases in any given survey, the simulations must have realistic galaxy models as 
well as observing conditions.}

When constraining systematic errors in ground-based lensing data, we may wish to use space-based data as the
basis of our simulations, due to its much higher resolution.  Indeed,
one might ask why simulations are needed at all: can we
simply rely on comparison with PSF-corrected shapes on space-based
images?  This approach was taken by \cite{2008ApJ...684...34K}, who compared shape
measurements using a KSB-based method \citep{1995ApJ...449..460K,2003ApJ...597...98H} on Subaru data against
shape measurements using the RRG method \citep{2000ApJ...536...79R} on COSMOS data.
However, there are some limitations to this approach.  First, 
\newtext{if one wishes to avoid complications due to a non-negligible
  PSF and therefore the need for substantial PSF correction in the space-based data, the
  comparison must be restricted to a subset of larger galaxies, as in
  \cite{2008ApJ...684...34K}.} 
%while the
%space-based PSF is small, it is not entirely negligible for the faint,
%small galaxies that dominated the lensing sample used for that
%comparison; thus, one must trust 
%the PSF correction on the space-based
%images to very high precision if one wants to draw conclusions about
%the calibration of the ground-based shapes.  
%Second, since different
%methods of PSF correction were used for the space- and ground-based
%data, the conclusions that were drawn may in fact have more to do with
%the different PSF correction than the different resolution of the data.
\newtext{More importantly}, such a direct 
comparison of the derived shapes or shears is not possible at all for some shape measurement methods
if the ellipticities are defined in incompatible ways, as will be
explained at greater detail later in this paper \newtext{(Sec.~\ref{SS:shapecatalog})}.   We thus conclude
that, rather than carrying out a catalogue-level comparison, we should use the space-based
data to make realistic simulations of ground-based lensing
data, to which a shear can be added and shear recovery can be tested. 

In this paper, we therefore have three goals.  First, we outline a
method for simulating ground-based lensing data using
higher-resolution data from space, including a careful treatment of
the original space-based PSF and pixel sampling, and conversion to the
new ground-based PSF and pixel sampling, inspired by
\cite{2000ApJ...537..555K}.  This method will allow the galaxies to be
sheared, so that we can test the recovery of gravitational shear,
including the many types of selection biases and PSF effects such as
those described in \cite{2004MNRAS.353..529H} and
\cite{2005MNRAS.361.1287M}.  This method is a model-free
generalization of that used in \cite{2010PASP..122..947D}, which
relies on shapelets decompositions (Gauss-Laguerre basis functions)
and therefore depends on the galaxy profiles and PSFs being 
well-described by sums of these functions to some finite order.
Given that  this assumption is not necessarily
true for realistic galaxies and PSFs (e.g.,
\citealt{2010A&A...510A..75M}), the advantage 
of a model-independent image simulation method is clear.  
Second, we describe a publicly-released implementation of this method
in IDL.  We emphasize that, while this paper focuses on weak gravitational
lensing, this simulation pipeline is equally applicable to many other
science analyses that are commonly done using ground-based data, for
example modeling of galaxy radial profiles.  Finally, we
demonstrate the method by simulating Sloan Digital Sky Survey (SDSS) lensing data, and show how
these simulations can be used to estimate shear systematics in SDSS to
high precision.  This will be of practical use for interpreting past
lensing measurements that used the SDSS shape catalogue we simulate,
and eventually for
reducing the systematic error budgets in future papers using that catalogue.

We begin in Section~\ref{S:lensing} by describing weak gravitational
lensing and its effects on galaxy shapes.
Section~\ref{S:psfcorrection} describes the process of removing the
effects of the PSF from measured galaxy shapes, so that lensing can be
measured.  The space-based data that are used as the basis of these
simulations, and the ground-based data that we simulate in this paper,
are described in Section~\ref{S:data}.  In Section~\ref{S:methodology}, we describe the methodology
that will be used to create accurate simulations of ground-based
data.   We describe steps that we took to prepare the space-based data
for this purpose in Section~\ref{S:imageprep}, and our specific
implementation of the simulation method in Section~\ref{S:implementation}.
\reftext{Technical tests of this implementation are presented in
Section~\ref{S:techvalidation}, and an example of how the {\sc shera} 
outputs can be used to test galaxy shape measurements is in Section~\ref{S:results}. }
We discuss these results in Section~\ref{S:discussion}.  

\section{Weak lensing basics}\label{S:lensing}

Gravitational lensing is the deflection of light from distant objects
(`sources') by all mass, including dark matter, along the line of
sight (`lenses').  Typically, it results in a weak but coherent
distortion in the shapes of distant galaxies (weak lensing).  This
distortion can be quantified by considering the true source position
${\bmath \beta}$ and the observed position ${\bmath \theta}$ with respect to
the lens; instead of the intrinsic surface brightness profile
$I({\bmath \beta})$, we observe a perturbed profile
$I({\bmath \theta}({\bmath \beta}))$ described by the following Jacobian in
the linear approximation:
\beq\label{E:jacobian}
 \frac{\partial{\bmath \beta}}{\partial{\bmath \theta}} = \left(
    \begin{array}{cc}
      1-\kappa-\gamma_1 & -\gamma_2 \\ 
      -\gamma_2 & 1-\kappa+\gamma_1 \\
    \end{array}
  \right)
\eeq
which includes shear $\gamma\equiv\gamma_1+\mathrm{i}\gamma_2 =
|\gamma|\mathrm{e}^{2\mathrm{i}\varphi}$ and convergence $\kappa$.
These are in turn related to the deflection potential
\beq
  \psi({\bmath \theta}) = \frac{1}{\pi}\int\rmd^2\theta'\,
  \kappa({\bmath \theta'})\,\ln|{\bmath \theta}-{\bmath \theta'}|\;,
\eeq
via
\beqa
\kappa({\bmath \theta})&=&\frac{1}{2}\nabla^2\psi({\bmath \theta}) \nonumber \\
 \gamma_1 &=& \frac{1}{2}(\psi_{,11}-\psi_{,22}) \nonumber \\
  \gamma_2 &=& \psi_{,12}
\eeqa
and to the projected lens mass distribution via 
\beq
  \kappa({\bmath \theta}) = 
  \frac{\Sigma(D_\mathrm{l}{\bmath \theta})}{\Sigma_\mathrm{c}}.
\eeq
Here we have used a critical surface density (geometric factor) defined as\footnote{\newtext{Eq.~\eqref{eq:sigmacrit} is written in physical coordinates for simplicity; in comoving coordinates additional 
factors appear.}}
\beq
  \Sigma_\mathrm{c} = \frac{c^2}{4\pi G}\,
  \frac{D_\mathrm{s}}{D_\mathrm{l}\,D_\mathrm{ls}}
\label{eq:sigmacrit}
\eeq
in terms of the angular diameter distances to the source
($D_\mathrm{s}$), lens ($D_\mathrm{l}$), and between the two
($D_\mathrm{ls}$). 

\reftext{The vast majority of the weak lensing
  measurements to date have focused on the measurement of shear (shape
  distortions) rather than convergence (magnification), and therefore
  require extremely accurate measurement of the shapes of source galaxies.}
% We also 
%defined $\gamma_t$ (the tangential shear component, perpendicular to the
%radial vector connecting lens and source) and the surface density contrast $\Delta\Sigma(R) =
%\langle\Sigma(<R)\rangle - \Sigma(R)$, where $R =
%D_\mathrm{l}\theta$.

%There
%are two main types of lensing shear measurements: cosmic shear, which
%involves correlating the shapes of pairs of distant galaxies to reveal the
%large-scale structure at lower redshifts, and galaxy-galaxy (or
%cluster-galaxy) lensing, which is the cross-correlation between
%distant source galaxy shapes and the positions of massive foregrounds
%such as galaxies or clusters.  
%Both types of lensing measurements rely,
%crucially, on robustly measuring the shape of the source galaxy images.

\section{PSF correction}\label{S:psfcorrection}

A complicating factor in lensing measurements is that in practice, the galaxy shape that is observed has not just been
lensed; it has also passed through an atmosphere (for a ground-based
telescope), telescope optics, and a detector.  This results in a
convolution of the intrinsic galaxy profile with the point-spread
function, or PSF.  In this paper, we
define the PSF as including not just atmospheric blurring, optic, and
detector effects, but also pixelisation (the `effective PSF' or `ePSF'
 in \citealt{2002AJ....123..583B}).  Fortunately, the images of
the stars provide a measurement, albeit a noisy one, of the PSF.

In order to measure the weak lensing shear, we must remove the effects
of the PSF on the source galaxy shape.  There are many methods
of doing so; see \cite{2007MNRAS.376...13M} or
\cite{2010MNRAS.405.2044B} for summaries of the 
common methods of PSF correction.  There are several types of bias
that can arise when trying to extract the lensing shear using the
PSF-corrected shapes \citep{2004MNRAS.353..529H}, for example:
\begin{enumerate}
\item {\em PSF dilution:} The PSF tends to round galaxy
  shapes.  If this rounding is not fully accounted for, it leads to a
  multiplicative calibration bias that may depend on the galaxy
  profile, $S/N$, resolution, or PSF characteristics.
\item {\em Systematic shear:} If the PSF has some nonzero ellipticity, then
  imperfect removal of that ellipticity can manifest as a small
  ellipticity added to each galaxy shape.  Since PSF shapes tend to
  have coherent correlations across the sky, this leads to a spurious
  systematics signal in the lensing measurement.
\item {\em Selection bias:} There are several types of selection
  bias.  For example, for galaxies of a given apparent size, it may be
  easier to distinguish the more flattened ones from stars, and so the
  more flattened ones may be more likely than rounder ones to end up
  in a source galaxy sample.  A selection bias that goes in the
  opposite direction is that some methods may have trouble
  extracting robust shapes for more flattened galaxies, thus selecting
  against them.  
\item {\em Noise rectification bias:} For sparsely sampled
  realizations of an elliptical density distribution, the ellipticity
  tends to be over-estimated.  This is an example of noise
  rectification bias, which is known to affect weak lensing
  measurements both in introducing spurious additive biases and
  affecting the calibration of the shears
  \citep{2000ApJ...537..555K,2002AJ....123..583B,2004MNRAS.353..529H}.
  The details of how it affects shape measurements depends on the
  details of how the shape measurements and PSF correction are
  performed.  For an example calculation of noise rectification bias
  as a function of detection significance, for methods that measure shapes using adaptive galaxy
  moments, see \cite{2004MNRAS.353..529H}.
\item {\em Population uncertainties:} Some methods of shape
  measurement rely on calibration factors that depend on the
  {\em intrinsic} properties of the galaxy population being studied,
  such as its ellipticity distribution.  Since we only observe the
  ellipticities after the PSF and noise have been added, we are
  necessarily limited in how well we can infer the intrinsic
  properties of the sample.  For some methods, e.g. as in
  \cite{2004MNRAS.353..529H}, this results in the {\em shear
    responsivity uncertainty}.  For other methods, such as Lensfit
  which uses a prior on the ellipticity distribution for Bayesian
  inference of galaxy shapes,
  the use of the wrong prior can result in an incorrect inference of
  the shear \citep{2007MNRAS.382..315M}.
\end{enumerate}

Because of the forms that these systematics take, shape measurement
systematics have typically been quantified
\citep{2006MNRAS.368.1323H,2007MNRAS.376...13M} using two numbers: a
multiplicative calibration bias\footnote{\cite{2007MNRAS.376...13M} also allowed for the
  possibility of a nonlinear response to shear, $\propto \gamma^2$.}
$m$, and an additive calibration bias $c$,
relating the estimated shear $\hat{\gamma}$ to the true one
$\gamma$:
\beq
\hat{\gamma} - \gamma = m\gamma + c.
\eeq
An ideal method would have $m=c=0$ for all galaxy types, PSFs, and
observing conditions (size of the galaxy relative to the PSF, and
$S/N$).  Unfortunately, even for existing methods that have $m$ and
$c$ that 
average to nearly zero for some galaxy populations, $m$ and $c$ typically 
vary with all of these factors \citep{2007MNRAS.376...13M,2010MNRAS.405.2044B}, so that in conditions other than
the ones they were tested on, these biases may be significant.

\reftext{It is worth explicitly contrasting the approaches to galaxy
  models that are commonly used for testing software used for weak
  lensing shear estimation.  One common method is to use analytic
  formulae such as S\'ersic profiles, either individually or as
  multi-component models with a bulge and disk.  The clear advantage
  of this approach is that one is in principle only limited by
  computer processing power and storage space in how many simulations
  to generate.  The disadvantage is that these models do not, in
  detail, represent galaxy profiles well.  For example, spiral arms
  are completely unrepresented by such an approach, and higher
  redshift galaxies ($z\gtrsim 1$) are more likely to be highly irregular and
  therefore unrepresented.  \citet{2007MNRAS.380..229M}
  and \citet{2011MNRAS.414.1047Z} showed that there is no
  stable shape measurement algorithm on finite-resolution data that
  has a shear calibration factor that is independent of the galaxy
  population, because the shear operation inevitably couples the
  lower-order moments of a sheared galaxy to higher-order moments
  (which include not just the radial profile, but spiral arms,
  irregularity, etc.).  A simulation containing simple models does not
  capture the higher-order moments of real galaxies, and so we do not
  expect that it will fully test for the (always present) dependence
  of shear calibration on the higher-order structure.  Another example is that the single-component
  models lack ellipticity gradients (changes in the projected
  ellipticity and/or twists of the isophotes).  These features are known to
  exist at a non-negligible level in real
  galaxies (e.g., \citealt{1985MNRAS.216..429L}, \citealt{2006MNRAS.370.1339H}, \citealt{2009ApJS..182..216K}), and cause biases in shear estimation for most extant
  shape measurement methods \citep{2010MNRAS.406.2793B}.  Finally,
  pure-S\'ersic simulations do not test for ``underfitting''
  biases\footnote{These are biases in an $M$-parameter fit to an image
    that arise when the true image has $N>M$ parameters, and some of
    the $N-M$ additional parameters are correlated with parameters of
    interest; see, e.g., \citealt{2010MNRAS.406.2793B}.} in shear measurement methods that fit S\'ersic
  profiles to galaxies.}

\reftext{Another approach is represented by the STEP2 simulations,
  which used shapelets decompositions of COSMOS galaxies (including
  the PSF, which was not removed).  These simulations are therefore 
  intrinsically limited by the cosmic variance in the COSMOS field.
  However, the clear advantage is that in principle, shapelets can (as
an orthonormal basis set) represent {\em any} galaxy features.
Unfortunately, due to the finite signal-to-noise of real data, it is
necessary to truncate the shapelets expansion at some finite order.
The consequence of this limitation has been documented in the
literature (e.g., \citealt{2010A&A...510A..75M}), and results in 
difficulty accurately representing galaxies with high
S\'ersic index, because of the need to represent both the strong inner
cusp and the large-scale wings of the profile.  As shown there, this
modeling difficulty can cause $\sim 20$ per cent biases when 
recovering the shear.  Furthermore, the lower pre-seeing RMS galaxy
  ellipticity in the STEP2 simulations at faint magnitudes, $e_\mathrm{rms}=0.20$ at $r=26$ versus $0.35$
  at $r=22$ \citep{2007MNRAS.376...13M} might arise primarily from the fact
  that the shapelets expansion was limited to a lower order for the
  fainter galaxy sample,  For a circular shapelets basis, the
  restriction to lower order tends
  to give rounder galaxies\footnote{\reftext{In practice, we also expect some
    contribution to this rounding from the fact that the ACS PSF was
    not removed from the COSMOS galaxies.}}.  Support for
  this statement comes from the fact that PSF-corrected COSMOS galaxy
  shapes using a non-shapelets based method \citep{2007ApJS..172..219L} have a roughly flat RMS
  ellipticity as a function of magnitude.  Given that nearly all extant PSF-correction
methods unfortunately have galaxy
population-dependent (and ellipticity-dependent) shear calibrations, we cannot blindly use
simulations that may not accurately represent some non-negligible
part of the
galaxy population to calibrate the shears coming from these methods to
very high precision.} % Thus, using shapelets-based simulations that
%cannot capture those problematic galaxy features may cause
%shapelets-based PSF correction methods (or any others that have
%difficulty with high S\'ersic index galaxies) to appear to perform
%unrealistically well at recovering shears.} 

\reftext{Finally, we consider the approach advocated here, using {\sc
    shera} to represent realistic galaxies.  The current limitation set by the
  cosmic variance in the COSMOS field is unfortunate, but we can
  ameliorate this problem in the future by using 
  the {\it HST} archive to expand the set of galaxies that can be used
  as the basis for simulations, both to other ACS fields and to
  include images from other {\it HST} instruments.  Also, since we are
  using realistic galaxies without modeling assumptions, we are free
  from concerns that some method of shape measurement might appear to
  perform unfairly well because the galaxies were constructed using
  the same set of models used for PSF correction.  More generally, we
  do not have to worry that we have missed features of the galaxy
  population that are problematic for some or all methods of PSF correction. The results of
  \cite{2010MNRAS.405.2044B}, and others cited in Sec.~\ref{S:intro}, 
  strongly suggest that if we want to precisely estimate the bias due to use of 
  some particular shape measurement method in real data, we must include
  realistic galaxies and observing conditions.  However,
this conclusion does not undermine the utility of the STEP and GREAT
simulations.  For example, the GREAT simulations provide a very
well-defined way to test the performance of shape measurement methods
as a function of particular parameters ($S/N$, PSF size,
galaxy radial profile) while keeping other parameters fixed. This test
provides valuable insight into the failure modes of particular
methods, facilitating method development, whereas {\sc shera}
effectively integrates over the parameters of the galaxy profile, PSF
size, and other survey parameters to provide a good overall bias
estimate, without necessarily revealing the details provided by the
GREAT simulations.}

%The multiplicative bias $m$ enters as $m^2$ into cosmic shear
%measurements, since they involve shape auto-correlations, and as $m$
%into galaxy-galaxy or cluster-galaxy lensing.  Furthermore, the
%additive correction $c$ is problematic for cosmic shear, generating an
%additional term in the shear auto-correlations; but for 
%galaxy-galaxy lensing its effects can typically be removed
%using cross-correlation techniques \citep{2005MNRAS.361.1287M}.

\section{Data}\label{S:data}

In this section, we describe the space-based data used as inputs to
the simulation pipeline, and the SDSS data that are being simulated as
a test case.

\subsection{COSMOS}\label{SS:cosmos}

The COSMOS {\it Hubble Space Telescope} ({\it HST}) Advanced Camera for Surveys (ACS) field
\citep{2007ApJS..172..196K,2007ApJS..172....1S,2007ApJS..172...38S} is
a contiguous 1.64 degrees$^2$ region centred at 10:00:28.6, +02:12:21.0
(J2000).  Between October 2003 and June 2005 ({\it HST} cycles 12 and 13),
the region was completely tiled by 575 adjacent and slightly
overlapping pointings of the ACS Wide Field Channel. Images were taken through the
wide F814W filter (``Broad I''). In this paper we use the
`unrotated' images (as opposed to North up) to avoid rotating the
original frame of the PSF. By keeping the images in the default
unrotated detector frame, they can be stacked to map out the observed
PSF patterns. The raw images are corrected for charge transfer
inefficiency (CTI) following \citet{2010MNRAS.401..371M}. Image
registration, geometric distortion, sky subtraction, cosmic ray
rejection and the final combination of the dithered images are
performed by the multidrizzle algorithm \citep{2002hstc.conf..337K}. As
described in \citet{2007ApJS..172..203R}, the multidrizzle parameters
have been chosen for precise galaxy shape measurement in the co-added
images. In particular, a finer pixel scale of $0.03\arcsec/$pix was
used for the final co-added images ($7000\times 7000$ pixels). The source
catalogue used in this paper is constructed following the methodology
outlined in \citet{2007ApJS..172..219L} and then matched to \newtext{an updated version
(v1.7 dated from the 1$^\mathrm{st}$ of August 2009) of} the COSMOS
photometric redshift catalogue presented in \citet{2009ApJ...690.1236I}. For the
purposes of this paper, the following cuts are then applied:
\begin{itemize}
\item $F814W< 22.5$: This cut allows us to start with a parent sample of
galaxies that is deeper than what can be seen in SDSS, but still with
very high $S/N$ in the COSMOS images.  
\item MU\_CLASS $=1$: This requirement \newtext{uses the relationship
    between the object magnitude and peak surface brightness} to select
  galaxies, and to reject stars \newtext{and junk objects such as
    residual cosmic rays} (the exact definition of \textsc{mu\_class} can be found in
  \citealt{2007ApJS..172..219L}).  
\item CLEAN $=1$: As in \citet{2007ApJS..172..219L}, this cut
  is required to eliminate galaxies with defects due to very nearby bright stars, or other similar issues.
\item GOOD\_ZPHOT\_SOURCE $=1$: This cut requires that there
  be a good photometric redshift, which typically is equivalent to
  requiring that the galaxy not be located within the masked regions
  of the Subaru $BVIz$ imaging used for photometric redshifts, and that it have a successful match against an
  object in the Subaru imaging.%, and that $z_{\rm phot}<9$.
\end{itemize}

\begin{figure}
\begin{center}
\includegraphics[width=\columnwidth,angle=0]{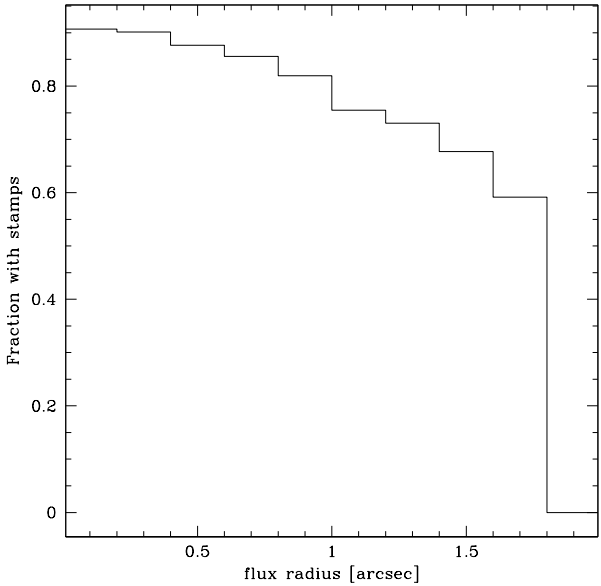}
\caption{\label{F:fracstamps}The fraction of galaxies in the parent
  sample of clean $F814W<22.5$ galaxy detections for which a postage
  stamp was generated, as a function of the COSMOS SExtractor
  FLUX\_RADIUS.}
\end{center}
\end{figure}

The first two cuts give an ideal parent sample of 33~517 galaxies, and
the latter two cuts (which are necessary in practice for manipulating
the images) reduce that to 30~225. Some of these galaxies
are too faint to be detected in SDSS, and some are too small to be
resolved given the size of the SDSS PSF.  For each of these galaxies,
an ideal postage-stamp size is estimated as 
\beq
L {\rm (pixels)} = 11 \sqrt{(1.5 r_{1/2})^2 + \left(\frac{1.2}{0.03
      \times 2.35}\right)^2}.
\eeq
This estimate uses the SExtractor\footnote{\texttt{http://www.astromatic.net/software/sextractor}} \citep{1996A&AS..117..393B} FLUX\_RADIUS (calculated with
PHOT\_FLUXFRAC$=0.5$) as an estimate of the
half-light radius $r_{1/2}$, and (\newtext{in a Gaussian approximation, with
$r_{1/2}\sim 0.7\sigma$}) adds it
in quadrature with an SDSS PSF of FWHM $=1.2$\arcsec, a typical
value. The factor of 0.03 converts the FWHM in arcsec to the COSMOS
pixel scale, and the 2.35 is required to express \newtext{the
  typical SDSS 1.2\arcsec\ PSF FWHM as a Gaussian $\sigma$}. It
then requires that the postage stamp go to more than $\pm 5\sigma$ in
the predicted galaxy size after convolution with the target
PSF.\footnote{Note that FLUX\_RADIUS is an azimuthally-averaged
  quantity.  Thus, for highly flattened objects, we may be concerned
  that PSF convolution will cause them to become so large that they do
  not fit on the generated images.  We test explicitly whether
  this is the case before using the resulting postage
  stamps.}  If this postage stamp size causes the
postage stamp to hit the CCD edge, then the galaxy is eliminated.
Likewise, those galaxies for which the ideal postage stamp size
exceeds 
$L=1000$ pixels were eliminated (119 objects), resulting in an
intrinsic upper limit in the values of FLUX\_RADIUS for which postage
stamps were generated.  
Consequently, the probability of a galaxy in our
parent sample having a postage stamp is a weak function of the observed
galaxy size, specifically the FLUX\_RADIUS.  This probability is shown
in Fig.~\ref{F:fracstamps}; when computing statistics of the sample,
we must weight by the inverse of this curve to remove this artificial
selection effect and obtain a
fair galaxy sample.

After these cuts to ensure that the postage stamp can be generated,
the sample for which there are postage stamps contains 26~113 galaxies.

\subsection{SDSS}\label{SS:sdss}

The SDSS \citep{2000AJ....120.1579Y} imaged roughly $\pi$ steradians
of the sky, and followed up approximately one million of the detected
objects spectroscopically \citep{2001AJ....122.2267E,
  2002AJ....123.2945R,2002AJ....124.1810S}. The imaging was carried
out by drift-scanning the sky in photometric conditions
\citep{2001AJ....122.2129H, 2004AN....325..583I}, in five bands
($ugriz$) \citep{1996AJ....111.1748F, 2002AJ....123.2121S} using a
specially-designed wide-field camera
\citep{1998AJ....116.3040G}. These imaging data were used to create
the cluster and source catalogues that we use in this paper.  All of
the data were processed by completely automated pipelines that detect
and measure photometric properties of objects, and astrometrically
calibrate the data \citep{2001ASPC..238..269L,
  2003AJ....125.1559P,2006AN....327..821T}. The SDSS I/II imaging surveys were completed
with a seventh data release \citep{2009ApJS..182..543A}, though this
work will rely as well on an improved data reduction pipeline that was
part of the eighth data release, from SDSS III \citep{2011ApJS..193...29A}.

%\textbf{Update with new redux:} 
%The catalog of source galaxies with shape measurements that we are
%simulating in this work was originally described in
%\citet{2005MNRAS.361.1287M}.  This source sample has over 30 million
%galaxies from the SDSS imaging data with $r$-band model magnitude
%brighter than 21.8, with shape measurements obtained using the REGLENS
%pipeline, including PSF correction done via re-Gaussianization
%\citep{2003MNRAS.343..459H} and with cuts designed to avoid various
%shear calibration biases.  The overall calibration uncertainty due to
%all systematics was originally estimated to be eight per cent
%\citep{2005MNRAS.361.1287M}, though the redshift calibration component
%of this systematic error budget has recently been decreased due to the
%availability of more spectroscopic data \citep{2008MNRAS.386..781M}.
%In this work, we hope to more tightly constrain the shear calibration
%(including PSF rounding and selection biases) than was originally
%possible.

\subsection{Shape catalogue}\label{SS:shapecatalog}

The catalogue of source galaxies with shape measurements that we are
simulating in this work is described in Reyes et al. (2011, in prep.),
and is an update of that originally described in
\citet{2005MNRAS.361.1287M} with additional area and several technical
improvements.  This source sample has over 42 million galaxies from
the SDSS imaging data with $r$-band model magnitude brighter than
21.8, with shape measurements obtained using the REGLENS pipeline,
including PSF correction done via re-Gaussianization
\citep{2003MNRAS.343..459H} and with cuts designed to avoid various
shear calibration biases. Among those cuts are a flux limit of
$r<21.8$, and the requirement that the PSF-corrected shape be measured
in both $r$ and $i$ bands with sufficient resolution (to be defined
more quantitatively later in this section). 

Using the software developed in this work,
we hope to more tightly constrain the shear calibration, including the
full list of possible biases from Sec.~\ref{S:psfcorrection}, than was
originally possible in \citet{2005MNRAS.361.1287M} \newtext{(which had allowed
for an overall $2\sigma$ shear calibration uncertainty of $[-5, +12]$
per cent for $r<21$ galaxies, and $[-8, 18]$ per cent for $r>21$ galaxies).}

%The old source catalog was based on SDSS rerun 137 ({\sc Photo
%  v}5\_4), and the new is based on rerun 301 ({\sc Photo v}5\_6).  The
%paper describing the catalog (Reyes et al. 2011, in prep.) will detail
%the changes that impact the shear systematics.  For this paper, we
%simply state briefly that the multiplicative shear calibration bias
%due to PSF dilution changes by $\sim 1.5$ per cent when changing from the
%old catalog to the new one.  Because the changes in shear calibration
%are well-understood, in this paper we simply simulate the SDSS images
%and process them using the software used to generate the new shape
%catalog, keeping in mind that this is equivalent to the old shape
%catalog to within certain well-understood limits.

One of the technical difficulties that complicates any direct comparison of
these shape measurements with those from the COSMOS catalogue is the
definition of the shapes.  Both the RRG method and
re-Gaussianization define ellipticity in terms of a moment matrix
${\bf M}$ via
\beqa
e_1 &=& \frac{M_{xx}-M_{yy}}{M_{xx}+M_{yy}} \nonumber \\
e_2 &=& \frac{2M_{xy}}{M_{xx}+M_{yy}}\;,
\eeqa
which translates to an ellipticity definition
\beq\label{E:ellipdef}
|e| = \frac{1-q_\mathrm{eff}^2}{1+q_\mathrm{eff}^2}
\eeq
for some effective minor-to-major axis ratio $q_\mathrm{eff}\equiv
b_\mathrm{eff}/a_\mathrm{eff}$.  However, the determination of the
moment matrix ${\bf M}$, and therefore the $e_1$, $e_2$, and $q_\mathrm{eff}$, 
is done differently for the two methods.

 In general, the definition of moments
according to each method uses
\beq\label{E:xxmom}
M_{ij}^{\rm (method)} = \int I({\bmath x}) w_{\rm method}({\bmath x})
({\bmath x}-{\bmath x_0})_i ({\bmath x}-{\bmath x_0})_j \rmd{\bmath x}.
\eeq
The RRG method, used to PSF-correct the COSMOS galaxy shapes, defines
second moments with a weight
function 
\beq
w_{\rm RRG}({\bmath x}) = \exp{\left(-\frac{x^2 + y^2}{r_w^2}\right)},
\eeq
\newtext{where $r_w$ is a galaxy size estimate calculated from the
  SExtractor detection area.  Thus, this method uses a
{\em circularly-weighted} moment with a fixed radius.}  In contrast, the re-Gaussianization
method uses adaptive moments, which entails minimizing the integral
\beq
E = \frac{1}{2} \int \left| I({\bmath x}) - A\exp\left[
-\frac{1}{2} ({\bmath x}-{\bmath x}_0)^T {\bf M}^{-1}
({\bmath x}-{\bmath x}_0) \right] \right|^2 \rmd^2{\bmath x}
\eeq
over the quantities $(A,{\bmath x}_0,{\bf M})$.  This procedure amounts to
weighting by $w^\mathrm{(adapt)}({\bmath x})$ corresponding to the
best-fitting {\em elliptical} Gaussian that represents the image itself,
which in practice is determined iteratively.   

Analytical calculations show that for an elliptical Gaussian 
profile, the difference between the circular vs. elliptical weight
functions means that the two ellipticities $|e^{\rm (RRG)}|$ and $|e^{\rm
  (adapt)}|$ will be related as
\beq\label{E:relateellip}
|e^{\rm (adapt)}| = \frac{2 |e^{\rm (RRG)}|}{1+|e^{\rm (RRG)}|}.
\eeq
Furthermore, for non-Gaussian light profiles, $e^{\rm (adapt)}$ does
not depend on any assumed radius whereas $e^{\rm (RRG)}$ does; more
problematically, for a profile with fixed axis ratio, changing
the profile from Gaussian to a more general profile with elliptical
isophotes (e.g. S\'ersic profiles) does not modify 
$|e^{\rm (adapt)}|$ whereas it does change $|e^{\rm (RRG)}|$.
Finally, in the presence of ellipticity gradients with radius, the different way
of choosing the effective radius of the weight function will result in
different measured ellipticities.  So, we
cannot compare individual estimates of the galaxy shapes from the two
PSF correction methods in any obviously model-independent way.   We therefore conclude that
the way forward is a simulation of the ground-based image using the
high resolution COSMOS image, in order to directly test the accuracy
of shear
recovery. 

 For the purpose
of this work, we define the `resolution factor' $R_2$ using the
trace of the adaptive moment matrices, 
\beq
T = M_{xx} + M_{yy}
\eeq
where $T^{(P)}$ and $T^{(I)}$ are the traces for the PSF and for the
PSF-convolved galaxy image, respectively.  Then the resolution factor
is 
\beq
R_2 = 1 - \frac{T^{(P)}}{T^{(I)}}\;,
\eeq
which approaches $1$ in the limit that the galaxy is perfectly
resolved, and $0$ in the limit that it is completely unresolved.  Our
requirement on the resolution factor is $R_2 > 1/3$. In the
limit of Gaussian PSF and galaxy \reftext{with standard deviations
$\sigma_\mathrm{gal}$ and $\sigma_\mathrm{PSF}$, respectively, this
resolution factor cut} corresponds to $\sigma_\mathrm{gal} >
\sigma_\mathrm{PSF}/\sqrt{2}$.

\section{Simulation methodology}\label{S:methodology}

In this section, we discuss the principles behind simulations of
lower-resolution (ground-based) data from higher-resolution
(space-based) data.  First, we define some notation.

We assume that a galaxy is described by an intrinsic unknown surface brightness function $f({\bf x})$ as a function of angular position ${\bf x}$.
We are given a high-resolution image (COSMOS) with some effective PSF $G_1({\bf x})$, i.e. the observed surface brightness is
\beq\label{E:highresim}
I_1({\bf x}) = [f\star G_1]({\bf x}) = \int f({\bf x}')G_1({\bf x}-{\bf x}')\,\rmd^2{\bf x}'.
\eeq

We would like to generate a low-resolution image $I_2$ (corresponding to
what would be observed by some ground-based telescope with PSF $G_2$),
\beq\label{E:lowresim}
I_2({\bf x}) = [f\star G_2]({\bf x}) = \int f({\bf x}')G_2({\bf
  x}-{\bf x}')\, \rmd^2{\bf x}'.
\eeq
While this equation may initially appear to represent a trivial
PSF-matching process (Sec.~\ref{SS:psfmatching}), there is a
complicating factor that arises if we want to represent a sheared
image (Sec.~\ref{SS:shear}), since shearing and convolution by the
space-based PSF do not commute.

\subsection{PSF matching}\label{SS:psfmatching}

First, we consider the case in which we simply wish to generate a
low-resolution image without any added shear.  In that case, the
task of generating a low-resolution image $I_2$ is simply a matter of
PSF-matching.  The simplest model-independent way to do this is to
work in Fourier space\footnote{We indicate 
Fourier-space quantities with a tilde, and the distances
in pixels in real and Fourier space are ${\bmath x}$ and ${\bmath k}$,
respectively.}. 
In that case, the convolutions with the PSF are multiplications:
\beq
\tilde{I}_1({\bf k}) = \tilde{f}({\bf k}) \tilde{G}_1({\bf k})
\eeq
and likewise for the low resolution image $\tilde{I}_2$.

\begin{figure}
\begin{center}
\includegraphics[width=\columnwidth,angle=0]{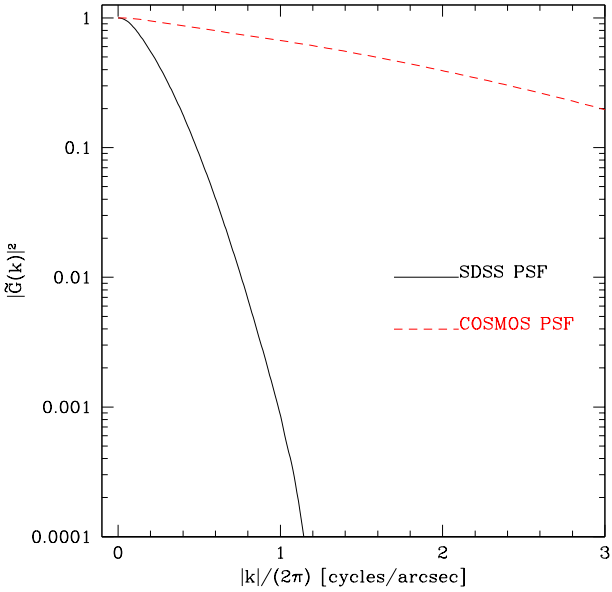}
\caption{\label{F:ftscales}An example of the relevant scales for the
  PSF in SDSS and COSMOS at a randomly-selected point in the COSMOS field.  The
  plotted quantity is the azimuthally-averaged PSF power
  $|\tilde{G}|^2$, as a function of wavenumber.  As shown, the
  band limit of the SDSS data is on scales where the Fourier transform
  of the COSMOS PSF is still close to $1$. 
  %The cycles/arcsec I think actually refers to k/2$\pi$ -- C.H.
  }
\end{center}
\end{figure}

\begin{figure}
\begin{center}
\includegraphics[width=\columnwidth,angle=0]{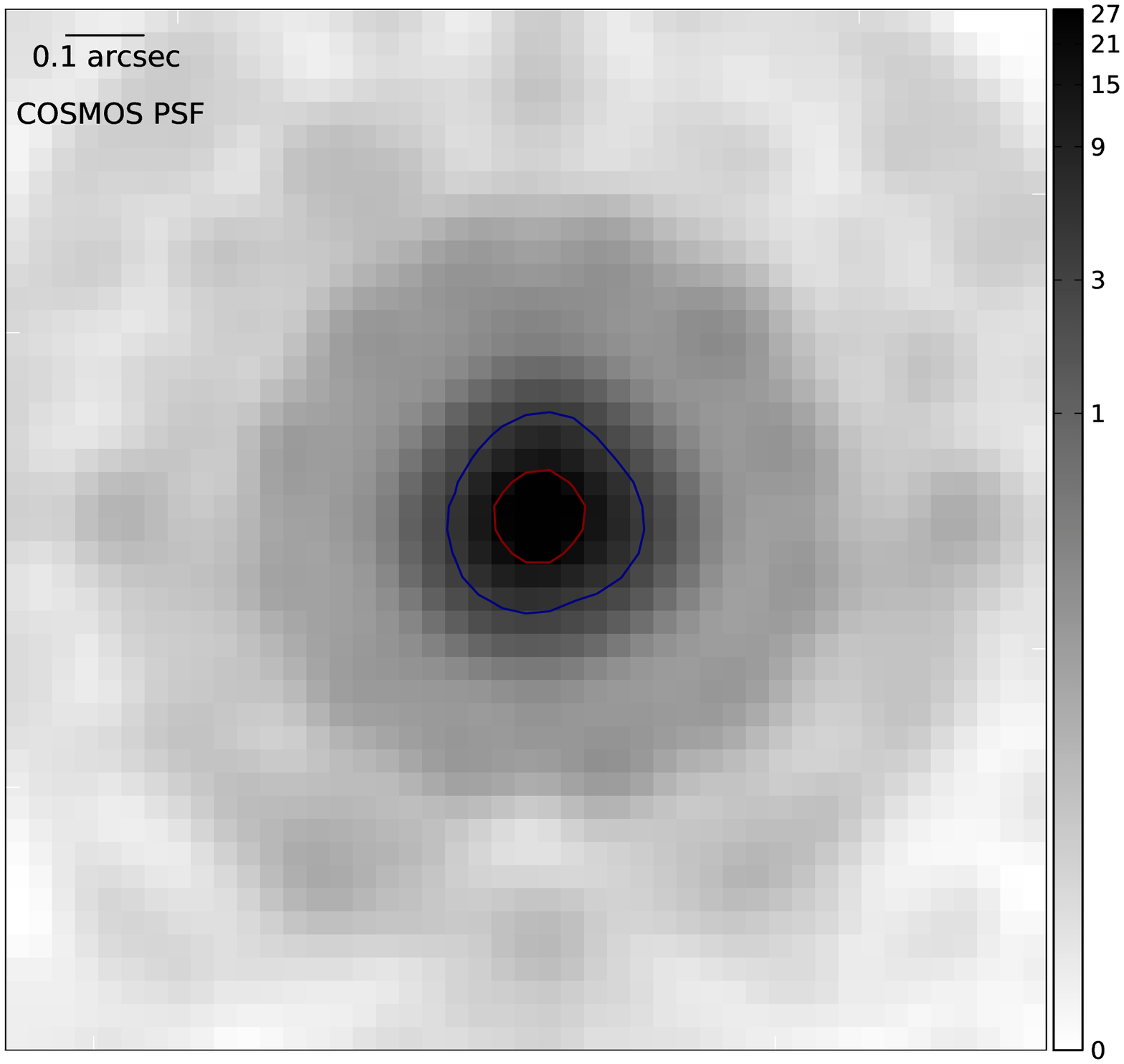}
\includegraphics[width=\columnwidth,angle=0]{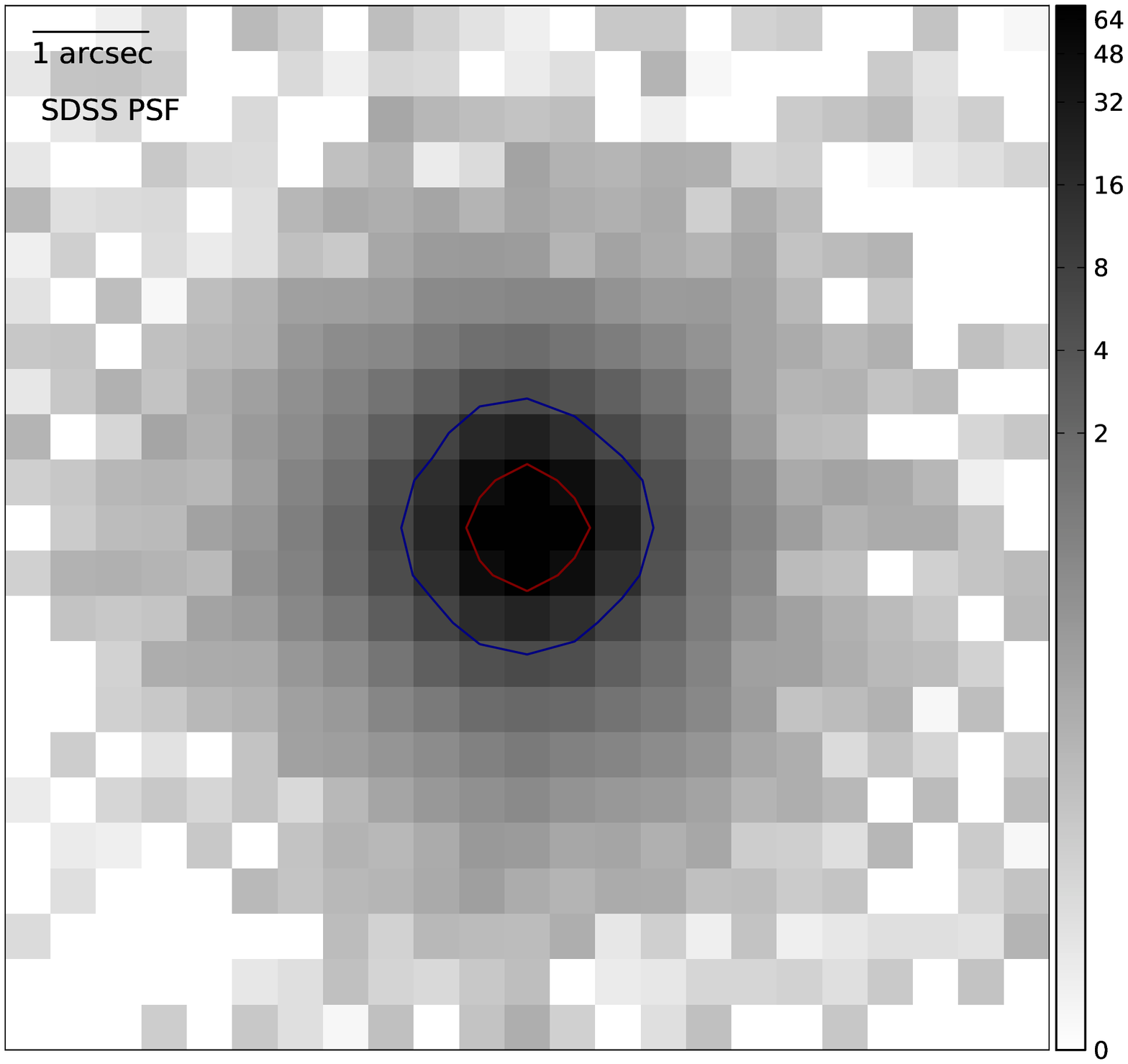}
\caption{\label{F:psfs}\newtext{{\em Top:} Real-space image of the
    Tiny Tim COSMOS PSF
  for which the azimuthally averaged Fourier space power was shown in
  Fig.~\ref{F:ftscales}.  The image is shown on a logarithmic stretch,
  in order to display the diffraction rings.  Contours are shown for a
  flux level equal to $0.5$ and $0.1$ times the maximum flux level. {\em Bottom:} Same as top, for
  the SDSS PSF; the low-level patterns in the outer regions are due to
  a small amount of noise  in the
   PSF model.}
  }
\end{center}
\end{figure}

In that case, once we have a Fourier-space image $\tilde{I}_1$, original PSF
$\tilde{G}_1$ and target PSF $\tilde{G}_2$, we can simply generate the
low-resolution image via
\beq\label{E:psfmatch}
\tilde{I}_2 = \left(\frac{\tilde{G}_2}{\tilde{G}_1}\right) \tilde{I}_1.
\eeq
Naturally, we must place some conditions on the low- and
high-resolution PSFs in order to carry out this PSF matching.  As a
rule, PSFs are band-limited at some wavenumber $k_c$ above which there
is essentially no power (this corresponds to the small scales below
which there is no information about the galaxy profile).  By
definition, the PSFs in high and low resolution data tend
to satisfy $|\tilde{G}_1| >|\tilde{G}_2|$ at all wavenumbers $k$, with
the band limit $k_{c,1} > k_{c,2}$; for an example of how this
relation is satisfied for typical COSMOS and SDSS data, see
Fig.~\ref{F:ftscales}, \newtext{and the corresponding real-space PSF images in Fig.~\ref{F:psfs}}.  In fact, this inequality is a requirement for
numerically stable and model-independent PSF-matching; if it is not satisfied, then
the operation in Eq.~\eqref{E:psfmatch} corresponds to deconvolution
for at least some wavenumbers, which 
will lead to undesired image properties such as
ringing.  This deconvolution can be done in the context of
model-fitting methods; however, the meaning of the small-scale power
recovered in the process of such a deconvolution is unclear \citep{2010MNRAS.406.2793B}.

\subsection{Introducing a shear}\label{SS:shear}

Now, we consider the less trivial case where we want to simulate a ground-based
image with an added shear, for the purpose of testing PSF correction.
We denote this sheared ground-based image $I_2^{(\gamma)}({\bf x})$ to
distinguish it from the unsheared ground-based image $I_2({\bf x})$
considered in the previous subsection.

To define the problem more clearly, we imagine a Jacobian matrix ${\bf
  J}$ that transforms the observed (post-shear) coordinates ${\bf
  x}_{\rm o}$ to the intrinsic (pre-shear) coordinates ${\bf x}_{\rm
  i}$ on a galaxy:
\beq
{\bf x}_{\rm i} = {\bf Jx}_{\rm o}.
\eeq

The Jacobian ${\bf J}$ is thus a $2\times2$ matrix.  Usually ${\bf J}$
will be close to the identity; indeed, to first order in the shear,
${\bf J}$ is simply given by Eq.~\eqref{E:jacobian}.  We will denote the singular values of ${\bf J}$ by $S_\pm$, with
$S_-\le S_+$, and $\det{\bf J}=S_-S_+$.  If ${\bf J}$ is
symmetric, then $S_\pm$ are also the eigenvalues.

We then have a sheared galaxy image
\beq
f_{\rm o}({\bf x}_{\rm o}) = f({\bf Jx}_{\rm o}),
\eeq
and the observed sheared image is
\beq
I_2^{(\gamma)}({\bf x}_{\rm o}) = [f_{\rm o}\star G_2]({\bf x}_{\rm o}) = \int f({\bf Jx}') G_2({\bf x}_{\rm o}-{\bf x}')\,\rmd^2{\bf x}'.
\eeq

It is assumed that the $S/N$ of the input images is large enough that
the  noise is negligible.  (Clearly the output image may be made
to have arbitrary levels of noise by adding noise at the end.)  For
the situation considered here, we thus limit ourselves to relatively
bright detections in COSMOS ($S/N \gtrsim 50$).  \newtext{In
Sec.~\ref{SS:cosnoise}, we demonstrate the effects of this low
level of noise in the input images 
on the simulated images, given that it is sheared and
therefore could contribute to an estimated shear.}  We defer the development of a
formalism to account for non-negligible noise levels in the input
high-resolution data to future work
that may use lower $S/N$ space-based data.

We assume that the PSFs $G_1$ and $G_2$ satisfy the following band-limiting constraints: first, that
$|\tilde G_2({\bf k})|=0$ (or at least is negligible) for $|{\bf
  k}|>k_c$ for some $k_c$ (the band limit); and second,
that $|\tilde G_1({\bf k})|$ is nonzero (and in practice we assume it
is far from zero, e.g. $\gtrsim 0.5$) for $|{\bf k}|<k_d$ for
some $k_d$.  We impose the assumption that
\beq
k_c < S_-k_d,
\eeq
which in practices amounts to requiring a significant range of scales
on which the ground-based PSF erases all information that is still
resolvable in the space-based images.

While the PSF-matching problem can be simply formulated as trying to
take $I_1({\bf x})$ and obtain $I_2({\bf x})$, the case where we want
to introduce a shear for testing purposes (a simulated gravitational
shear, which changes the galaxy image {\em before} the imposition of
the PSF) instead requires us to infer $I_2^{(\gamma)}({\bf x}_{\rm o})$.

The principle behind the solution is to attempt a partial
deconvolution of $I_1$.  We now define a filter $T({\bf x})$ that has
a Fourier transform satisfying
\beq
\tilde T({\bf k}) = \frac1{\tilde G_1({\bf k})} \,\, {\rm for}\,\,
|{\bf k}|<k_d.
\eeq
(Here we have used our band-limiting condition on $G_1$.)  For $|{\bf k}|\ge k_d$ an arbitrary choice of $\tilde T({\bf k})$ may be made, e.g. it may be taken to 
decrease smoothly to 
zero so that the real-space function $T({\bf x})$ does not have tails at large radii.

Then we define the {\em pseudo-deconvolved} image $P({\bf x})$ by
\beq\label{E:pdconvimg}
P({\bf x}) = [I_1\star T]({\bf x}) = \int I_1({\bf x}')T({\bf x}-{\bf x}')\,\rmd^2{\bf x}',
\eeq
or in Fourier space,
\beq\label{E:fpdconvimg}
\tilde P({\bf k}) = \tilde I_1({\bf k})\tilde T({\bf k}) = \tilde f({\bf k}) \tilde G_1({\bf k}) \tilde T({\bf k}),
\eeq
so that $\tilde P({\bf k}) = \tilde f({\bf k})$ for $|{\bf k}|<k_d$.

Our second step is to shear the pseudo-deconvolved image, i.e. we create
\beq
P_{\rm o}({\bf x}_{\rm o}) = P({\bf Jx}_{\rm o}).
\eeq
The Fourier transform is given by the usual rule for the transform of a quantity with a linear shear,
\beq\label{E:linshearft}
\tilde P_{\rm o}({\bf k}_{\rm o}) = \frac1{|\det{\bf J}|}\tilde P({\bf J}^{T\,-1}{\bf k}_{\rm o}).
\eeq
The singular values of ${\bf J}^{T\,-1}$ are $S_\pm^{-1}$.
Therefore, we see that if $|{\bf k}_{\rm o}|<k_c$, then:
\beq
|{\bf J}^{T\,-1}{\bf k}_{\rm o}|\le S_-^{-1}|{\bf k}_{\rm o}|<S_-^{-1}k_c<k_d.
\eeq
Therefore,
$\tilde P_{\rm o}({\bf k}_{\rm o}) =\tilde f_{\rm o}({\bf k}_{\rm o})$ for $|{\bf k}_{\rm o}|<k_c$.

Finally, we convolve $P_{\rm o}$ with the target low-resolution PSF $G_2$ to get
\beq
H({\bf x}_{\rm o}) = [P_{\rm o}\star G_2]({\bf x}_{\rm o}) = \int P_{\rm o}({\bf Jx}') G_2({\bf x}_{\rm o}-{\bf x}')\,\rmd^2{\bf x}'.
\eeq
In this case, we have
\beq
\tilde H({\bf k}_{\rm o}) = \tilde P_{\rm o}({\bf k}_{\rm o}) \tilde G_2({\bf k}_{\rm o}).
\eeq
There are now two cases: $|{\bf k}_{\rm o}|$ is either (i) $<k_c$ or (ii) $\ge k_c$.  We consider each of these:
\begin{list}{$\bullet$}{}
\item If $|{\bf k}_{\rm o}|<k_c$, then $\tilde P_{\rm o}({\bf k}_{\rm o})=\tilde f_{\rm o}({\bf k}_{\rm o})$ so $\tilde H({\bf k}_{\rm o}) = \tilde I^{(\gamma)}_2({\bf k}_{\rm 
o})$.
\item If $|{\bf k}_{\rm o}|\ge k_c$, then $\tilde G_2({\bf k}_{\rm o})=0$ so $\tilde H({\bf k}_{\rm o}) = \tilde I^{(\gamma)}_2({\bf k}_{\rm  
o})=0$.
\end{list}
In either case, we have $\tilde H({\bf k}_{\rm o}) = \tilde I^{(\gamma)}_2({\bf k}_{\rm  
o})$, so it follows that $H({\bf x}_{\rm o})=I^{(\gamma)}_2({\bf x}_{\rm o})$.  Therefore, the function $H$ that we have constructed is exactly
the sheared image that would be observed with the low-resolution telescope.

\subsection{Other observational issues}

To demonstrate that the PSF-matching is accurate, we will
determine the target PSF and noise level using the observing
conditions at the position of the COSMOS galaxy in the SDSS imaging.
This procedure will allow us to compare the simulations with the real
SDSS imaging in the COSMOS field, as a basic test of the {\sc shera}
code.  
Likewise, we will use these simulations to determine the shear
calibration bias, as a demonstration of the method.  However, in order
to make a fair test of the shear calibration of the entire SDSS shear catalogue, it
would be necessary to draw random points from within the SDSS area and
use the observing conditions from those points.  The difference
between these two approaches would not be significant if the quality
of the SDSS imaging in the COSMOS field was representative, but as discussed in Appendix~\ref{S:sdssprops}, this is not
the case.

Also, in order to fairly test the shear calibration, we must impose
any selection criteria from the real data on the simulations.  This
will also allow for a test of selection biases, since the input galaxy
sample is a fair sample of all galaxies brighter than some flux limit
($F814W<22.5$) that is deeper than the SDSS flux limit in the single epoch
images ($r<21.8$).

Finally, one limitation of these COSMOS images that we use as the
basis for our simulations is that they only are in $F814W$ (broad $I$).
The existence of colour gradients means that the galaxies would look
morphologically different in other passbands, and the effect is probably the strongest
for galaxies with a reasonable-sized bulge and disk, for which the
bulge is more prominent in red bands and the disk in blue bands.
However, since most lensing analyses use $r$ or $I$ for shape
measurement, this does not represent a significant limitation \citep{2011arXiv1105.5595V}.  It is
an argument, however, for using another field with multi-band data,
provided that (a) the CTI effects (see section \ref{SS:cosmos}) are well-understood and corrected
for properly, and (b) the field was chosen in some fair way (e.g., it
is not a galaxy cluster field, which would have an atypical morphology
distribution).

For our purposes in SDSS, we can use simulations with a fair
distribution of observational conditions to precisely
determine the $i$-band shear calibration.  Since our science analyses
use averaged $r$ and $i$ band shapes, we can then use the data itself
to determine the ratio of the measured signal to that with just $i$
band.  The measurements in $r$ and $i$ are highly correlated because the shape noise
is the same, so this allows the shear calibration for the actual
science analyses to be determined very precisely.

\section{Image preparation}\label{S:imageprep}

Before we can actually carry out the image simulations, there are a
number of steps that must be done to process the galaxy postage stamp
images described in Section~\ref{SS:cosmos}.  These steps
have been carried out already for the images that are being
publicly
released\footnote{\texttt{http://irsa.ipac.caltech.edu/data/COSMOS/images/\linebreak
    galaxy\_postage\_stamps/}}.
\newtext{Thus, these newly released images differ from the already-public
version 2 COSMOS images in that they are restricted to a bright
subsample of galaxies, they have
a smaller post-processing pixel scale (0.03''), they include the pixel-based CTI
correction, and include the post-processing described in the remainder
of this section.} 

\subsection{Catalogue of galaxy properties}\label{SS:catalogue}

We have used the COSMOS data to prepare a number of inputs that can be
used for simulations of SDSS $i$-band images.

First, we need a total galaxy flux in $i$ band, whereas the COSMOS
images are in $F814W$.  A significant fraction of the galaxies that
constitute our parent sample are detected in SDSS, and therefore have
measurements of the $i$-band flux.  However, these measurements are
far noisier than the flux measurements from the COSMOS data, so we
will use the COSMOS $F814W$ fluxes to determine the normalization of the
flux in the simulated images.\footnote{If we use the SDSS magnitudes,
  then we include noise in the measurement twice, since the SDSS
  magnitude measurements are noisy due to the sky noise that we then
  put into the simulations.}  We begin with the reported SExtractor
MAG\_AUTO magnitudes in $F814W$, which we
correct for galactic extinction using the dust maps from
\cite{1998ApJ...500..525S} and the extinction-to-reddening ratios from
\cite{2002AJ....123..485S}.  These magnitudes are designed to
precisely determine the total magnitudes for galaxies, similar to Kron
magnitudes \citep{1980ApJS...43..305K}, and are superior to aperture
magnitudes in recovering all the galaxy flux.

In order to account for slight differences in the two filters, we then compare the extinction-corrected COSMOS
MAG\_AUTO and SDSS model magnitudes for reasonably bright galaxies
($i<20$), and determine a mean offset of $0.03$ mag, $i = F804W-0.03$.  This mean offset is then
subtracted from the $F814W$ magnitudes for all galaxies in the parent
sample.  %One important subtlety for comparisons of COSMOS
%and SDSS photometry is that in the SDSS reduction used for this work, about 33 per cent
%of the COSMOS area was considered to be non-photometric according to
%the ubercalibration method\footnote{As indicated by a calibration
%  status flag CALIB\_STATUS$\ne 1$.} \citep{2008ApJ...674.1217P}.  Thus, we are careful
%to restrict only to those regions for which the photometric
%calibration was considered reliable in SDSS to determine the mean
%offset between the photometric systems. 

\subsection{Postage stamp preparation}\label{SS:psprep}

There are several types of processing that must be done to the
original CTI-corrected galaxy
postage stamps before using them as inputs to the simulations.  For
this purpose, we use SExtractor \citep{1996A&AS..117..393B} {\sc
  v2.5.0}.  First, we run SExtractor with the COSMOS multidrizzle
weight map for each postage stamp, instructing it to subtract
off a flat background level equivalent to the residual background
after the original multidrizzle processing was completed.  We do not
allow it to carry out its own sky determination because the size of
the postage stamps is not sufficient for it to do so without being
unduly influenced by the light of the galaxies in the postage stamp.

The results of this step provide us with a list of detected objects in
the postage stamp.  Ideally, if deblending is properly done, then one
of those (our target galaxy) will be at the
centre of the postage stamp.   We use the SExtractor output
segmentation image to identify all pixels belonging to 
objects other than the target galaxy, and replace those pixels with a noise field having the same
noise characteristics as the rest of the postage stamp (including
overall noise amplitude as well as pixel-to-pixel correlations).
\newtext{The number of masked objects is $<18$ ($<60$) for 50 (95) per
cent of target galaxies, resulting in $0.9$ ($6$) per cent of the
pixels being masked; the majority of the masked objects are quite
small and faint, with some appearing to be misidentification of the
correlated noise as actual objects.}  
This object masking procedure is dependent upon SExtractor correctly
identifying all pixels belonging to other objects; it is not fully
successful with some very bright objects, leaving a halo of pixels
containing a low, but visually noticable, level of light surrounding
the masked regions.  Fortunately the incidence of such cases is low. 
These processed postage stamps are included with the code and data release.

While carrying out this procedure, we compute additional statistics
for each postage stamp, including (a) the distance of the nearest
object to the postage stamp centre (which we require to be $\leq 5$
pixels, after visual inspection of cases failing this cut suggested
that those cases suffered from poor deblending), and (b) various noise
statistics such as the median and the mean pixel value for those
pixels not included in objects (which can differ significantly if
there is some very large bright object in the postage stamp that did
not get properly masked).  Imposing cuts based on these statistics
reduces the size of our sample from 26~113 to 25~527 postage stamps, a
decrease in sample size of 2.2 per cent.

If the masked objects are sufficiently close to the central
galaxy, then in the SDSS it will not be possible to distinguish
between them.  This fact will allow us to quantify the level of
undeblended projections in our shape catalogue: we identify those cases
for which the apparent size of the galaxy in the SDSS is significantly
larger than its counterpart in the PSF-matched images from COSMOS.

\subsection{COSMOS PSF estimation}\label{SS:cospsf}

In order to remove the effects of the COSMOS PSF, we must determine
the COSMOS PSF at the galaxy position.  We follow the same procedure
as in \cite{2007ApJS..172..219L} and \cite{2007MNRAS.376...13M}, who use PSF models
from a modification of version 6.3 of the Tiny Tim ray-tracing
program\footnote{\texttt{http://www.stsci.edu/software/tinytim/}}.
These models represent PSFs for different \newtext{primary/secondary 
separation, since that separation 
is the main determinant of the PSF ellipticity.  
They are known to be a bit too small because they neglect the
`red halo' that enlarges real {\em HST} PSFs at long wavelengths \citep{1998SPIE.3355..608S}, possibly
due to backscattering off the front surface of the CCD. We 
quantify whether the deviation between these models and the real
stellar images represents a problem for using the models to simulate
ground-based data in Sec.~\ref{SS:testtinytim}.}

Our procedure is to use the previous determination
\citep{2007ApJS..172..219L} of the primary/secondary separation for each exposure
based on the ellipticity of $\sim 20$ bright stars, then to use that
particular Tiny Tim model extrapolated to the CCD $(x, y)$ position of
each galaxy.    The estimated Tiny Tim PSF for each galaxy position is
included with the data release associated with this work.

%Other PSF estimation schemes have been suggested using dense stellar fields
%\citep{2007A&A...468..823S} and indeed one might wonder, for this
%particular application in which we use the COSMOS PSF to achieve
%PSF-matching to some desired ground-based PSF, how much does a deviation between
%the Tiny Tim PSF models and the true observed PSF constrain how well we
%can accurately simulate a particular ground-based PSF.  For several
%example cases with dense star fields, we found that around the
%band-limit of the SDSS data, the Tiny Tim models consistently
%described the PSF to within $\sim 1$ per cent, typically being
%slightly more compact by this amount (i.e.,
%$|G_\mathrm{real}(k=1/\mathrm{arcsec})/G_\mathrm{TT}(k=1/\mathrm{arcsec})|^2
%\approx 1.01$.  The deviations were more significant on scales that
%are more than $3$ times smaller than that; however, these are
%irrelevant for the purpose of mimicking ground-based data, since they
%cannot be resolved anyway.

\subsection{Determination of target imaging properties}

\newtext{In this section, we describe what information must be specified about
imaging conditions in the data that is to be simulated.  Here, we
restrict ourselves to a general discussion; however, for the purpose
of testing and demonstrating the code, we attempt to simulate the
SDSS data in the COSMOS field.  Appendix~\ref{S:sdssprops} contains details of exactly what SDSS pipeline
outputs are used.}

\newtext{\textbf{PSF:} The desired PSF for the simulated data must be provided as a
postage-stamp image.  It should be the desired PSF including the pixel
response function; e.g., if a Gaussian PSF is of interest as a test
case, the Gaussian should be integrated within pixels, rather than
sampled at the pixel centres.}

\newtext{\textbf{Photometric calibration:} The most basic case would be to
simulate data using the $F814W$ magnitudes.  To do so, the code
requires the total flux normalization in counts, as determined from
the COSMOS total galaxy magnitude and the flux normalization for the
ground-based survey of interest.}  %However, the catalogs released with
%the code also provide the magnitudes with a small offset to account
%for the difference between $F814W$ and SDSS $i$.}

\newtext{\textbf{Noise level:} The default noise model is a spatially constant,
uncorrelated Gaussian random field.  For most ground-based surveys,
the sky level is sufficiently high that the Poisson noise due to the
sky is effectively Gaussian; and most galaxies used for weak lensing
are sufficiently faint that the sky noise dominates over the noise due
to the galaxy flux.  Thus, the code simply requires a single noise
variance to work in this basic mode.  However, it also allows the
choice of Poisson noise, and the user may optionally input a gain in
order to also add the noise due to the galaxy flux (important for
relatively bright galaxies).  
Future versions of the code may
allow the user to simulate a correlated noise field
with a user-defined noise power spectrum, which will be important for
assessing the impact of correlated noise on shear estimation.}

\newtext{The simulated postage stamps have a sky level of zero.  Any constant
or varying sky should be added by the user after running {\sc shera}.
}

\section{Implementation}\label{S:implementation}

For compatibility with many common astronomical image manipulation
packages, we have implemented this simulation method in IDL.  The data
release includes a detailed description of the code options; here, we
limit ourselves to a basic description of how the code carries out the
procedure from Sec.~\ref{S:methodology}.

The {\sc shera} code operates on a set of input postage stamps: the
original COSMOS image and PSF, and a target PSF to which we want to
match.  The manipulation of the images to create the simulated
galaxies is performed using double precision arithmetic.

The first step in the image manipulation is to change the sizes of the
input postage stamps of the COSMOS galaxy, COSMOS PSF, and target PSF
due to several considerations.  When doing the
Fourier space manipulations, it is convenient to have the ratio of the
COSMOS and the target PSF postage stamp sizes be equal to the ratio of
the target and COSMOS pixel sizes, which is
$ \rpix = 0.396\arcsec/0.03\arcsec$ in the case of SDSS
simulations (however, the code allows for nearly arbitrary choice of
target pixel size, such that $\rpix>1$).  Likewise, it will be most convenient when working
in Fourier space if the COSMOS PSF and COSMOS galaxy postage stamp
sizes are equal.

Thus, we begin by padding the arrays until they achieve the
appropriate size ratios.  While the default is to pad with zeros, we
also provide the option of padding with a realistic COSMOS noise
field.  
Once we have a target PSF postage stamp of size $\nt\times \nt$
and COSMOS galaxy and PSF postage stamps of size $\nc\times \nc$, with
$\nc=\rpix \nt$, we can proceed with the analysis.  To begin, we
renormalize the flux in the PSF postage stamps so that the sum of the
flux in all pixels is $1$.

In the description that follows, we denote the observed galaxy images
using $I$ (with subscript C for their image in COSMOS and T for the
simulation of the target dataset), and PSFs using $G$ (again with subscripts to
indicate which PSF).  Thus the images we begin with are \ic, \pc, and
the target PSF \ps.  All three images are Fourier-transformed  using the IDL routine {\tt fft}, after which we
multiply them by $\nc^2$ or $\nt^2$ for proper normalization.  The
result of the Fourier transform is a double-precision complex array
with the same dimensions as the original.\footnote{While this
  resulting array would seem to have twice as much information as the
  original real-space arrays, in fact the real part of the result is
  even and the imaginary part is odd, so the amount of information is
  preserved.}  %Because of the peculiar indexing scheme of the IDL {\tt
%  fft} routine, it is necessary to shift the resulting array so that 
%the zero wavenumber pixel will be at the centre.  

With the Fourier-space COSMOS PSF \tpc, we can now construct the
pseudo-deconvolution kernel $\tkpd({\bmath k})$.  Unlike a pure deconvolution
kernel, $1/\tpc$, \tkpd\ has an additional factor that avoids division
by small numbers (i.e., where the COSMOS PSF has erased most
information).  We define this factor as
\beq\label{E:f}
\tilde{Y}({\bmath k})= \frac{1}{1 + |0.5/\tpc({\bmath k})|^{20}}
\eeq
and thus \tkpd\ by  
\beq\label{E:pdkern}
\tkpd({\bmath k}) =  \frac{\tilde{Y}({\bmath k})}{\tpc({\bmath k})}.  
\eeq 
The $\tilde{Y}$ factor has been chosen to be very close to $1$ for all
scales where $|\tpc| \gtrsim 0.5$,
and zero when $|\tpc| \lesssim 0.5$, with a smooth and rapid
transition between these two regimes.  Thus, it approximates a pure
deconvolution at wavenumbers where such an approach is possible, and
removes all power at smaller scales.  In practice, comparison with
Fig.~\ref{F:ftscales} demonstrates that this kernel gives a pure
deconvolution for all scales within a factor of 2 of the SDSS band
limit. 

The pseudo-deconvolved image (Eq.~\ref{E:fpdconvimg}) can then be formed directly by 
multiplication of the elements of the two arrays at a given ${\bmath k}$, using
\beq
\ticpd({\bmath k}) = \tkpd({\bmath k}) \,\tic({\bmath k}).
\eeq
The PSF for this pseudo-deconvolved image is simply Eq.~\eqref{E:f}.
Examination of these pseudo-deconvolved images in real-space suggests
that they very frequently include some ringing, always at higher
wavenumbers (smaller scales) than the band limit of any reasonable
ground-based PSF.  In practice this ringing is not relevant, since we
do not work explicitly with the real-space pseudo-deconvolved images,
and the step of convolving to match a ground-based PSF will remove 
the ringing. 

At this stage, in order to reduce the effects of shape noise, we also
define a 90 degree rotated galaxy image.  Since we would like this
image to be rotated before shearing or applying the target PSF, we
rotate \ticpd\ by 90 degrees about its central pixel to create
\ticpdr, and note that its effective PSF is the 90 degree rotation of
Eq.~\eqref{E:f}.  

The next step is to shear both \ticpd\ and \ticpdr.  As described in
Section~\ref{S:methodology}, the shearing can be carried out simply in
Fourier space using Eq.~\eqref{E:linshearft}.  We take advantage of the fact
that the coordinates before and after shearing are linearly related to
each other, and use the IDL routine {\tt poly\_2d} that is designed to
perform polynomial warping of images with various interpolation
methods.  For the level of accuracy that we wish to achieve, we
require the most precise (and time-intensive) interpolation allowed by
that routine, cubic interpolation \citep{1983CGIP...23..258P}.  This interpolation method approximates sinc
interpolation, which is in principle exact if the image is Nyquist
sampled.  Instead of the sinc function, this routine uses cubic polynomials to
make a function that is very similar, and that goes to zero (and 
 has a derivative that goes to zero) at the
edge of the window used for the interpolation.  Since the arrays we
want to shear are complex, we separately interpolate the norm and
phase using this routine in order to reconstruct the sheared and
pseudo-deconvolved images \ticpds\ and \ticpdrs.  \reftext{In
Sec.~\ref{SS:interpolation}, we will present tests demonstrating that
the interpolation routine we have used is sufficiently accurate for
our purposes.} % \footnote{\newtext{We
  %  have tested the capabilities of the cubic interpolation routine in
  %numerous ways to ensure that when we use it to shear images, the
  %output shear is really the one we desire, to better than 0.1 per
  %cent.  These tests have included producing galaxy images with
  %arbitrary rotations (since we know precisely how their ellipticities
%should transform under a rotation) and magnifications, for a variety
%of galaxy morphologies and apparent sizes.}}

The final Fourier-space manipulation is to match the desired target PSF by constructing a
kernel \tks, or \tksr\ for the 90 degree rotated image.  To do this,
we must divide the target PSF \tps\ by the effective PSF for the
pseudo-deconvolved image 
(Eq.~\ref{E:f} or its 90 degree rotation).  Once we have the
matching kernel, the PSF-matching is then performed in Fourier space
via multiplication of each element of \ticpds\ and \ticpdrs\ by the
corresponding element of \tks:
\beq
\tis({\bmath k}) = \ticpds({\bmath k}) \times \tks({\bmath k})
\eeq
and
\beq
\tisr({\bmath k}) = \ticpdrs({\bmath k}) \times \tksr({\bmath k}).
\eeq

The transformation to real space is again carried out using {\tt fft}.
The real part of the resulting double-precision complex array is taken
(in practice, the imaginary part, which should be precisely zero, is
very small but nonzero due to negligibly small numerical
inaccuracies).  To achieve the target pixel scale, these images are
resampled, which requires interpolation since the ratio of the pixel
sizes is not necessarily an integer.  For this purpose, we use the IDL
routine {\tt congrid} with cubic interpolation (the same interpolation
used for the shearing, which approximates a sinc function).  The
Fourier-space PSF-matching procedure implicitly accounts for the
different pixel response functions, which is why we resample rather
than rebinning the images.

At this point, the total number of counts in the image  is
renormalized, and noise is added as desired.  The total processing
time per typical galaxy is approximately 1 second.  For the purpose
of our basic testing, we save four images per galaxy: the original
orientation and the 90 degree rotated orientation, both before adding
noise and after adding noise.  For the purpose of the discussion that
follows, we will refer to these as `noiseless' (ignoring the low
level of noise from the original COSMOS data) and `noisy,' respectively.

\begin{figure*}
\begin{center}
\hspace{-0.5in}
$\begin{array}{c@{\hspace{0.6in}}c}
\includegraphics[width=3.2in,angle=0]{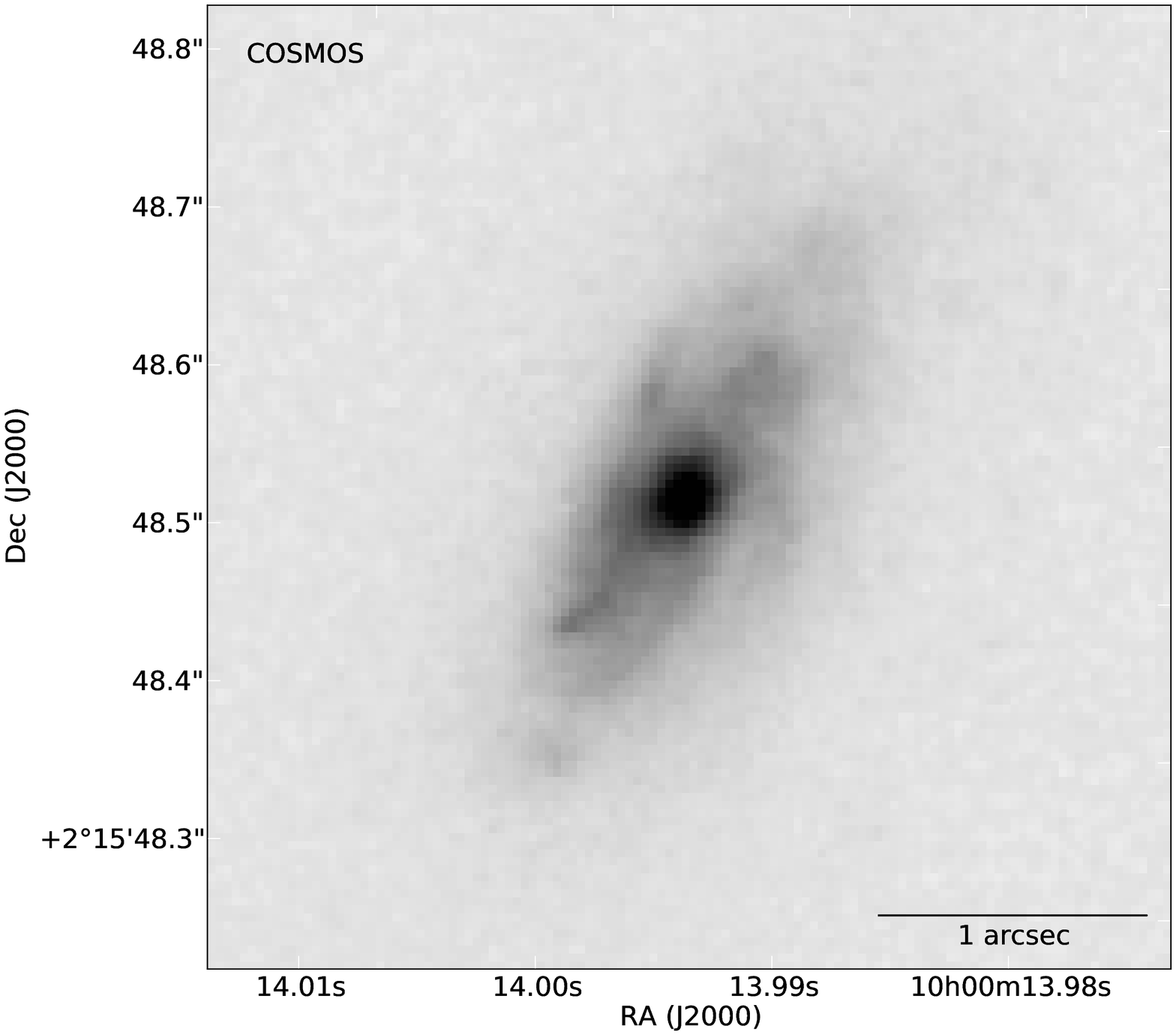} &
\includegraphics[width=2.8in,angle=0]{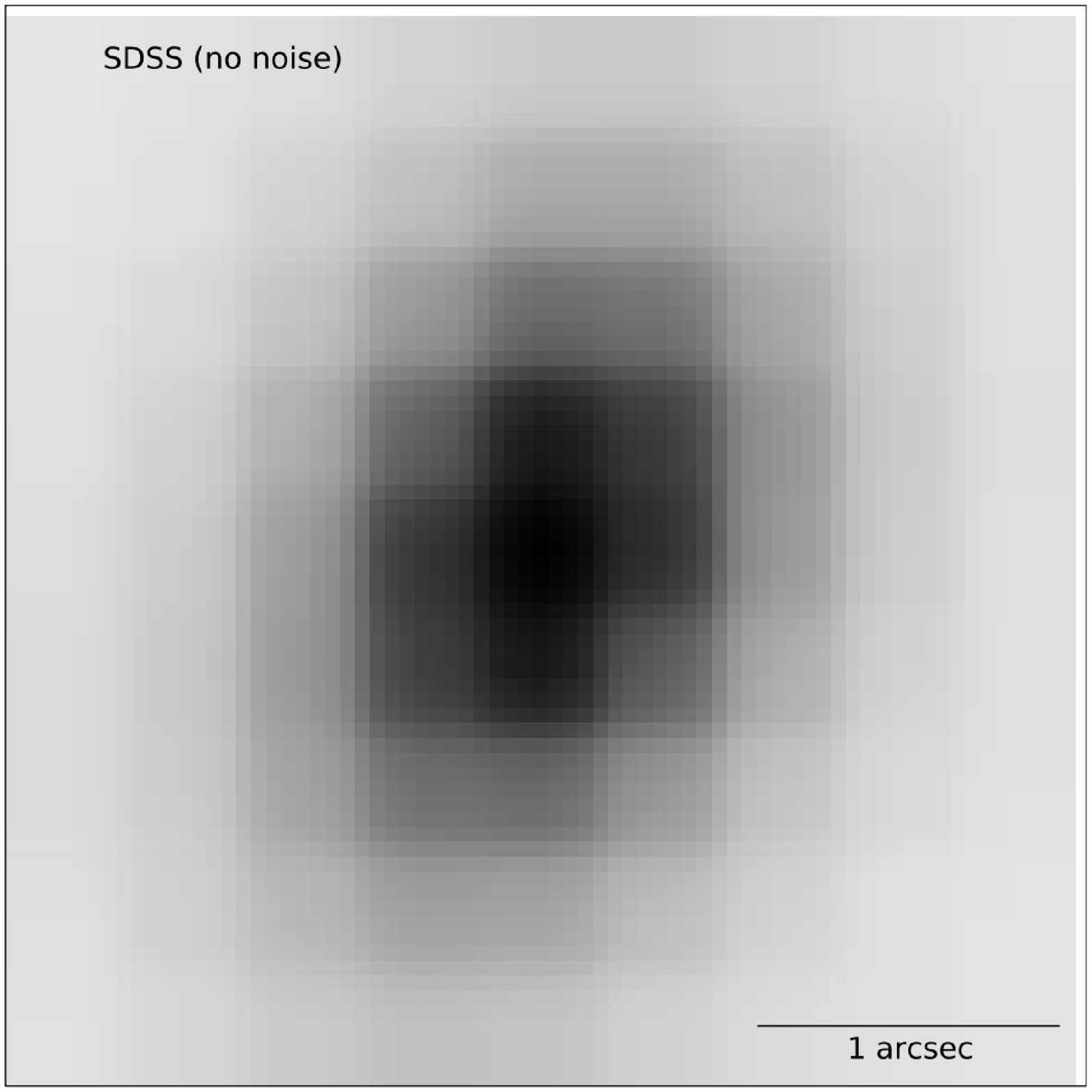}\\
\includegraphics[width=2.8in,angle=0]{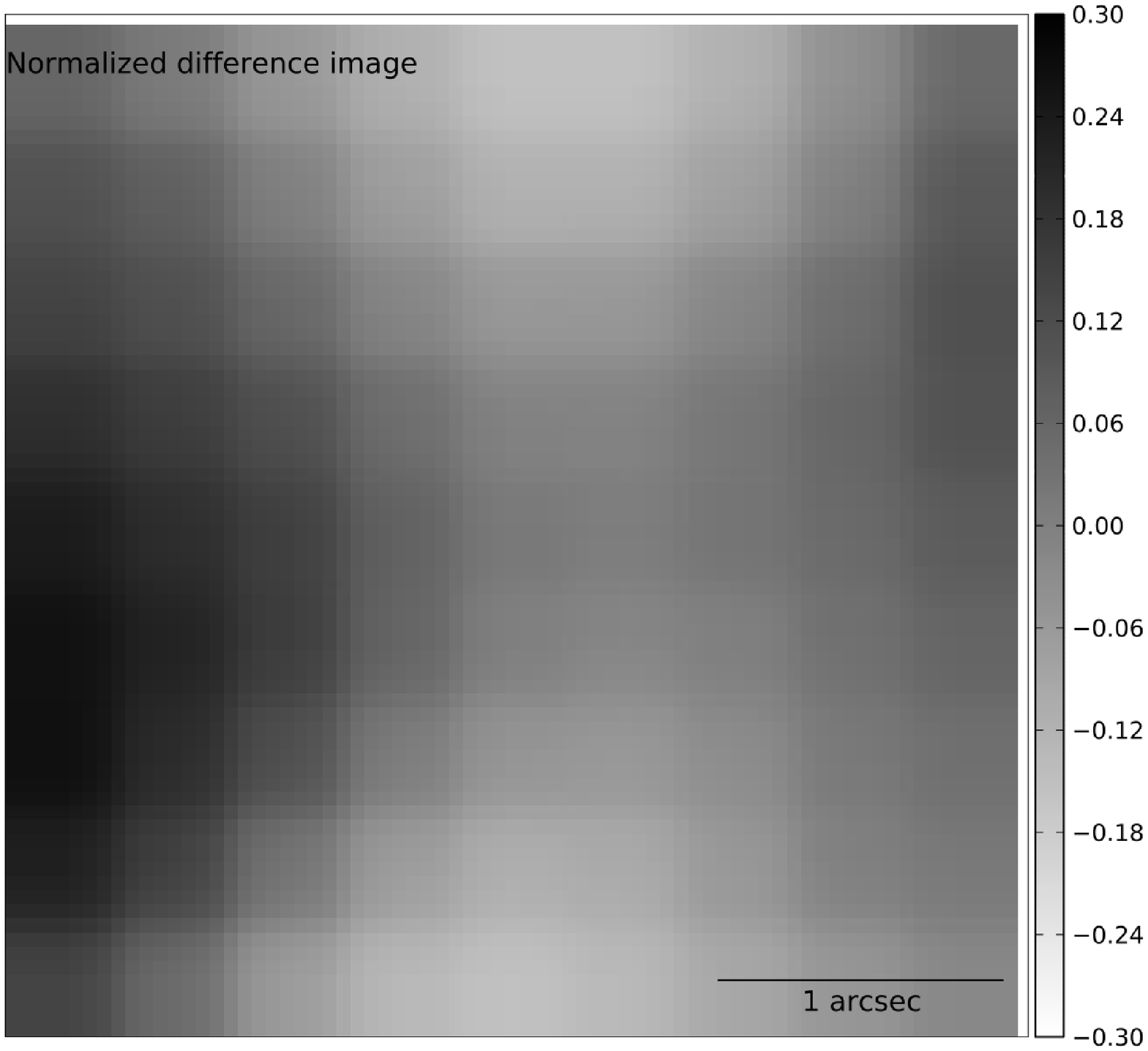} &
\includegraphics[width=2.8in,angle=0]{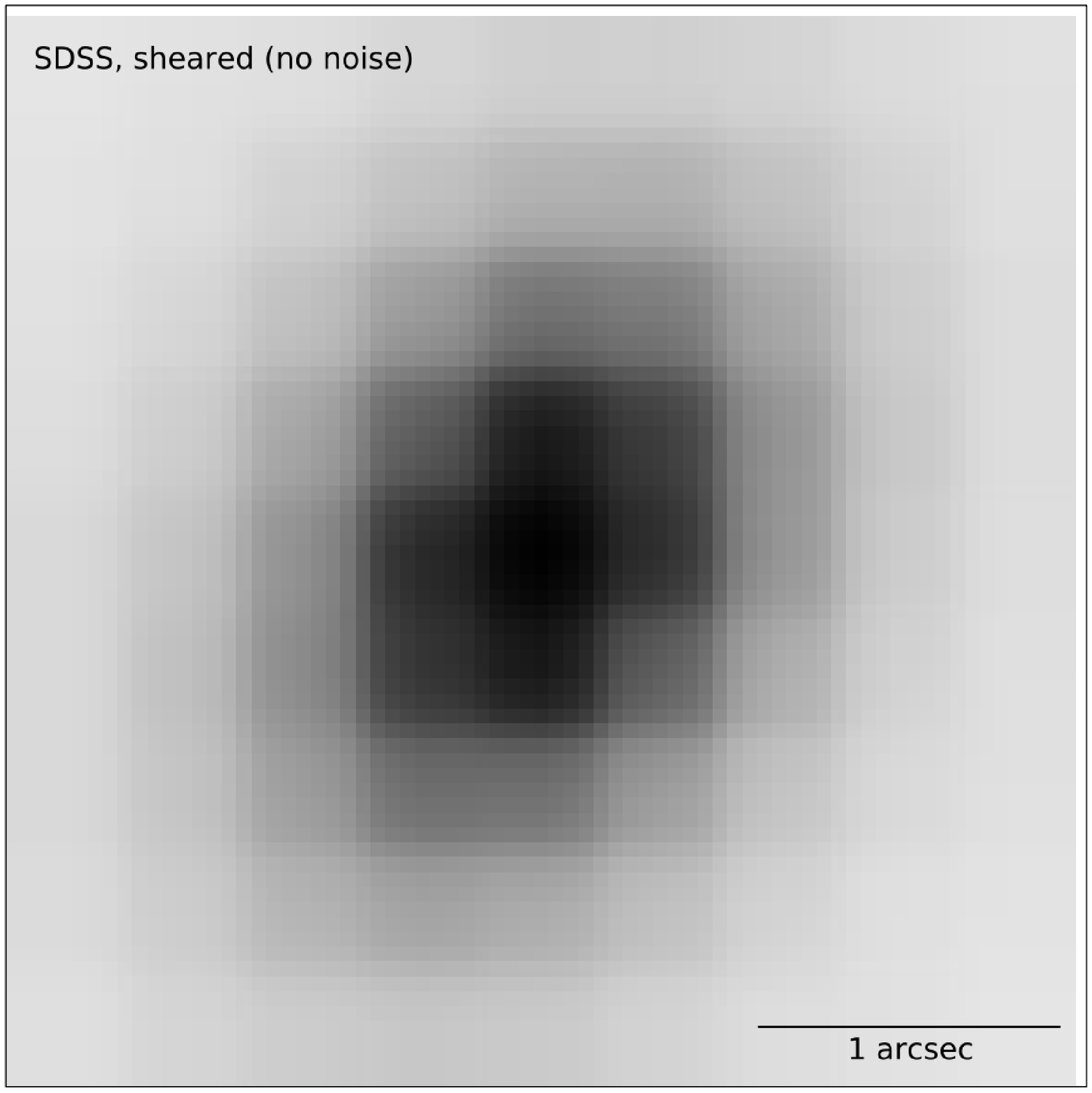} \\
\end{array}$
\caption{\label{F:example}\newtext{Example of how the {\sc shera}
    processing changes the galaxy images.  {\em Top left:} The original
    galaxy image in COSMOS, on a linear scale.  {\em Top right:} The
  appearance of the galaxy in SDSS after PSF-matching, before adding
  levels of sky noise consistent with SDSS.  {\em Bottom right:} Same
  as bottom left, but with a significant shear in the horizontal direction,
  $\gamma_1=0.1$.  {\em Bottom left:} Difference image between the
  sheared and unsheared simulated image, normalized by the unsheared
  image.  Because the differences are at most a few tens of per cent,
  they are difficult to pick out visually by comparing the top and
  bottom right images.}
}
\end{center}
\end{figure*}

\newtext{An example of how this processing works for one particular COSMOS
galaxy is shown in Fig.~\ref{F:example}, which shows the original
COSMOS image (with some fine
structure), and the degraded SDSS image without and with a
gravitational shear of $\gamma_1=0.1$, i.e. along the horizontal
axis.}

\section{Technical validation of {\sc shera}}\label{S:techvalidation}

\reftext{This section includes the results of several tests of the
  technical aspects of {\sc shera}, to demonstrate that the procedure
  outlined in Sec.~\ref{S:implementation} works as intended.}

\subsection{Accuracy of interpolation}\label{SS:interpolation}

While most of the mathematical operations carried out by {\sc shera}
are simple and easy to carry out to extremely high accuracy (e.g., a
Fourier transform), two of the operations are non-trivial because they
involve interpolations.  As described in Sec.~\ref{S:implementation},
we use the IDL cubic interpolation routine both for shearing the
pseudo-deconvolved images, and for resampling the PSF-matched images
to the target pixel scale.

Here we present the results of tests that demonstrate that the IDL
cubic interpolation routine is sufficiently accurate for both shearing
and resampling.  While we carried out numerous tests of the pipeline
using both real galaxies and analytic models, here we focus on tests
that use the real COSMOS galaxies, under the assumption that they
provide a more stringent test than analytic galaxy and PSF profiles.

The first test is of the shearing of the pseudo-deconvolved image.  In
principle, we could carry out this test by transforming the
pseudo-deconvolved image to real-space both before and after shearing,
and then comparing the observed adaptive moment matrices.  Since there
is no PSF in these images, we can simply check that the moments
transform under shear according to equation (2-13) of
\cite{2002AJ....123..583B}.  However, as
stated in Sec.~\ref{S:implementation}, the pseudo-deconvolution leads
to ringing in real-space.  The ringing is relatively high frequency
and therefore difficult to accurately interpolate, which might lead us
to conclude that our shearing is not very accurate.  However, this
ringing does not
affect the accuracy of the shearing on the final ground-based image
since convolving with the ground-based PSF will eliminate the
ringing.  Thus, we restrict our tests of the pseudo-deconvolved images
to the Fourier-space images used for the actual shearing, and check that the Fourier counterpart of
equation (2-13) from \cite{2002AJ....123..583B} is satisfied.

For this test, we use a random 5 per cent of the COSMOS images, and
apply a shear $(\gamma_1,\gamma_2)=(0.02,0)$ before matching to SDSS.
We do not add noise to the images, in order to allow for the highest
possible precision in these tests.  The motivation behind
applying zero shear in one component is that it allows us to test that
shearing one component does not lead somehow to spurious shearing in
the component that we do not intend to shear.  For each galaxy, we compute
the adaptive moments of the pseudo-deconvolved Fourier space image
before shearing $(e_1,e_2)$ and after shearing
$(e_1^{(\gamma)},e_2^{(\gamma)})$.  We then compare the latter with
the expected ellipticities $(e_{1,\mathrm{exp}}^{(\gamma)},e_{2,\mathrm{exp}}^{(\gamma)})$, to
get the error in the observed shear $\Delta\gamma_i =
e_i^{(\gamma)}-e_{i,\mathrm{exp}}^{(\gamma)}$ for $i=1,2$.  We can use
the values for this random subsample of the COSMOS galaxies to study
the distribution of $\Delta\gamma_1/\gamma_1$ and $\Delta\gamma_2$.
We find that this distribution is mildly non-Gaussian (with positive
kurtosis), and has a median value of $\Delta\gamma_1/\gamma_1=1.6\times
10^{-5}$ and $\Delta\gamma_2=1.5\times 10^{-6}$.  This result suggests
that on average, when simulating a sample of
$\gtrsim 1000$ galaxies, the simulated shears are equal to the
requested ones to extremely high accuracy.   Moreover, the act of
shearing one component does not lead to any significant spurious shear
in the other component.

However, we should also consider the width of the distributions of $\Delta\gamma_1/\gamma_1$ and $\Delta\gamma_2$.  If
the distribution is broad, then when
simulating a few individual galaxies, there could be some systematic
deviation from the desired shear value which does not average out as
it would when simulating many galaxies.  The ensemble 68 per cent confidence
intervals are $-0.0015 < \Delta\gamma_1/\gamma_1 < 0.0032$ (the median
is not at the centre of this range because it is a skewed
distribution) and $|\Delta\gamma_2| < 3.5\times 10^{-4}$ (this
distribution is not skewed, presumably because no shear was actually applied). We therefore
conclude that for any individual simulated galaxy, (a) when applying a
shear, there is a 68 per cent chance that the actual applied shear
will be within $[-0.15, 0.32]$ per cent of the desired shear,
and (b) shear components to which we do not intentionally apply a
shear remain unsheared at the level of a few $\times 10^{-4}$.  Thus,
the interpolation is sufficiently accurate 
to precisely shear the galaxies on average (that
is, that there is no systematic problem with the applied shears) and
 even for single galaxies, the applied shears are correct at the
level of a few tenths of a per cent.

The other operation for which we must use interpolation is the image
resampling.  There may be a concern that the tests described above are not an adequate test of the interpolation for resampling, for
the following reason: when we apply a (typically small) shear, the
pixel grid is not highly distorted near the centre of the galaxy.
This means that the interpolation is being used to estimate values
very close to being on the pixel grid, which should not be too
difficult.  However, when we resample to some arbitrary pixel grid, we
might end up interpolating (even near the image centre) to some
locations that are not close to lying on the pixel grid.
We therefore require a test of the interpolation that samples the
image in a more general way than the above test.

Here we present a test of the images after
pseudo-deconvolution, shearing, PSF-matching, and returning to real
space - in other words, the actual point in {\sc shera} where
resampling takes place.  The test is that instead of resampling the
images, we apply a random rotation, and compare the change in the
moments with the expected change.  We know exactly how the galaxy
ellipticities should transform:
\beqa
e_{1,\mathrm{rot}} &=& [\cos{(2\theta_\mathrm{rand})}] e_1 +
[\sin{(2\theta_\mathrm{rand})}] e_2 \\ \notag
e_{2,\mathrm{rot}} &=& -[\sin{(2\theta_\mathrm{rand})}] e_1 +
[\cos{(2\theta_\mathrm{rand})}] e_2.
\eeqa
An additional test is to ensure that the area implied by the adaptive
moments ($(M_{xx}M_{yy}-M_{xy}^2)^{1/2}$) is unchanged by rotation.  

When we carry out this test, we find that there is a non-zero but
extremely small systematic change in the ellipticities (compared to
the expected ellipticities after rotation): typically $-1.3$ and $+2.6\times
10^{-6}$ for $e_1$ and $e_2$, respectively.  We also find a tiny but
statistically significant change in the areas implied by the adaptive
moments, at the level of $-8\times 10^{-7}$.  These numbers are
sufficiently small that they are truly subdominant to other
systematics issues
that arise in any realistic lensing analysis.

\subsection{Simulated galaxies compared to real data}\label{SS:validation}

\reftext{Another important test for {\sc shera} is to compare the simulated SDSS
images of COSMOS galaxies with the real SDSS images of those
galaxies.   Here, we will describe several tests.}

%In this Appendix, we describe some basic tests that were used to
%validate the outputs of the {\sc shera} pipeline, via a comparison of
%the SDSS imaging in the COSMOS region with the simulated data in that
%region. 

\subsubsection{Galaxy numbers}

As described previously, the sample of galaxies for which we have
generated COSMOS postage stamps contains 26~113 galaxies
(Sec.~\ref{SS:cosmos}).  This number was reduced to 25~527
(Sec.~\ref{SS:psprep}) when we require that the postage stamps go
successfully through our postprocessing (to mask out additional
objects, etc.) and then to 17~706 when we require that the galaxies
lie in regions considered photometric in the SDSS imaging
(Sec.~\ref{SSS:photometricity}).

There is one additional cut that we must impose after convolution to
match the SDSS PSF. This cut relates to the fact that the original
postage stamp sizes were estimated based on a circularly-averaged
characteristic radius, which means that for very large and flattened
galaxies, some flux might go off the edge of the convolved postage
stamp in the direction of the galaxy major axis.  To test for this
issue, we processed all of the `noiseless' simulated images,
comparing the flux in (a) the pixel with the maximum flux, versus (b)
the pixel with the maximum flux when considering only those pixels at
the very edge of the postage stamp. We then eliminated those galaxies
for which the latter was $>0.01$ times the former.  This left us with
a final galaxy sample for all tests of 17~667 galaxies (some of them
too faint to measure in the SDSS images).

\subsubsection{Comparison with real data}

The first comparison we make is between the simulated images (without
shear or 90 degree rotation) and the actual SDSS images, for those
that are detected.  In principle, the images should be the same except
for noise and the centroiding of each object within the central pixel (which
we have made no attempt to match).

\begin{figure*}
\begin{center}
$\begin{array}{c@{\hspace{0.5in}}c}
\includegraphics[width=2.82in,angle=0]{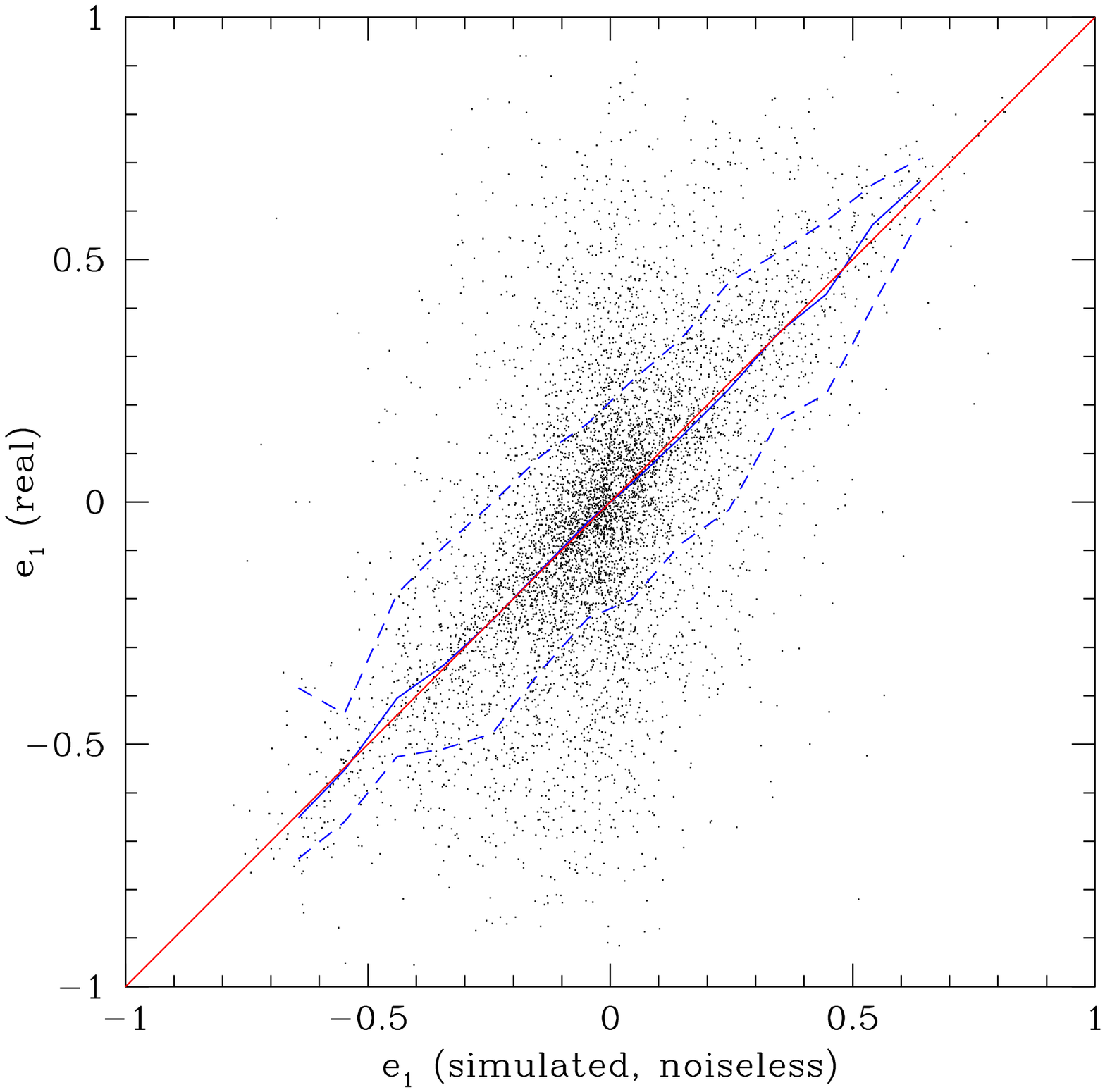} & 
\includegraphics[width=2.82in,angle=0]{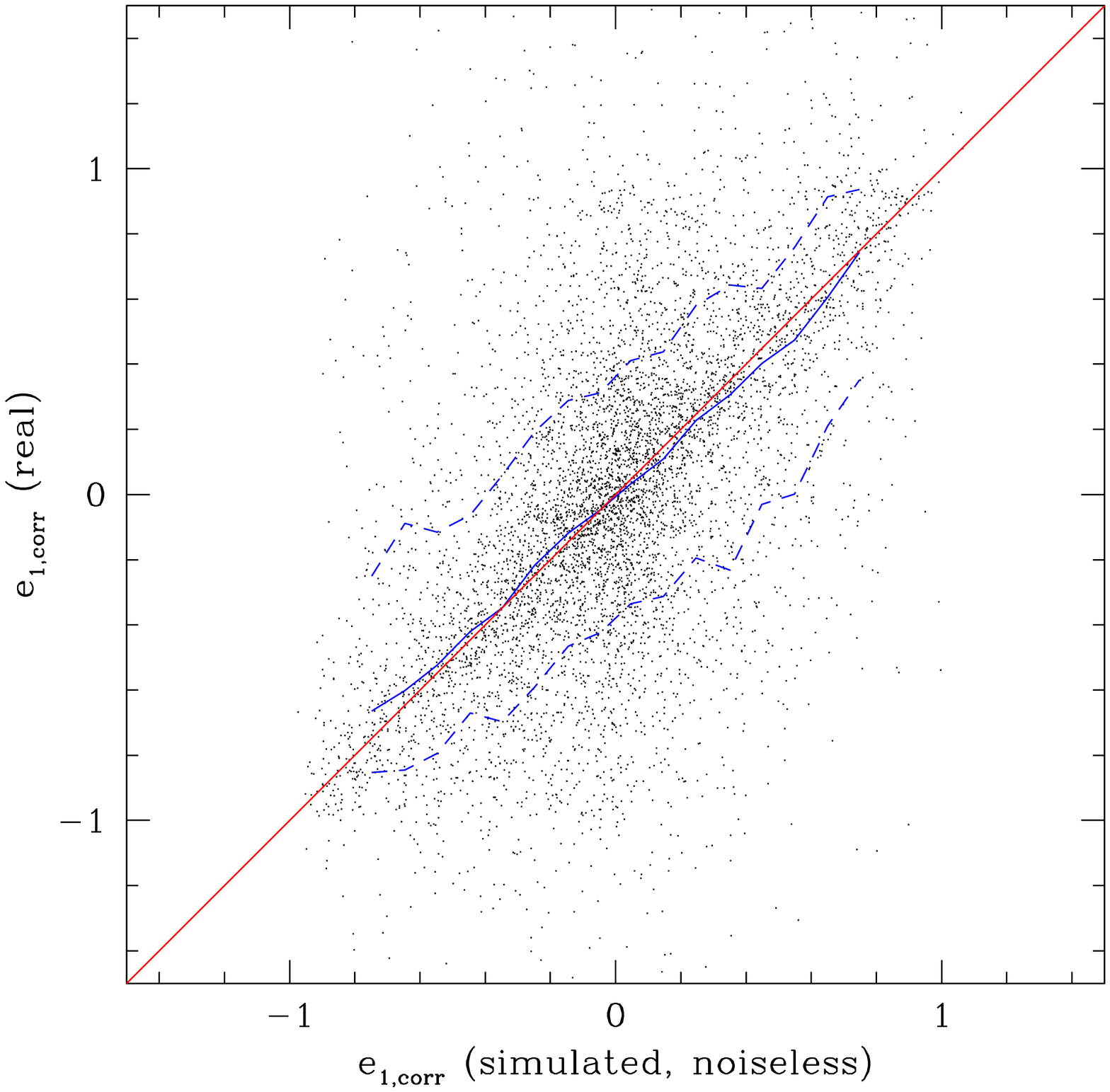} \\
\includegraphics[width=2.82in,angle=0]{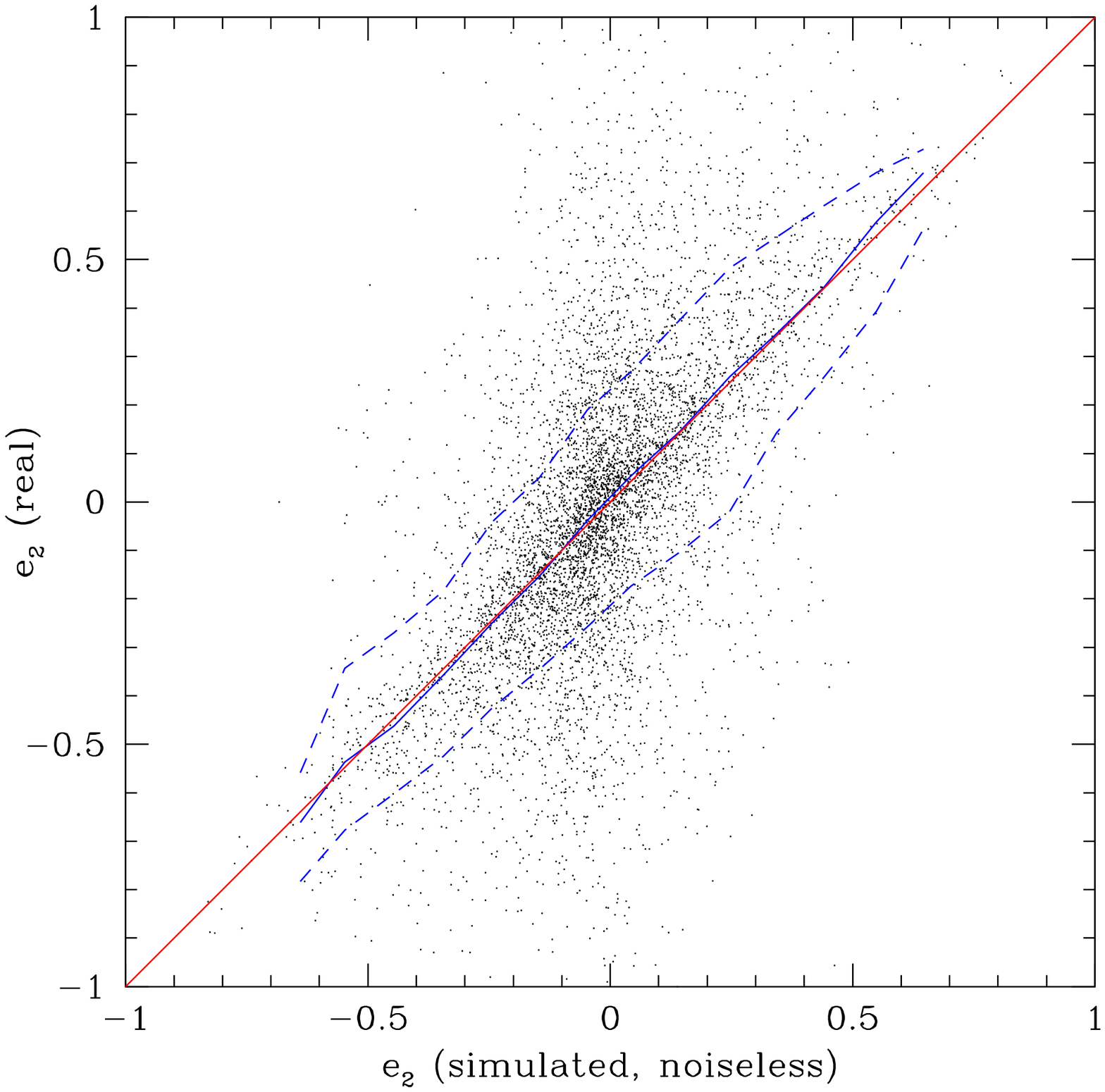} & 
\includegraphics[width=2.82in,angle=0]{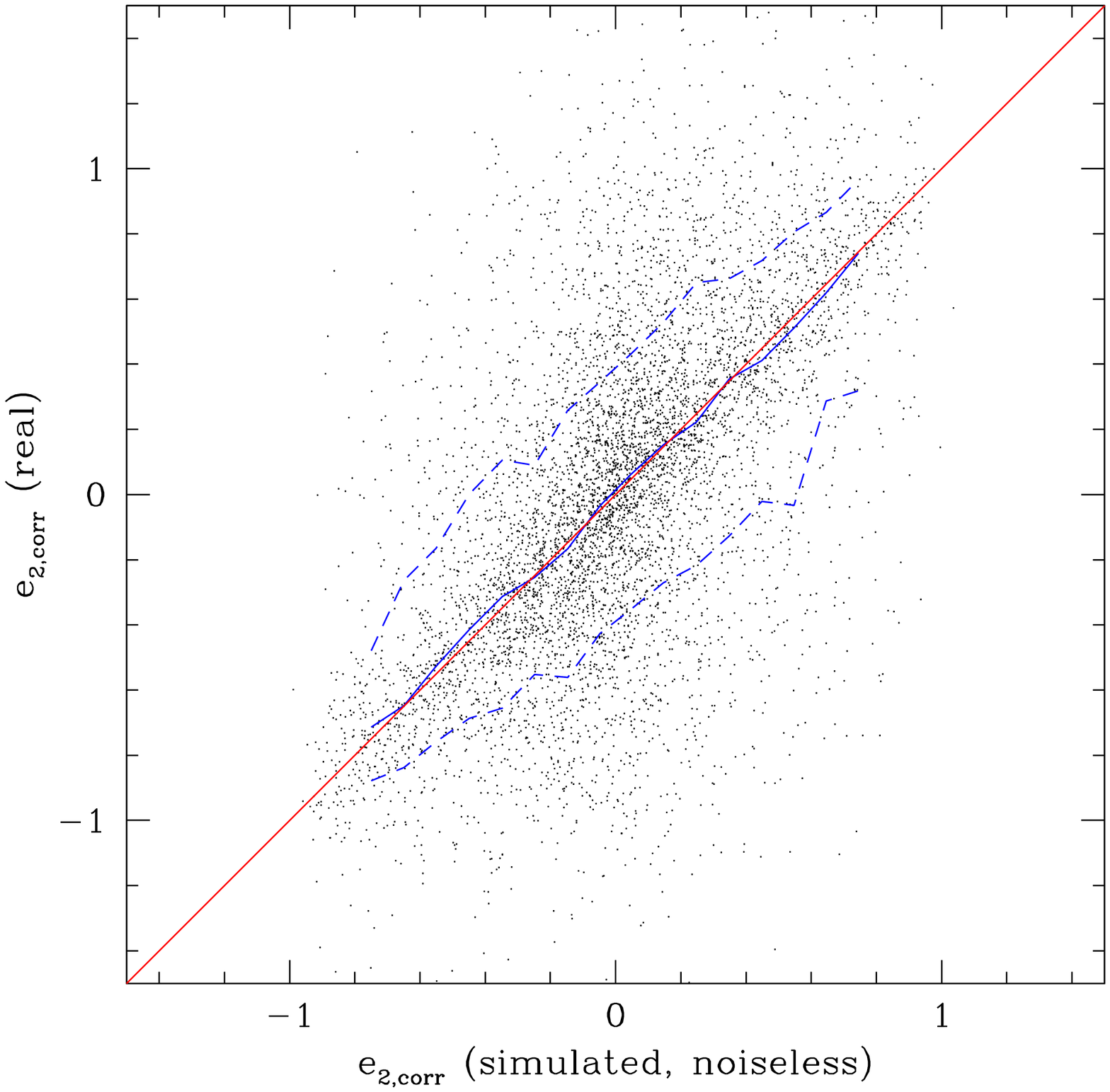} \\
\end{array}$
\includegraphics[width=2.82in,angle=0]{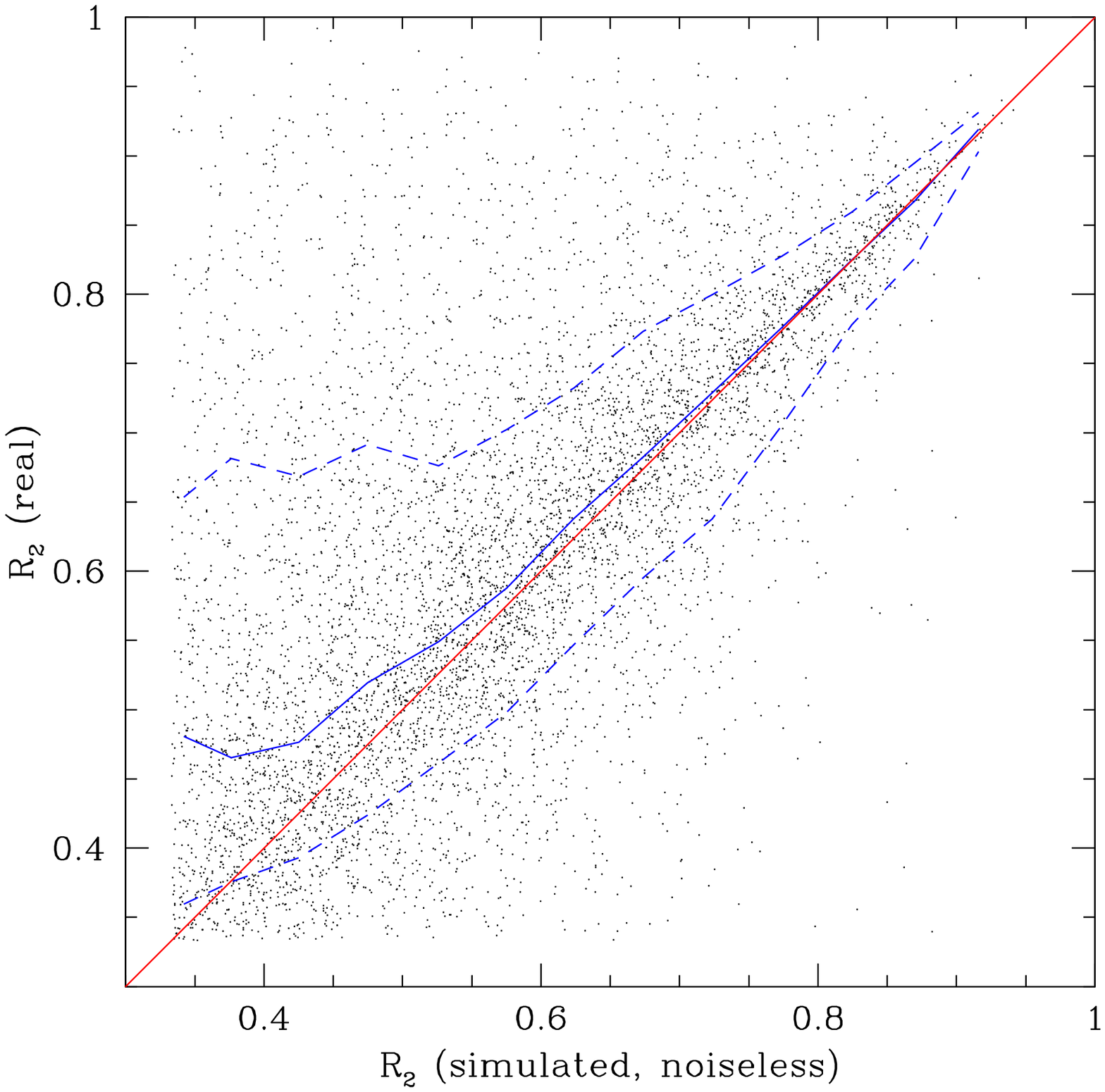} 
\caption{\label{F:basicsimresults}Basic shape comparison for real SDSS
  data versus (noiseless) simulated data, with points for each simulated galaxy
  and trend lines indicating the median (solid) and $68$ per cent CL
  (dashed).  For comparison, the 1:1 line is shown in all cases.  {\em Top left:} Observed
  $e_1$ shape component (along the pixel direction) of the
  PSF-convolved galaxy image in the simulation without added noise versus in the real
  data.  {\em Top right:} Same as left, but after PSF correction.
  {\em Middle row:} Same as top row, but for the $e_2$ ellipticity
  component.  {\em Bottom row:} Galaxy resolution compared to the PSF
  in the simulation versus in the real data.}
\end{center}
\end{figure*}

We begin by comparing the results for the \newtext{noiseless} simulations against
those with the original SDSS data.  In this case, there are 9469
galaxies that have measurable galaxy shapes (with re-Gaussianization)
both in the real data and in the simulations; of those, 6361 pass our
resolution cuts in both cases.  We restrict the comparison of shapes
to that sample of 6361, which should be fairly similar to the
source catalogue except without a cut on magnitude, which would remove
another $\sim 30$ per cent of the galaxies.  The results
are shown in Fig.~\ref{F:basicsimresults}, which shows that for both
the observed ellipticities $(e_1, e_2)$, and for
the PSF-corrected $(e_{1,\mathrm{corr}}, e_{2,\mathrm{corr}})$, the median trend is for the simulated values
to be equal to the ones in the real data (modulo noise, which tends to
cause scatter in the vertical direction, since it is present in the
real data but not in these simulations).  This finding suggests that
the simulation pipeline is indeed providing realistic data.  

\newtext{The apparent exception to the close comparison between
  simulated and real data is the $R_2$ comparison, which shows a
  distinct trend towards better resolution in the real data
  than in the simulations.  There are two causes for this finding: the
  first is that we are imposing the $R_2$ cuts in different ways in
  the simulated and real data, since the former lacks noise. This
  results in a form of Malmquist bias, given that we impose the
  $R_2$ cut on the `true' resolution in the simulation but on the
  noisy $R_2$ in the real data, which means that at the low resolution
  end of the sample, there is an induced bias when comparing the noisy
  versus the noiseless results.  Indeed, this bias essentially vanishes if we make the same plot using the $R_2$ for the simulations
  with added noise.}  

\newtext{However, even ignoring the very low
  resolution end, we can see that there is a weak (few per cent)
  tendency for galaxies to be scattered preferentially towards the
  upper left part of the plot.  Detailed examination of some galaxies
  in that part of the figure suggests that while some are noise
  fluctuations in the real data, others are due to deblending failure. There are multiple nearby objects in reality, which are
  resolved in COSMOS, so that the additional objects besides the
  central galaxy were masked in our processing of the postage stamps;
  but the blend was not resolved in SDSS and hence was treated as one
  larger object.  A careful study of these simulations can therefore
  be used to study deblending failures in ground-based data.}

\subsubsection{Comparison between original and rotated}

As an additional sanity check, we also show a comparison between the
shapes for the original and the rotated images, in the case of no
added shear, but with added noise.  For the sake of simplicity, we show only one shear
component; results for the other are comparable.  Here, we rely on the
fact that
\begin{itemize}
\item the intrinsic shapes should be the opposite of each other,
  i.e. $e_{\mathrm{orig, int}} = -e_{\mathrm{rot,int}}$; and
\item any systematic additive component to the shapes from the PSF
  should be the same, i.e. $e_{\mathrm{orig, sys}} \approx e_{\mathrm{rot,sys}}$.
\end{itemize}
As a rule these additive systematic components will not cancel out
over all the galaxies, because of the tendency for there to be a
coherent PSF ellipticity in any given field that results in the
systematic components of the galaxy ellipticities having the same sign.

\begin{figure}
\begin{center}
\includegraphics[width=2.7in,angle=0]{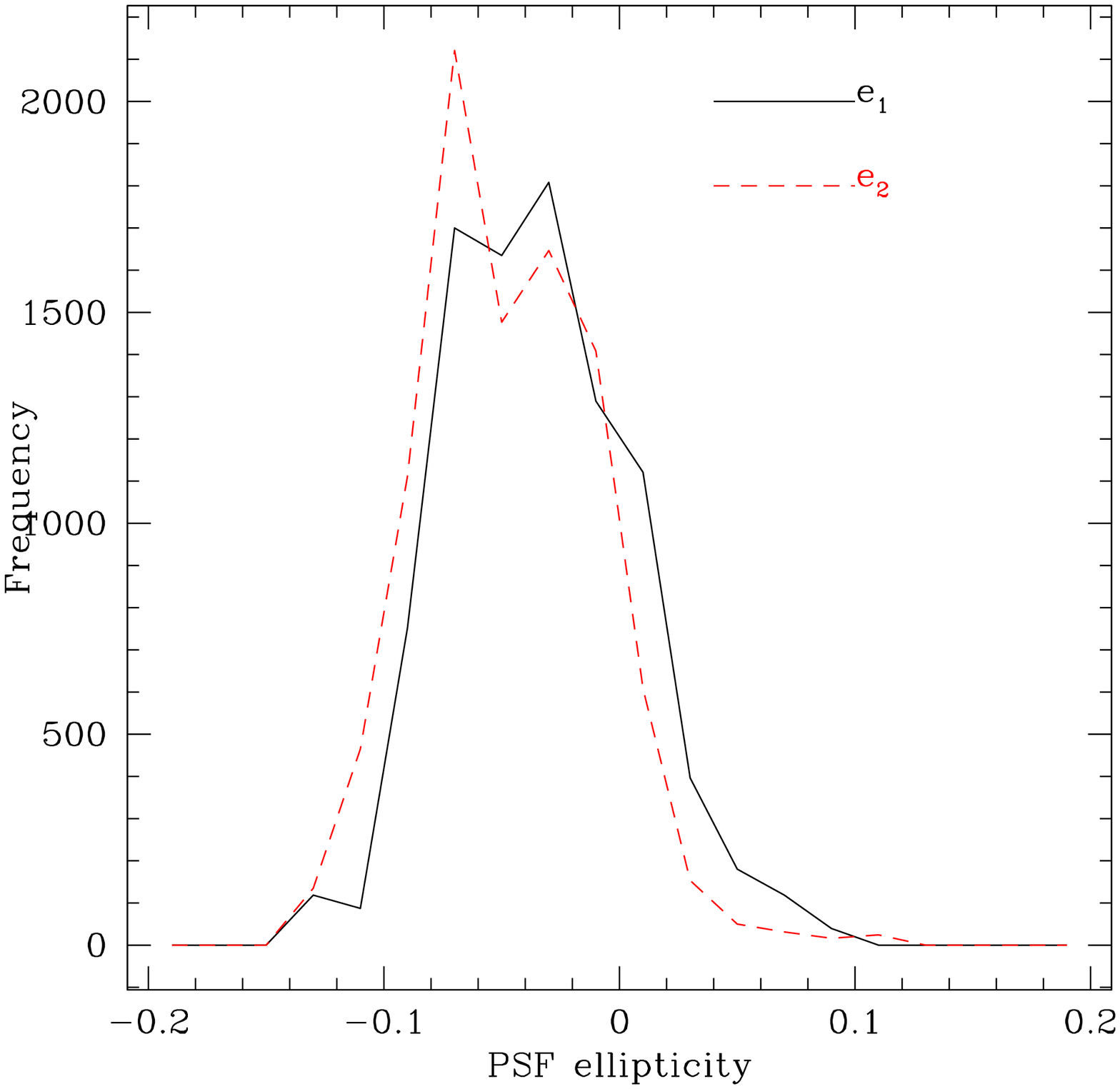}
\includegraphics[width=2.7in,angle=0]{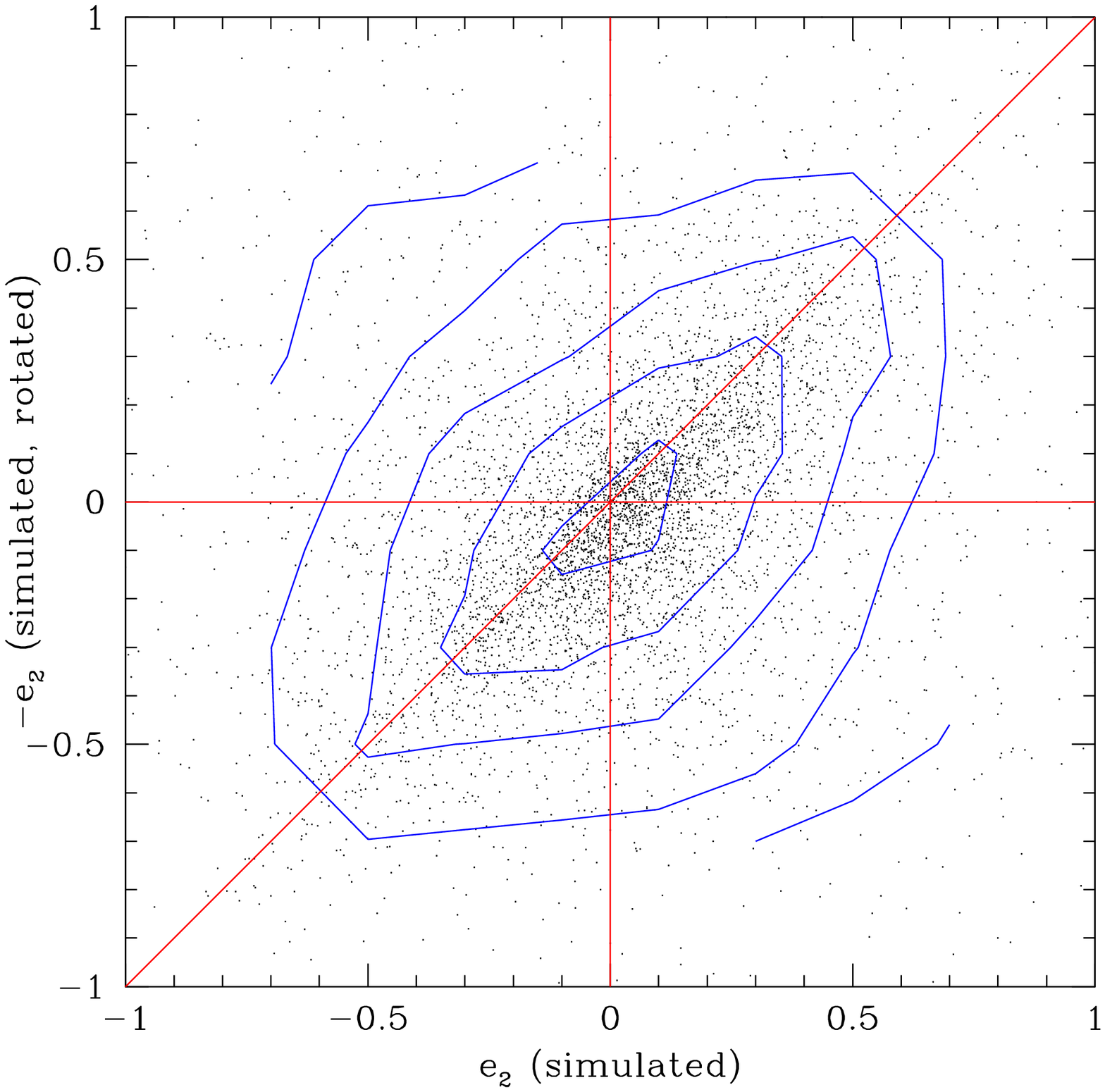}
\includegraphics[width=2.7in,angle=0]{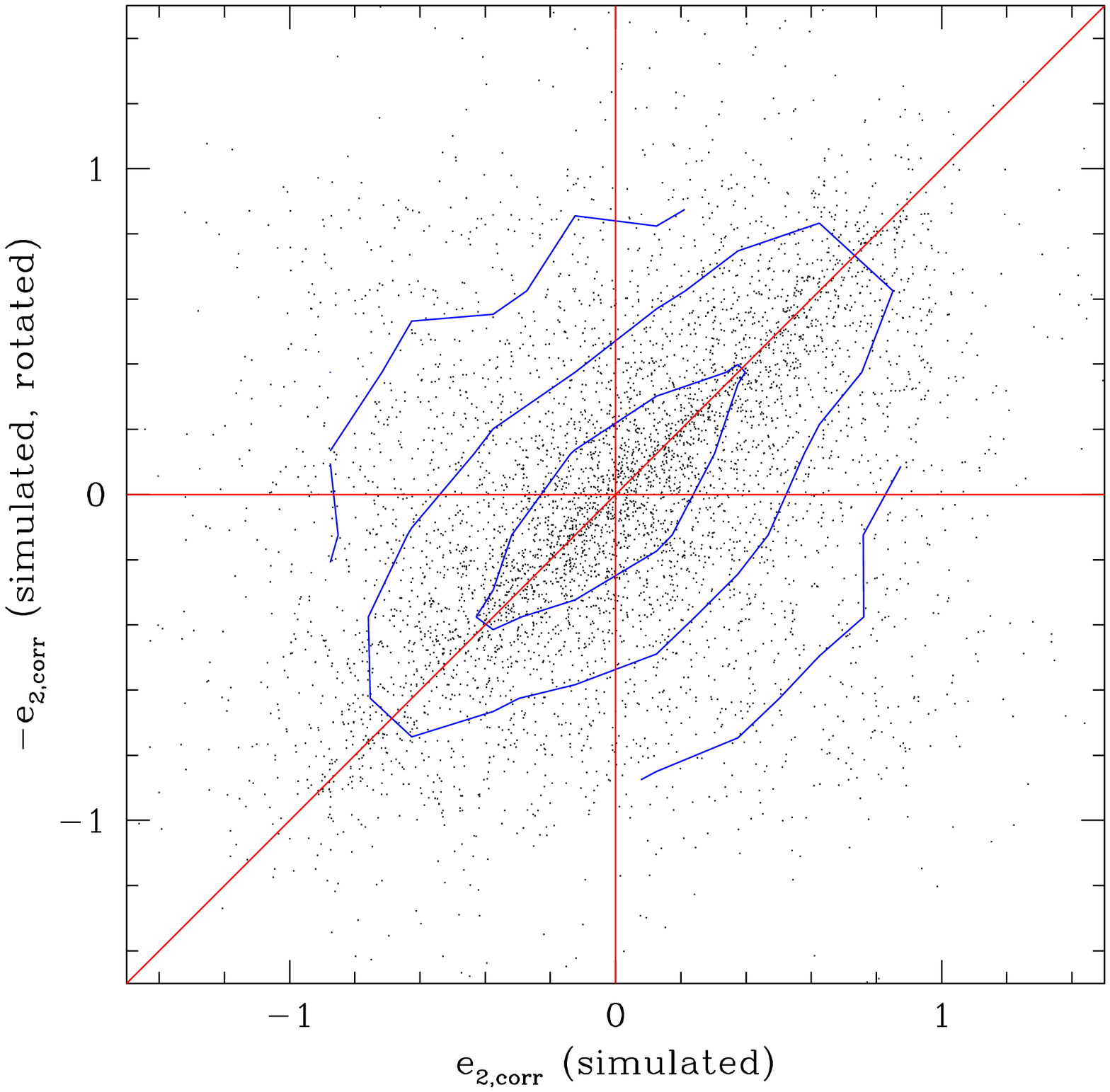}
\caption{\label{F:origvsrot}
Shape comparisons for the galaxy images in their original orientation,
and with a 90 degree rotation (using the same, non-rotated PSF).  {\em
  Top:} Histogram of PSF ellipticities in the COSMOS field, which
shows a coherent tendency towards $e_1<0$ and $e_2<0$.  {\em Middle:} Observed
  $e_2$ shape of the
  PSF-convolved galaxy image in the original and rotated simulation images.
{\em Bottom:} Same as middle, but after correction for the effects of
the PSF.}
\end{center}
\end{figure}

Thus, when we plot $e_\mathrm{orig}$ versus
$-e_\mathrm{rot}$, we should find that before PSF correction, the
results are offset from the one-to-one line (but are parallel to it),
and the results are returned to the one-to-one line after PSF
correction.  These results are shown in Fig.~\ref{F:origvsrot}, and
are entirely consistent with our expectations.

\subsection{The need for pseudo-deconvolution}\label{SS:pdconv}

\reftext{Before using our simulations to test the impact of
  pseudo-deconvolution (removal of the ACS PSF), we first consider the basic reasons why it may
  be important to include in a simulation pipeline that is meant to be
  able to accurately simulate ground-based data with a wide
  range of observing conditions.}

\newtext{Though we have focused on the ACS in this paper, the importance of
  pseudo-deconvolution is likely generic to all space telescope data
  (including other {\it HST} cameras) since it is a feature of
  diffraction patterns.  While the core of the PSF has a radius of
  $\sim\theta_{\rm D}=\lambda/D=0.07''$ for ACS/$F814W$ (where $\lambda$
  is the wavelength of observation and $D$ is the outer diameter of
  the telescope), the diffraction rings contain a significant amount
  of power.  For example, an Airy disc scatters a fraction $\sim
  2\pi^{-2}\theta_{\rm D}/\theta$ of the light to radii $>\theta$ (for
  $\theta/\theta_{\rm D}\gg 1$). Thus for a galaxy with scale radius
  $\gg\theta_{\rm D}$, the effect of the PSF on observed galaxy
  properties scales in proportion to $\theta_{\rm D}$ rather than
  $\theta_{\rm D}^2$ as one would expect based on Gaussian
  approximations or second moments. An equivalent effect can be seen
  in Fourier space: $\tilde G({\bmath k})$ for an Airy disc has the
  leading behavior $1-2\pi^{-2}\theta_{\rm D}|k|+$\dots\ rather than
  having the first nontrivial term be $k^2$. Thus diffraction even by
  a large aperture (small $\theta_{\rm D}$) has an effect even for
  long-wavelength features in the image.} 

\reftext{We can see one
  manifestation of this effect in Fig.~\ref{F:ftscales}, which shows the
  azimuthally-averaged Fourier space representation of the ACS PSF.
  While the ACS PSF is always above the SDSS PSF in that plot,
  indicating that the ACS PSF preserves more information than the SDSS
  PSF at all values of wavenumber $k$, it is nonetheless the case that
  it is tens of per cent below 1 at the band limit of the SDSS PSF,
  primarily because of the large-scale impact of the diffraction
  rings.}  

\reftext{The above argument provides the justification for including
pseudo-deconvolution as part of {\sc shera}, given our intention that
it should be useful for simulating ground-based imaging data under a
wide range of observing conditions.  However,}  it is not immediately obvious, for the case of data with typical
seeing of $\sim 1.2$\arcsec\ such as SDSS, that the
pseudo-deconvolution is necessary for accurate image simulations.  Here, we address this question using simulations for which the
pseudo-deconvolution was not performed.  That is, instead of removing
the ACS PSF on scales that we will want to use from the ground,
shearing, and then finding a matching kernel to the ground-based PSF,
we simply shear and then convolve the data directly with the SDSS PSF,
ignoring the ACS PSF entirely.  This procedure was used for the STEP2
simulations \citep{2007MNRAS.376...13M}.

\begin{figure}
\begin{center}
\includegraphics[width=2.6in,angle=0]{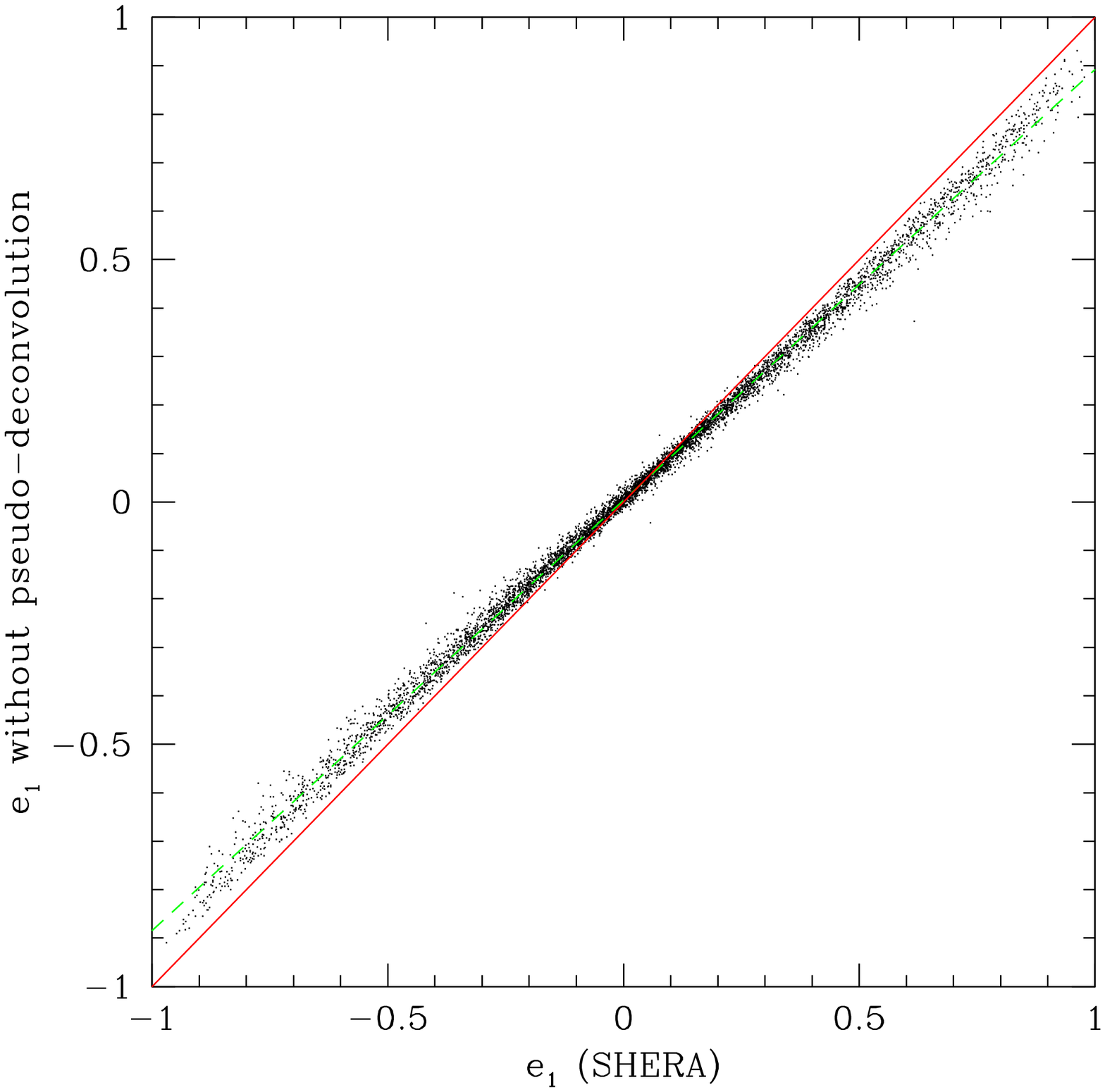}
\includegraphics[width=2.6in,angle=0]{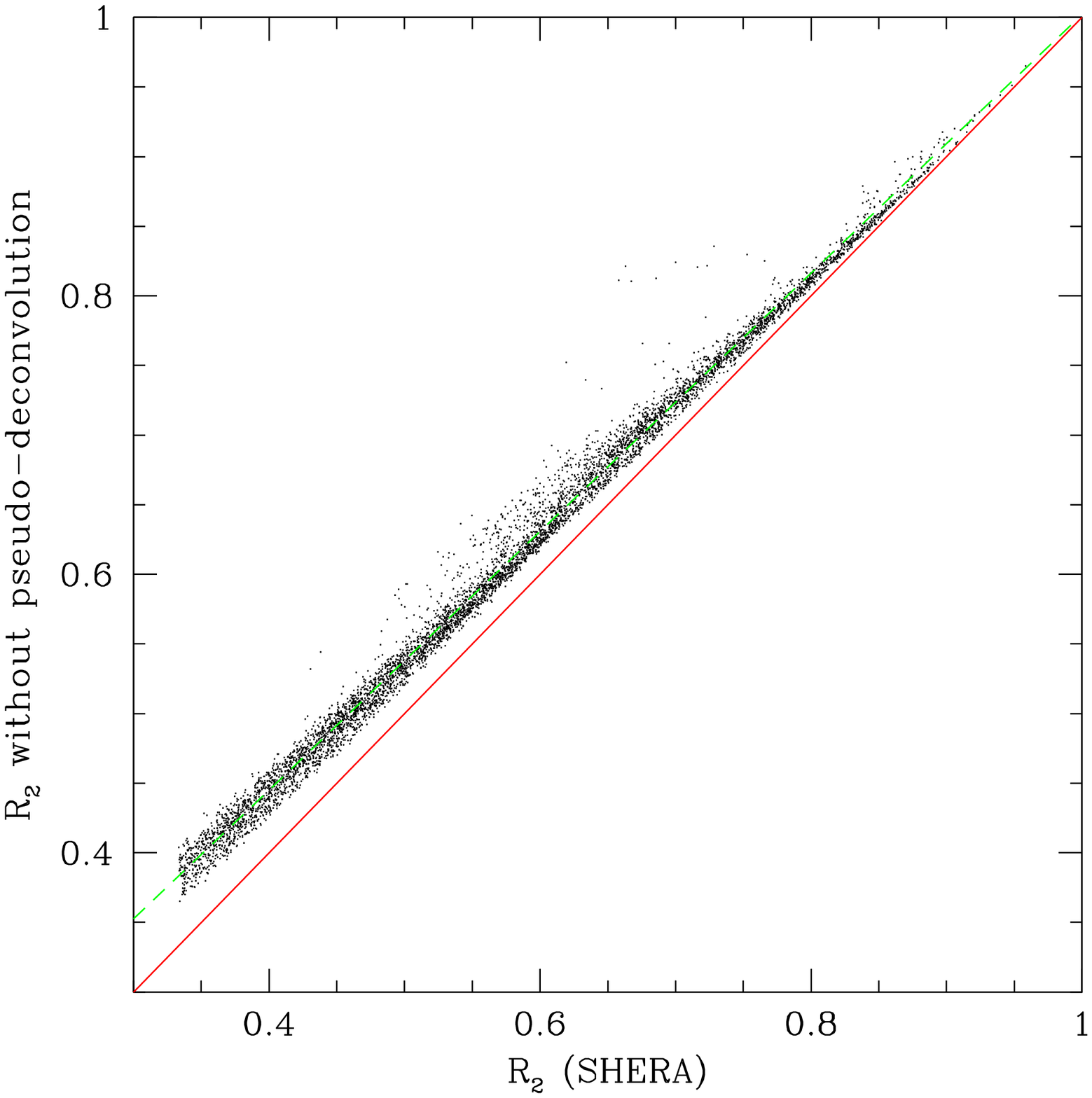}
\includegraphics[width=2.6in,angle=0]{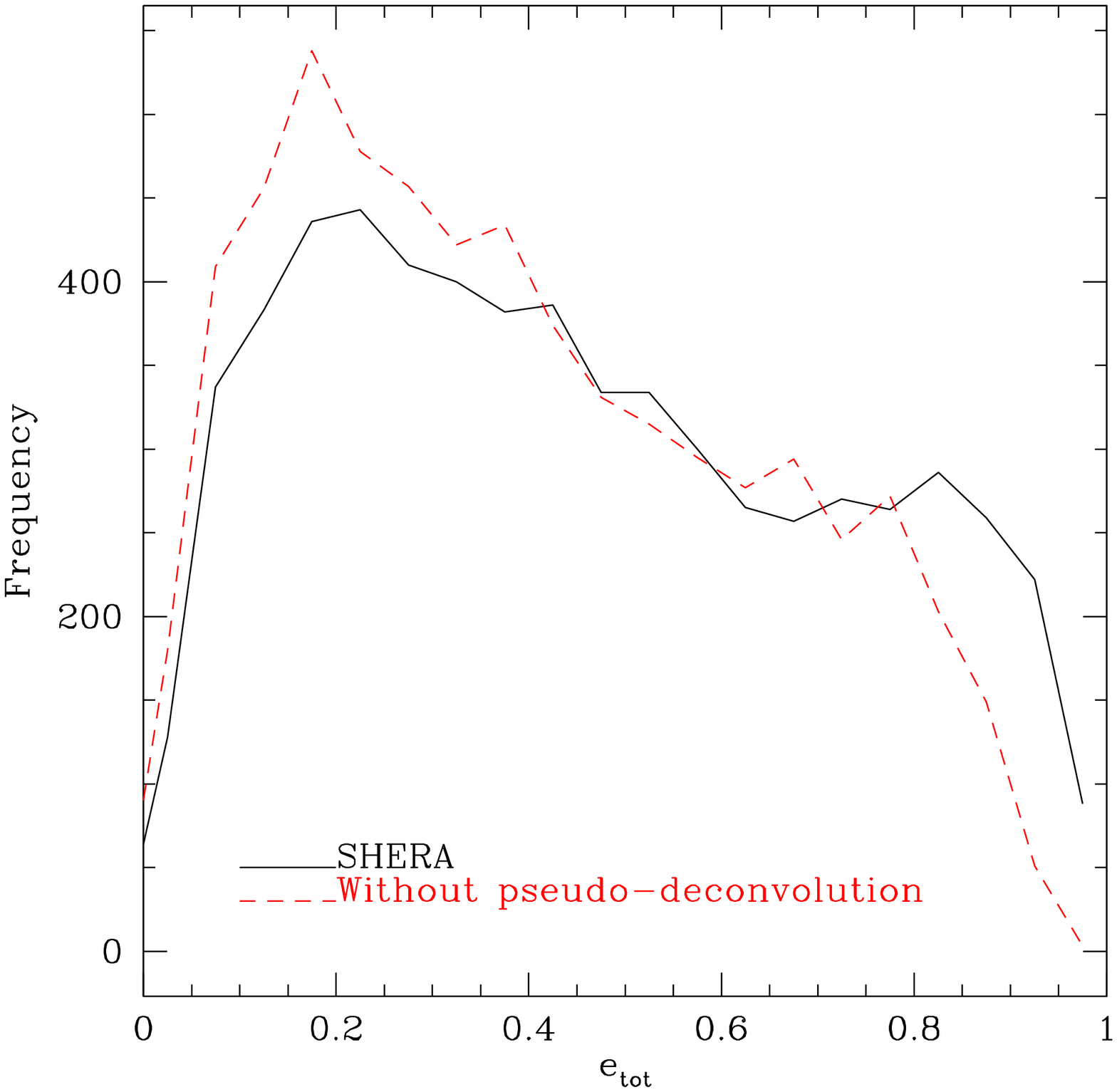}
\caption{\label{F:nopdconv}
Comparison of {\sc shera} simulations versus those that do not include
the pseudo-deconvolution of the ACS PSF.  {\em
  Top:} Scatter plot comparing the values of one ellipticity
component, $e_1$, for each galaxy. The solid line shows the $1:1$
line; \newtext{the dashed line shows the best-fit line, reflecting a
  factor of 0.9 multiplicative offset.} {\em Middle:} Scatter plot
comparing the values of resolution factor, $R_2$, for each galaxy.
\newtext{The dashed best-fit line reflects a $\sim 15$ per cent offset
  for the poorly-resolved galaxies.} 
{\em Bottom:} Histogram of $e_\mathrm{tot}=(e_1^2+e_2^2)^{1/2}$ values
in the two simulation sets.}
\end{center}
\end{figure}

The results of comparing these simulations without
pseudo-deconvolution, to the regular {\sc shera} simulations without
added noise, are
shown in Fig.~\ref{F:nopdconv}.
Statistical analysis reveals the same trends in the data with noise,
however they are less visually apparent.

As shown in the top panel of Fig.~\ref{F:nopdconv}, when we compare
individual galaxy ellipticities in the {\sc shera} simulations and those
simulations 
without pseudo-deconvolution, the latter tend to be rounder.  The
effect is, unsurprisingly, worse at large ellipticity.  In the
middle panel, we can see that the observed sizes are also affected:
the galaxies appear to be larger compared to the SDSS PSF than they do in
the simulations that account for the ACS PSF.  As expected, the trend
is more important for less well-resolved galaxies.  Finally, we can
see on the bottom that there is a concrete effect on the
histogram of total ellipticities, with the trend suggested in the top
panel that the simulations without pseudo-deconvolution yield a
rounder galaxy population overall.

As a consequence of the overall rounder galaxy population, the
inferred galaxy RMS ellipticity \erms\ is $\sim
0.31$, in contrast to the finding from simulations including
pseudo-deconvolution (Fig.~\ref{F:ermsmag}) that it is
$\sim 0.36$.  Moreover, we find that the inferred shear changes
slightly, becoming less negative by 1 per cent.  \newtext{Given the evidence that the
 PSF-matching is important in determining the observed galaxy
properties at the $\sim 5$--10 per cent level, and Sec.~\ref{SSS:shearcalibtrends-nonoise} showed
that the shear calibration for re-Gaussianization depends on galaxy
properties, the different derived shear calibration is likely due to
the fact that not deconvolving the ACS PSF is equivalent to
simulating a slightly larger, rounder galaxy population.  This fact
does {\em not} invalidate the utility of the STEP2 simulations for basic
testing of PSF-correction algorithms; it simply means that to
constrain shear calibrations in some hypothetical ground-based dataset
to better than 1 per cent for this particular galaxy population, one
should use simulations that (a) more closely mimic the imaging
conditions at that particular telescope, (b) include other steps in
the image processing, such as the need to combine multiple exposures,
and (c) include pseudo-deconvolution to more faithfully represent the
intrinsic properties of the galaxy sample.} 

The effects that are described in this section will be more important
for several limiting cases: (1) ground-based images with better
resolving power than SDSS, \reftext{such as Subaru Suprime-Cam (for which the
median seeing is 0.7\arcsec\ and the pixel size 0.2\arcsec, both
numbers a factor of $\sim 2$ smaller than for SDSS)} and (2) ground-based images that are
deeper, including more galaxies that are intrinsically small and
faint.  \reftext{We have explicitly tested the first scenario, using a typical
Subaru Suprime-Cam PSF, and found that the effects of ignoring
pseudo-deconvolution on the observed galaxy sizes are nearly twice as
large as for SDSS.}  Thus, for all upcoming surveys, image simulations that rely
on space-based data to precisely calibrate shears must
account for the PSF if they want to simulate a realistic galaxy
population using {\sc shera}.

\subsection{Impact of using Tiny Tim rather than observed PSFs}\label{SS:testtinytim}

\newtext{As pointed out in Sec.~\ref{SS:cospsf}, there are some known
  issues with the Tiny Tim PSFs that are used for PSF correction in
  \cite{2007ApJS..172..219L} and \cite{2007MNRAS.376...13M}, and that
  are used in this work and included with the associated data
  release.  Consequently, we must estimate the impact of using them
  rather than real stars (from dense stellar fields, with the same
  primary/secondary separation) when performing the
  pseudo-deconvolution step before matching to the target ground-based
  PSF.}

\newtext{For this test, we used a random subsample of COSMOS galaxies,
  and compared the observed resolution factors ($R_2$) with respect to
  the SDSS PSF when we used {\sc shera} with the same input
  parameters, only varying which COSMOS PSF was used for the
  pseudo-deconvolution.  We found that for well-resolved galaxies, the
  $R_2$ value was 0.4 per cent larger when using the real PSF stars
  than using the Tiny Tim models; at our lower resolution limit, they
  were typically 1.2 per cent larger.  For context, we saw in
  Fig.~\ref{F:nopdconv} that at the resolution limit, if we did not
  pseudo-deconvolve but instead ignored the ACS PSF entirely, the
  resolution factors in the simulated data differed by 16 per cent.
  Thus, crudely speaking, pseudo-deconvolution using the Tiny Tim PSF models effectively
removes $\gtrsim 93$ per cent of the impact of the ACS PSF in the final
simulated images.}

\newtext{The practical barrier at this time to simply using the
  stellar images for pseudo-deconvolution is that the stellar fields
  are typically not deep enough to get a very high $S/N$ PSF estimate
  on a per-star basis; the first diffraction ring is barely, if at
  all, visible when looking at a single star.  Thus, a reliable PSF
  interpolation routine would be necessary to fit for a high $S/N$ PSF
  model as a function of CCD position, which includes a non-negligible
  amount of development and testing to validate it.  While such
  development is ultimately necessary for very high precision tests
  using {\sc shera}, we defer it to future work.}

\subsection{Impact of noise in original COSMOS images}\label{SS:cosnoise}

\newtext{The noise in the original COSMOS galaxy postage stamps is sheared and
convolved with the target PSF.  In this section, we concretely
demonstrate the impact of that low noise level on the simulated
images.}

\newtext{To carry out this test, we take a subset of the simulated SDSS images,
and do the following operations:
\begin{itemize}
\item We start with the CTI-corrected and cleaned galaxy postage
  stamps from Sec.~\ref{S:imageprep}, but in addition to masking out
  all additional objects in the postage stamp, we also mask out the
  central object.  This gives us a standard galaxy-size postage stamp
  but with only a correlated noise field, no real objects.
\item We process it with a modified version of {\sc shera} using the
  same target PSF as when simulating the real SDSS data, including
  forcing the code to use the same normalizing factors for the flux as
  when making the simulations that do have the galaxy present.  This
  ensures that the normalization of the resulting sheared, PSF-convolved noise
  field is the same as that in the standard galaxy
  simulations.  To make it easier to detect systematic effects, we
  impose a relatively large shear, $\gamma_\mathrm{input}=0.1$.
\item We then add the resulting sheared, correlated noise fields to
  the (shear-free) galaxy simulations used in
  Sec.~\ref{SS:validation}, without and with noise added to match
  SDSS.
\end{itemize}
}

\newtext{We compare the original noiseless simulations versus
those that have the new correlated noise fields added, constructing $\Delta \hat{e}_1$, and $\Delta
\hat{e}_2$ (for the PSF-corrected galaxy shapes).  This comparison reveals systematic offsets in properties due to the
correlated noise fields that result from running a COSMOS noise field
through {\sc shera}.  The results, with different input shear values
($\gamma_\mathrm{input}$) 
for the noise fields, are consistent with 
}
\beq
\frac{\Delta\hat{\gamma}}{\gamma_\mathrm{input}} \sim 0.01.
\eeq

\newtext{As a practical consequence, this means that the original
  noise in the COSMOS images, when sheared, does not affect our
  ability to test shear recovery to the per cent level.  However, if we
wish to constrain shear calibration to well below the per cent level,
then this effect (in addition to the effects in
Sec.~\ref{SS:testtinytim}) must be accounted for.  We defer
consideration of this issue to future work.}

\section{Results: shear calibration}\label{S:results}

\newtext{In this section, we present one example usage of the {\sc
    shera} pipeline to assess the calibration of lensing shear
  measurements using the re-Gaussianization PSF correction method.
  %More basic technical results about the {\sc shera} pipeline,
  %including technical validation via a comparison between the
  %simulated and real SDSS imaging in the COSMOS region; an assessment 
  %of the necessity of including pseudo-deconvolution of
  %the ACS PSF; comparisons between simulations using the Tiny Tim versus real
  %PSFs; and a concrete demonstration of the impact of the low but
  %non-zero noise level in the original COSMOS image (which is also
  %sheared and convolved with the target PSF) are found in
  %Appendices~\ref{S:validation}--\ref{S:cosnoise}.
}

For these tests, we chose 14 sets of $(\gamma_1, \gamma_2)$ 
shears to simulate; these are shown in the top
panel of Fig.~\ref{F:basicshearcalib}.  For each of the 17~706 galaxies used for
the simulations to validate the pipeline
(Sec.~\ref{SS:validation}), we used {\sc shera} to generate 56 additional
simulated galaxies: 14 shear sets, 2 noise options (noiseless and
noisy), and 2 orientations (original and 90 degree rotated).  We then
select galaxies in various ways (to be described shortly), and defined
a weight function for each galaxy:
\beq\label{E:weightfunc}
w_i = \frac{w_{\mathrm{COSMOS},i}}{\sigma_{e,i}^2 + \erms^2}.
\eeq
Here the numerator $w_{\mathrm{COSMOS},i}$ is the inverse of the
postage stamp selection function (Sec.~\ref{SS:cosmos} and
Fig.~\ref{F:fracstamps}) meant to remove our selection bias against
physically large galaxies.  \newtext{The denominator only is
  significant for shear estimates using the simulations with sky
  noise, since the shape measurement error is negligible for those
  without added noise.}  For typical galaxy-galaxy lensing analyses,
there is an additional factor in this weight function:
$\Sigma_\mathrm{c}^{-2}$, which corresponds to optimal weighting in
the case that lens and source redshifts are both known.  For the
simulations, we cannot easily include such a factor, because
simulating the SDSS photo-$z$ would require simulating $ugriz$ data and
processing it with the SDSS {\sc Photo} pipeline.

To estimate the shear, we then defined
\beq\label{E:shearest}
\hat{\gamma}_\alpha = \frac{\sum_i w_i (\gamma_{\alpha,
    i,\mathrm{orig}} + \gamma_{\alpha, i, \mathrm{rot}})}{4 S_\mathrm{sh} \sum_i w_i}
\eeq 
in terms of the PSF-corrected shapes, for shear components $\alpha =
1, 2$ and galaxies $i$.  The $S_\mathrm{sh}$ factor, or shear
responsivity, represents the response of our particular ellipticity
definition Eq.~\eqref{E:ellipdef} to a shear; it is equal to
$1-\erms^2$.% \approx 0.86$ (in Sec.~\ref{SSS:shearcalib-noisy} we will
%address nuances related to the estimation of \erms\ in the presence of
%noise, and how that can bias the shear responsivity and therefore the shear).  

\subsection{Without sky noise}\label{SSS:shearcalib-nonoise}

\begin{figure}
\begin{center}
\includegraphics[width=2.7in,angle=0]{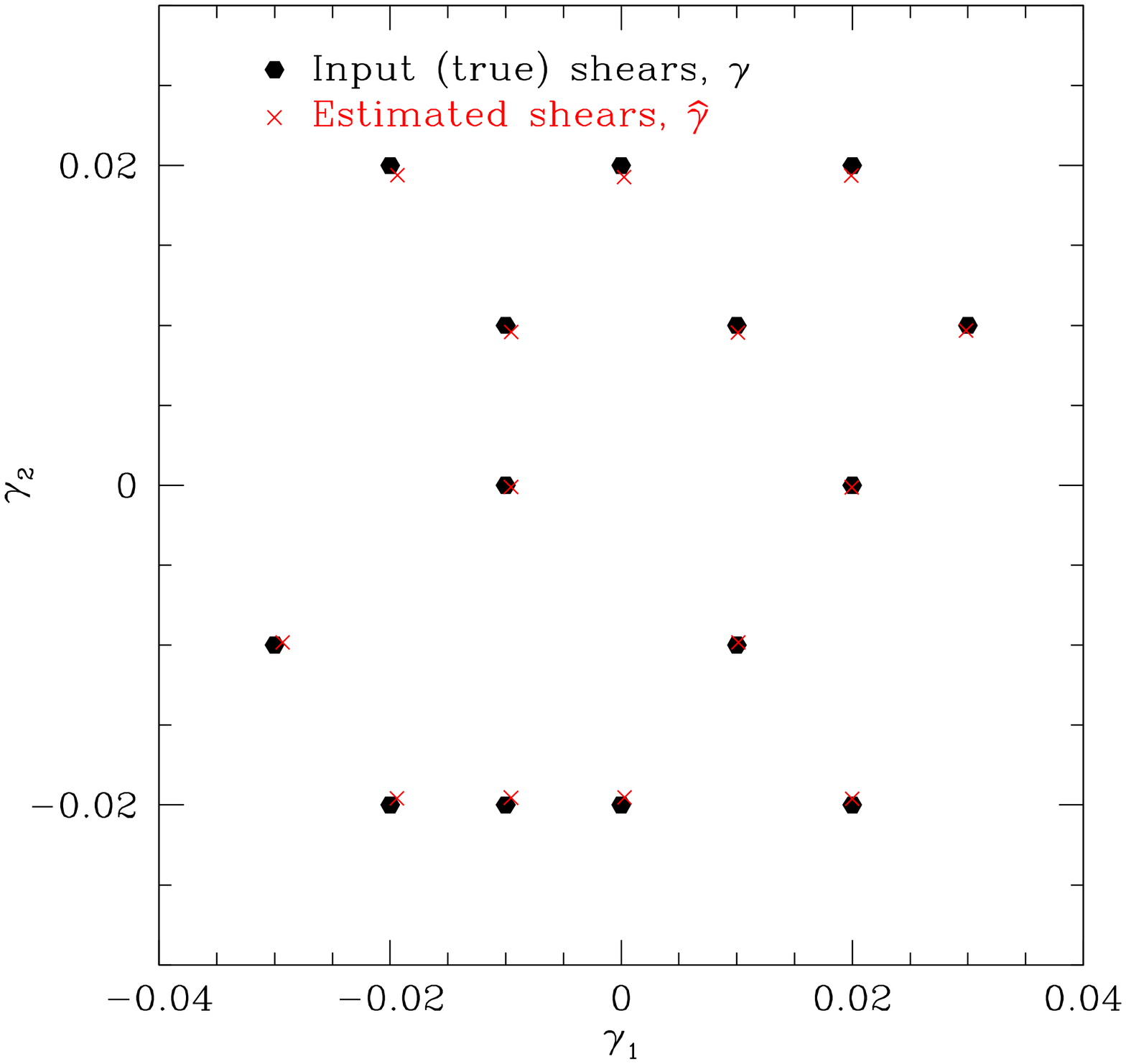}
\includegraphics[width=2.7in,angle=0]{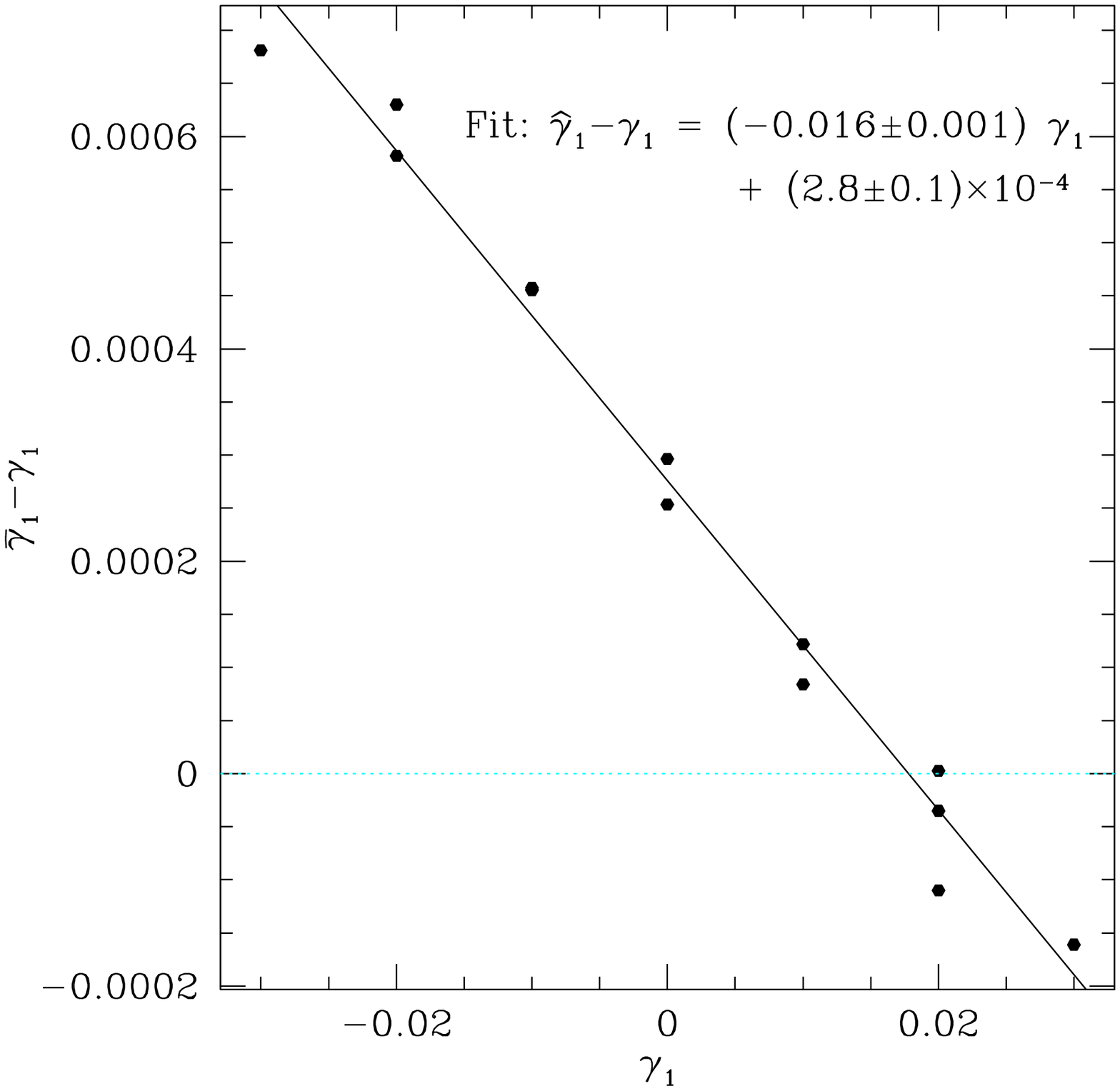}
\includegraphics[width=2.7in,angle=0]{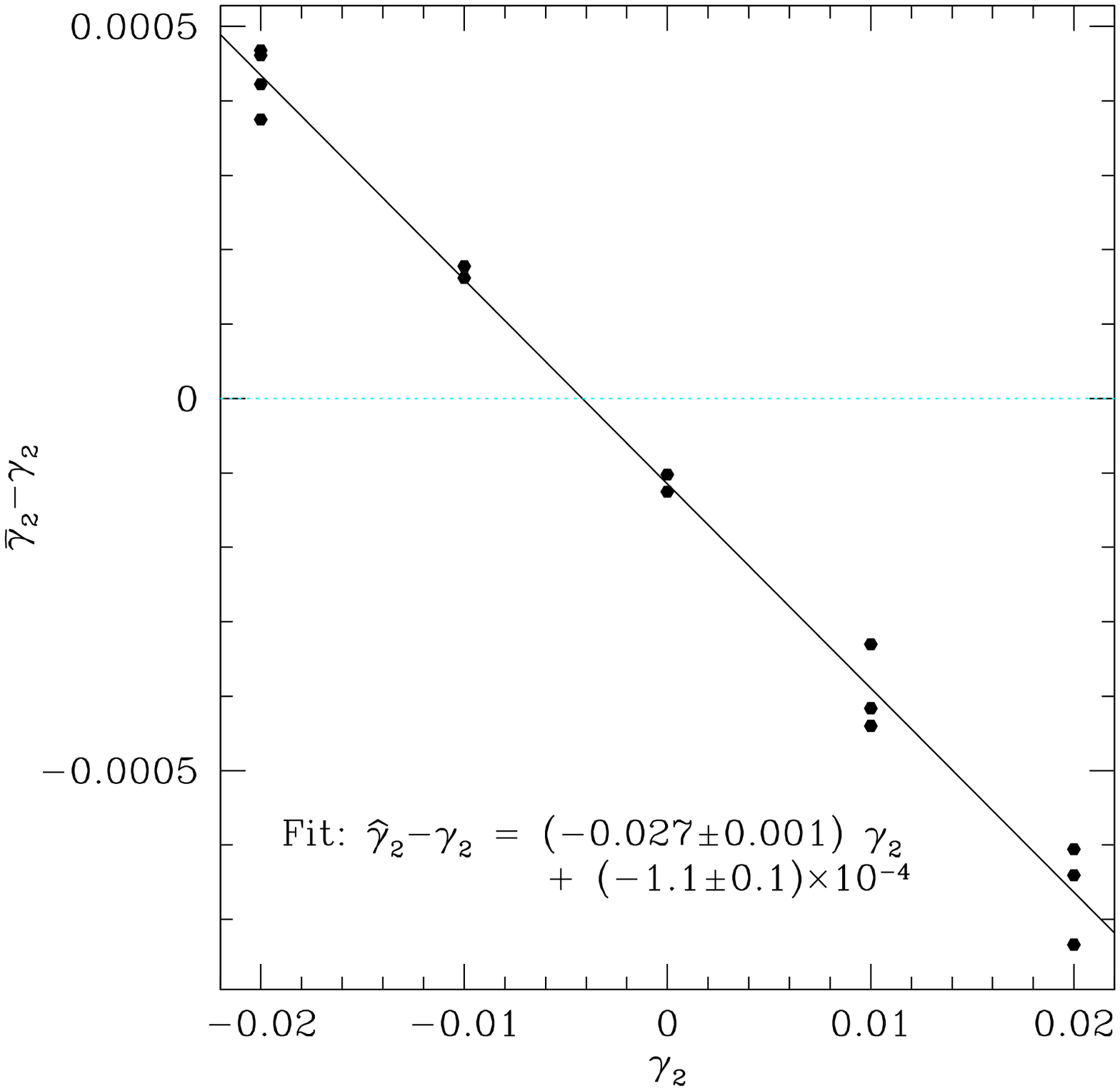}
\caption{\label{F:basicshearcalib}
Results of shear calibration tests for the re-Gaussianization
PSF-correction technique, using the `noiseless' COSMOS simulations
with flux and resolution cuts. 
{\em
  Top:} The true input shear values, and the estimated ones.  {\em
  Middle:} The error in recovered shear component 1,
$\hat{\gamma}_1-\gamma_1$, as a function of the input shear.  The
best-fitting line is also plotted.
{\em Bottom:} Same as middle, but for the other shear component.}
\end{center}
\end{figure}

We first present results using the `noiseless' simulations, which we
expect to do with very small statistical errors since there is essentially no
measurement error, and the use of original and rotated shapes should
effectively eliminate the shape noise.  In order to select an
approximately reasonable galaxy population for this test, we use the
COSMOS $F814W$ magnitudes and the typical SDSS colours to impose an
approximation of the $r<21.8$ cut (corresponding to that for the real 
source catalogue).  We also require a resolution factor $R_2>1/3$ for
both the original and rotated galaxy; this results in a sample of
$6~160$ galaxies.  The comparison between the estimated shears
$(\hat{\gamma}_1, \hat{\gamma}_2)$ and the true ones for each of the
14 simulations with shear is shown in
Fig.~\ref{F:basicshearcalib}. 

As shown, there is a clear detection
of non-zero calibration bias and additive PSF systematics.  When
fitting to 
\beq\label{E:calibbias}
\hat{\gamma}_\alpha - \gamma_\alpha = m_\alpha \, \gamma_\alpha + c_\alpha\;,
\eeq
both the slope (calibration bias) $m$ and the additive constant $c$
differ from zero. %in a statistically significant way 
%for both shear components.  
The calibration
biases are $-1.6 \pm 0.1$ and $-2.7\pm 0.1$ per cent for the two shear
components.  The fact that these biases differ for the two components,
and that the latter is worse than the former, is consistent with
results of \cite{2007PASP..119.1295H} and \cite{2007MNRAS.376...13M}.  The
standard explanation %for this result 
is that the pixel resolution is
effectively a factor of $\sqrt{2}$ worse along the diagonals of the
pixels than along the pixel direction.  \newtext{If we remove the
  weighting in Eq.~\eqref{E:weightfunc}, then the calibration biases
  change by $0.1$ per cent, the size of the $1\sigma$ error.  The
  sign and magnitude of this change can be explained by the weak but non-negligible correlation between the
  galaxy weights $w_\mathrm{COSMOS}$ (to account for the inability to
  create postage stamps for some of the larger galaxies) and the
  galaxy size, given the trends we will see in the calibration bias
  with galaxy size in Sec.~\ref{SSS:shearcalibtrends-nonoise}.}
  
%It is also apparent from the upper panel of Fig.~\ref{F:basicshearcalib} that the recovered
%shear for a given input shear has a slight dependence on the other
%input shear component; for example, at a given $\gamma_1$, a larger
%$\gamma_2$ results in a larger $\hat{\gamma}_1$.  
%The range of input shears that were used for this test is insufficient
%to check for some non-linear response to shear, i.e. higher order
%polynomials in $\gamma_\alpha$ in Eq.~\eqref{E:calibbias}.

The nonzero additive contamination ($c$ values) can be explained by the nonzero average PSF
ellipticity, which is imperfectly removed from the galaxy images by
re-Gaussianization.  For context, the typical PSF ellipticity in SDSS
is $\sim 0.05$, so the fractional contamination is 
$|c_1/e_{1,\mathrm{PSF}}| \sim 5\times 10^{-3}$.

\subsection{Dependence on sample properties (noiseless)}\label{SSS:shearcalibtrends-nonoise}

In figure 5 of \cite{2005MNRAS.361.1287M}, predictions were shown for
the shear calibration bias for the re-Gaussianization method using
noiseless simulations, for exponential and de Vaucouleurs galaxies,
using \newtext{a Kolmogorov turbulence-induced profile, i.e.
$\ln\tilde G({\bmath k})\propto -k^{5/3}$}.  As shown there, the calibration biases depend on
the galaxy profile, being more negative for de Vaucouleurs profiles
than for exponentials; on the resolution factor, being more negative
at intermediate resolutions ($R_2 \sim 0.6$) and closer to zero for
resolutions at the lower ($1/3$) and upper ($1$) limits; and on the
intrinsic ellipticity, being more negative for more intrinsically
circular galaxies. We now test all of these predictions, again in
the case without measurement noise.

\begin{figure}
\begin{center}
\includegraphics[width=2.6in,angle=0]{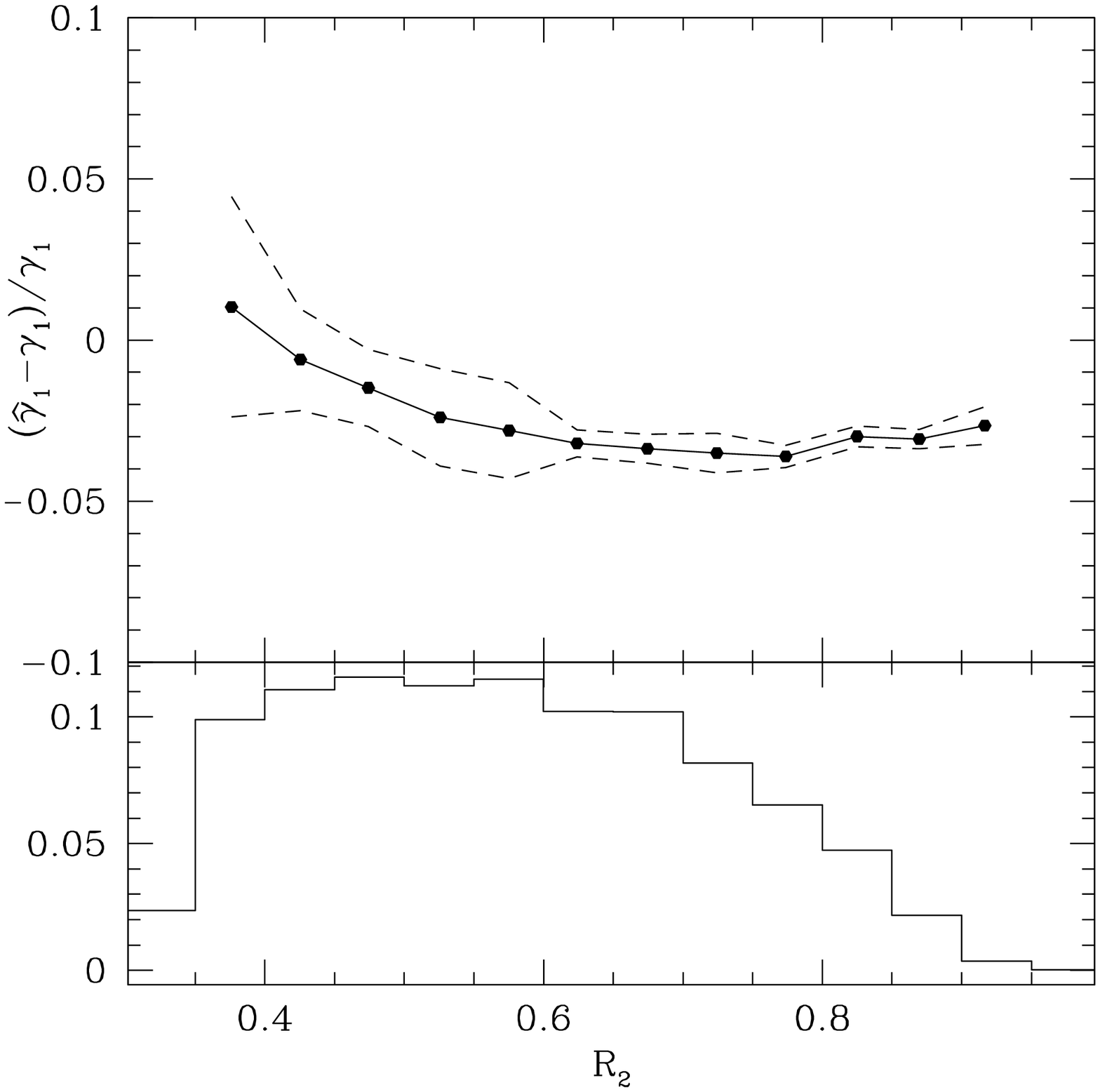}
\includegraphics[width=2.6in,angle=0]{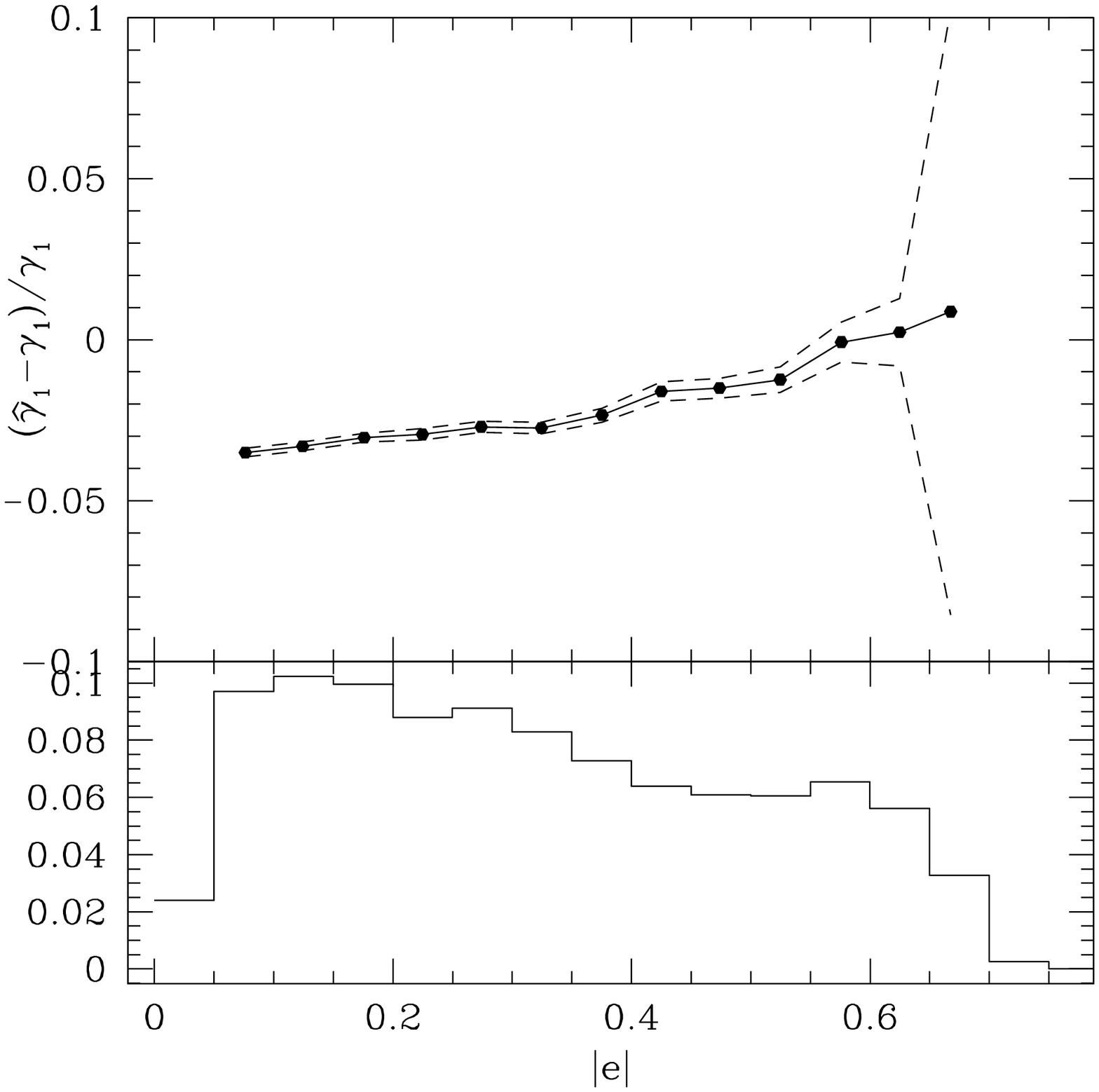}
\includegraphics[width=2.6in,angle=0]{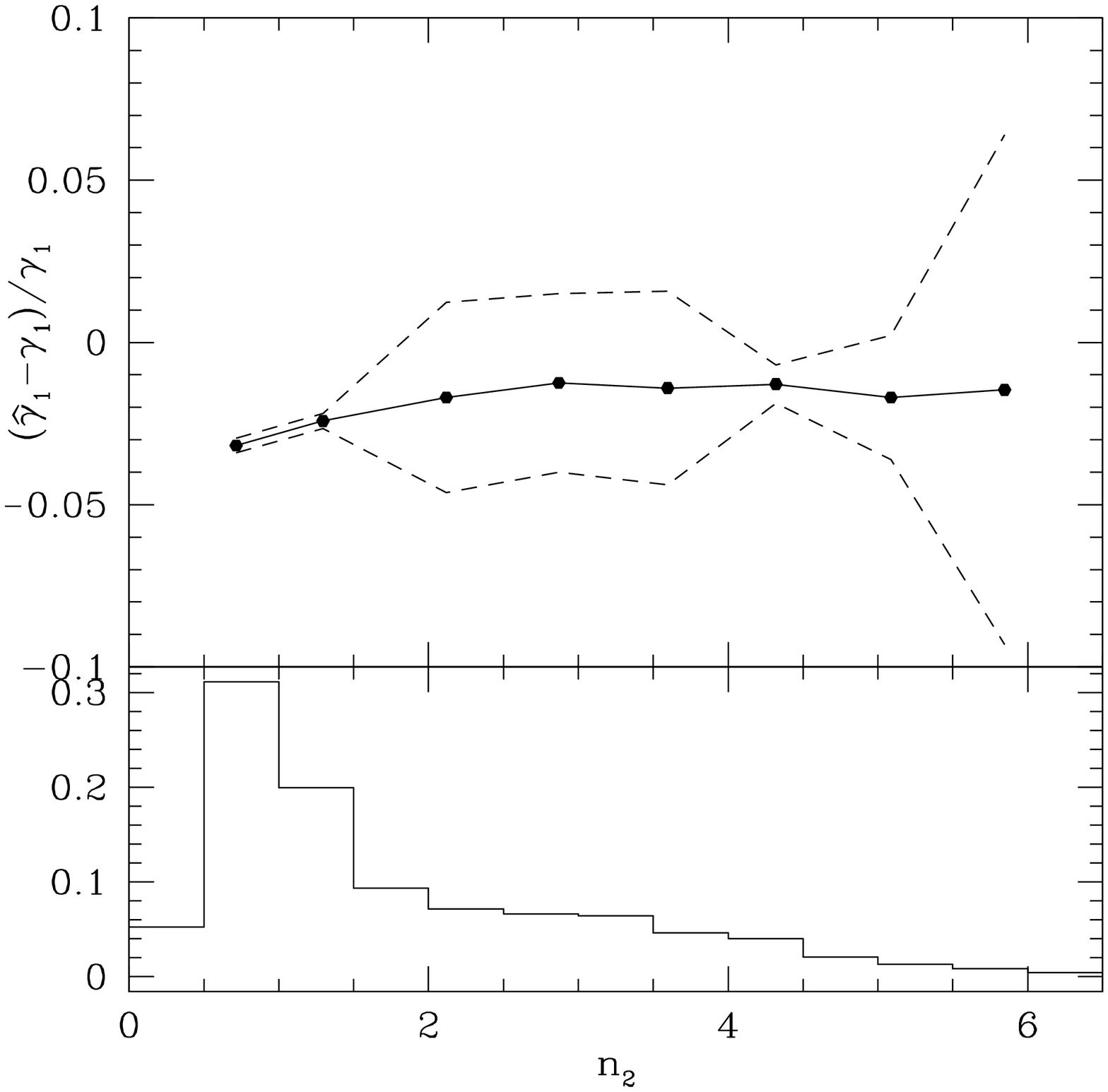}
\caption{\label{F:shearcalibtrends} Results of shear calibration tests
  for the re-Gaussianization PSF-correction technique, using the
  `noiseless' COSMOS simulations with flux and resolution cuts.
  {\em Top:} The fractional error in the recovered shear as a function
  of the galaxy resolution with respect to the PSF, $R_2$, along with
  the histogram of $R_2$ values for the noiseless simulations.  The points
  and solid line show the median trend; the dashed lines show the
  statistical uncertainty in the trend-line.  {\em
    Middle:} Same as top, but as a function of total ellipticity
  $|e|$.  {\em Bottom:} Same as top, but as a function of S\'ersic
  $n_s$ as determined from the COSMOS images.}
\end{center}
\end{figure}

To carry out these tests, rather than computing an ensemble
$\hat{\gamma}$ to compare with the input value via weighted summation
over all the individual shear values, as in Eq.~\eqref{E:shearest}, we
instead consider individual shear estimates for every single galaxy.
To estimate a shear for each galaxy, we use the original and rotated
shape measurements for each component (4 measurements), and write four
equations based on eq. (2-13) of \cite{2002AJ....123..583B} to express
those observables in terms of the intrinsic shape and the applied shear (4
unknowns).  These equations are nonlinear functions of the intrinsic
shape and applied shear; we solve this nonlinear system of equations
using the IDL implementation of Broyden's method to estimate
$\hat{\gamma}_{\alpha,i}$ (for shear component $\alpha$, galaxy $i$).  

Given the individual shear estimates, we can compute the fractional
error in each one, $(\hat{\gamma}_{\alpha,
  i}-\gamma_\alpha)/\gamma_\alpha$ in bins in galaxy properties.  The properties used for this
test are the resolution factor $R_2$; the total ellipticity
$\etot = \sqrt{e_1^2+e_2^2}$; and the S\'ersic $n_s$ derived from single
component fits to the
COSMOS images from \cite{2007ApJS..172..434S}, using {\sc GIM2D}
\citep{1998ApJ...507..585M}. The results are shown in
Fig.~\ref{F:shearcalibtrends}.

As shown in the top panel, we find a calibration bias that approaches
zero for the lowest and highest resolutions in our
sample, and goes as low as $-4$ per cent for the galaxies at $R_2\sim
0.7$.  This finding is qualitatively similar to that from the
noiseless simulations of \cite{2005MNRAS.361.1287M}, with the
difference being that these results intrinsically average over a more
realistic galaxy population (all morphologies rather than
single-component S\'ersic profiles with $n_s=1$ and $4$) and real PSFs. 

The middle panel of Fig.~\ref{F:shearcalibtrends} shows that the
calibration bias tends to be most negative ($\sim -3$ per cent) for
galaxies that are nearly circular, and increases to $0$ for $|e|\sim
0.6$, becoming slightly positive at even higher ellipticities (where, however,
there are very few galaxies and therefore the statistical significance
of the trend 
above $|e|>0.6$ is unclear).  This trend is again similar to that
from \cite{2005MNRAS.361.1287M}.%, suggesting that those earlier results
%were not dependent on the overly simplistic nature of the galaxy
%profiles simulated.

The bottom panel of Fig.~\ref{F:shearcalibtrends} shows the trend with
S\'ersic $n_s$, which gives a calibration bias that is most negative
for low $n_s$ (exponential galaxies) and is closer to zero for higher
$n_s$.  This finding is the opposite of that from
\cite{2005MNRAS.361.1287M}; however, it is important to note that in
this case the $n_s$ values are not exact representations of the real
galaxy morphologies, given that most galaxies show some deviation from
a perfect elliptical S\'ersic profile (either having more small-scale
structure, or being clearly composed of multiple components such as a
bulge and a disk).  Thus, it is unclear that the results shown here as
a function of $n_s$ can truly be directly compared with those from
\cite{2005MNRAS.361.1287M}, without first assessing which of the
COSMOS galaxies are in fact consistent with a featureless %smooth elliptical
S\'ersic model.  Moreover, the statistical significance of the
observed trend is fairly weak.

\subsection{Noisy simulations}\label{SSS:shearcalib-noisy}

We now consider the shear calibration bias in simulations with %that have
sky noise.  However, because the sky level in the SDSS
imaging of the COSMOS field is atypically high
(Sec.~\ref{SS:sdssobscosmos}), we use a set of simulations that are
otherwise identical to the ones from Sec.~\ref{SSS:shearcalib-nonoise}, but
with sky noise that is 15 per cent lower (in the standard deviation).  %We opt
The more typical noise level in these simulations makes them more like typical SDSS data.

There is an important subtlety when using a sample that has noise to
estimate shear calibration bias: %unlike in the previous subsection
%that used noiseless simulations, 
we must be very careful when
selecting galaxies to use for the shear estimation.  In Sec.~\ref{SSS:shearcalib-nonoise}, we simply
approximated an $r<21.8$ cut in order to get a roughly similar galaxy
population as in the real data.  Here, however, we know that galaxies
that have a noise fluctuation such that they are harder to detect
would realistically not be included in our sample, and that is a good
thing given that their shears should be unreasonably difficult to
measure.  In order to perfectly mimic our sample selection in the real
data, we would have to process the simulated images in the exact same
way as the real data, using {\sc Photo}.  However, there are other
practical obstacles to constraining the shear calibration in all of
SDSS using this simulation set (e.g., the fact that we have not sought
to carefully mimic the distribution of observing conditions throughout
the entire SDSS area).  So, we instead use a simpler
approximation of our selection in the real data, as a basic
demonstration of the power of the {\sc shera} code rather than as a
quantitative estimate of the SDSS shear calibration bias.% to be used to
%interpret science results.

Our crude selection relies on the estimated shear measurement errors $\sigma_\gamma$
for each galaxy.  The re-Gaussianization code estimates
$\sigma_\gamma$ (per component) by taking the input value of sky variance
$\sigma_\mathrm{sky}^2$ from
Eq.~\eqref{E:skyvar}, and using \citep{2002AJ....123..583B}
\beq
\sigma_\gamma = \frac{\sigma_e}{2} = \frac{\sqrt{4\pi n}\sigma_\mathrm{sky}}{f_w} = \frac{2}{S/N}\;,
\eeq 
where $f_w$ is the weighted flux (by the adaptive moment
matrix). This equation
is only approximate, and assumes Gaussianity of the PSF-convolved
galaxy image.

Thus, our approach to object selection in these simulations %, for which
%we do not have SDSS model magnitudes, is
is as follows:
\begin{enumerate}
\item We examine the real SDSS source catalogue to find what $S/N$ is implied by the
  $\sigma_\gamma$ values at our limiting magnitude $r=21.8$.  While
  there is a range of $\sigma_\gamma$ values at fixed magnitude,
  %depending on the galaxy size, shape, and radial profile, 
we find
  that $r<21.8$ corresponds to $\sigma_\gamma \lesssim 0.21$  (or $S/N > 9.5$). 
\item We impose that $\sigma_\gamma$ cut on 
  the
  original and rotated simulated images\footnote{Technically, in the real data,
    we impose our $S/N$ cut in $r$.  However, there is also a
    magnitude 
    cut in $i$ which, given the relation between the sky variances in
    the two bands and the typical galaxy colours, corresponds to a
    similar $S/N$ cut in that band.  Thus, requiring the magnitude
    and resolution cuts in $r$ and $i$ in the real data is parallel to
  our imposition of $S/N$ and resolution cuts in the simulations for
  both the original and the rotated images.}.  In practice, for a
given set of noisy simulations, typically 4100 galaxies pass this cut,
the resolution cut, and the $\etot <2$ cut.
\end{enumerate}

The first aspect of the shear estimation that we can test using the
noisy simulations is the shear responsivity calculation, which is in
the denominator of our shear estimator Eq.~\eqref{E:shearest} and is
\beq
S_\mathrm{sh} = {1-\erms^2}\;,
\eeq
where the rms ellipticity per component is ideally defined as a
weighted sum over the ellipticities of the source population, 
\beq\label{E:erms-nonoise}
\erms^2 = \frac{1}{2}\left[\frac{\sum_i w_i (e_1^2 + e_2^2)}{\sum_i w_i}\right]
\eeq
given true, noiseless, $e_1$ and $e_2$ values (the mean ellipticity per component, $\langle e_1\rangle = \langle
e_2\rangle = 0$).  In real data, we lack a noiseless
estimator of $e_1$ and $e_2$, so we must 
estimate $\erms$ using our noisy estimated $\hat{e}_1$ and
$\hat{e}_2$:
\beq\label{E:erms-noisy}
\hat{e}_\mathrm{rms}^2 = \frac{1}{2}\left[\frac{\sum_i w_i
    (\hat{e}_1^2 + \hat{e}_2^2 - 2\sigma_e^2)}{\sum_i w_i}\right].
\eeq
Errors in the estimated \erms\ from
Eq.~\eqref{E:erms-noisy} propagate into the shear estimates via 
Eq.~\eqref{E:shearest}.  

\begin{figure}
\begin{center}
\includegraphics[width=\columnwidth,angle=0]{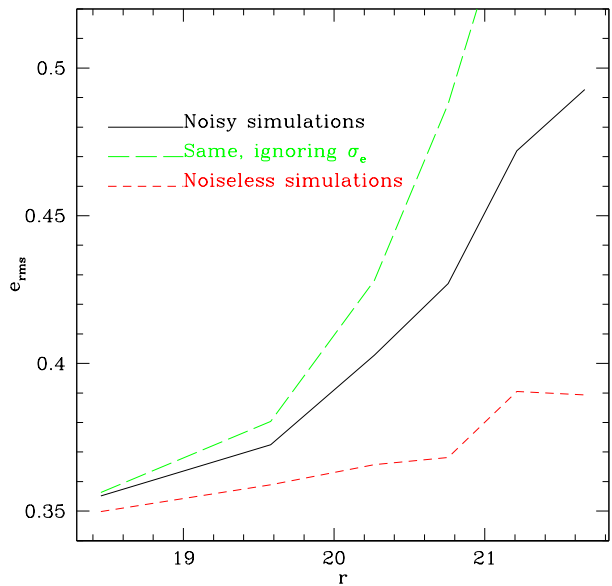}
\caption{\label{F:ermsmag} The RMS ellipticity \erms\ as a function of
  the apparent magnitude.  Results are shown for the simulated data
  with noise, before (long-dashed) and after (solid) subtracting off our estimates of the
  shape measurement error due to photometric noise.  The results are
  also shown for the same galaxies using the ellipticities from the
  noiseless simulations (dashed line).} 
\end{center}
\end{figure}

For our noiseless simulations, we come as close as possible to being
able to carry out a `true' \erms\ estimate,
Eq.~\eqref{E:erms-nonoise}.  In practice, these simulations do include
a low level of noise from the COSMOS simulations, but we at least know
that this will tend to increase the estimated \erms; thus, the \erms\
from the noiseless simulations serves as an upper limit on the true
\erms.  Hence, we first estimate \erms\ as a function of
magnitude for the noiseless and the noisy simulations, to estimate how
much the inaccurate $\sigma_e$ estimates might be
biasing \erms, $S_\mathrm{sh}$, and the estimated
shears. The results of this estimate are shown in
Fig.~\ref{F:ermsmag}. 

As shown, the RMS ellipticity is close to flat with magnitude in the
noiseless simulations.\footnote{In an upcoming paper, Reyes et
  al. (2011, {\em in prep.}), we will present evidence that the
  deviations from flatness in these simulations are due to small
  levels of nonlinearity in the PSF correction that lead to
  non-Gaussian error distributions.}  In contrast, in the noisy
simulations, it appears to increase significantly at fainter
magnitudes, even after our attempts to subtract off the measurement
error.  This result implies that our shape measurement errors are
underestimated.  %While we can use multiple noise realizations of
%galaxy images to more realistically estimate the shape measurement
%error, 
Fig.~\ref{F:ermsmag} gives us a way to assess how much the
estimated shear responsivities are incorrect due to our underestimated
shape measurement error.  We find that for the noisy simulations, the
responsivity is estimated as $0.84$, whereas in the noiseless ones, it
is $0.86$.  Thus, our responsivity is 2.5 per cent too low, so the
shears are overestimated by this amount in the noisy simulations.  In
the text that follows, when we quote shear calibration biases for
noisy simulations, they
include this shear overestimation due to shear responsivity error.

\newtext{The value of $\erms\sim 0.36$ in Fig.~\ref{F:ermsmag} may
  seem  inconsistent with that from the COSMOS data for bright
  galaxies, figure 17 of \citet{2007ApJS..172..219L}, which gives
  $\erms\sim 0.27$.  However, as described in
  Sec.~\ref{SS:shapecatalog}, the shape estimates in these two works
  differ in use of circularly weighted \citep{2007ApJS..172..219L}
  vs. elliptical-weighted adaptive moments (this work), such that we {\em should} find a larger \erms\ in
  this work.  While Eq.~\eqref{E:relateellip} relating
  the two ellipticity measurements is only exactly valid in certain
  unrealistic limits (Gaussian profiles), we can nonetheless use it to
  estimate the effect of this different shape definition.  If our 
  simulation ellipticities are transformed using
  Eq.~\eqref{E:relateellip} to those that are expected using RRG, then
  the resulting $\erms=0.28$, quite similar to that from
  \citet{2007ApJS..172..219L}.  Residual difference may be due
  to the cases where Eq.~\eqref{E:relateellip} fails to relate the two
  ellipticities correctly. 
} 

Next, we carry out a similar shear calibration bias calculation as in
Sec.~\ref{SSS:shearcalib-nonoise}, but with the noisy simulations,
using the $\sigma_\gamma$ cut
discussed earlier in this section to remove those galaxies that have
low significance detections.  %In this case, the shear estimates
%are naturally noisier than in the previous section (even with the
%rotated galaxy pairs to eliminate shape noise), and as a result 
We use
two independent noise realizations of each galaxy pair to reduce the noise.  In comparison with the
noiseless case shown in Fig.~\ref{F:basicshearcalib}, for which we had
found ensemble calibration biases of $m_1=-1.6 \pm 0.1$  and
$m_2=-2.7\pm 0.1$, the noisy case gives calibration biases of
$m_1=-3.8$ and $m_2=-4.3\pm 2.5$ per cent\footnote{\newtext{When calculating
    unweighted sums over the galaxies, without the weight factor in
    Eq.~\eqref{E:weightfunc}, the calibration biases worsen by $-2$
    per cent, just under $1\sigma$.  This difference reflects 
    the fact that the bias is worse when we include galaxies near the flux limit, which get downweighted due to their
    larger $\sigma_e$.}}.  This result is just $2$ per cent
worse than the noiseless case; however, the net $-4$ per cent
calibration bias also includes a $+2$ per cent bias due
to incorrect shear responsivity, implying a $-6$ per cent bias due to
insufficient dilution correction and noise rectification bias.%(a factor of 3 worse than the
%noiseless case, where the net $-2$ per cent calibration bias did not
%include any errors in the shear responsivity).

We can compare these results with noisy simulations against those from
the STEP2 simulations \citep{2007MNRAS.376...13M}. There are a number
of reasons why we do not expect the results to agree: for example, the
fact that a deeper sample population was being simulated in STEP2, with
intrinsically different properties; the fact that higher resolution
Subaru PSFs and pixel scale were used in STEP2; and the fact that the ACS PSF was not
pseudo-deconvolved before convolving with the ground-based PSF.
Nonetheless, we compare against the following STEP2 results: the shear
calibration bias for re-Gaussianization (denoted `RM' in \citealt{2007MNRAS.376...13M}) was typically $-2.5$ per cent; and the shear calibration
bias is more negative for fainter magnitudes.  Our $-4.0\pm 2.5$ per
cent calibration bias is statistically consistent with the STEP2
results.  Moreover, the fact that the calibration bias became more
negative as we moved from noiseless to noisy simulations is
qualitatively consistent with the STEP2 result that calibration bias
is more negative at fainter magnitudes (i.e., lower $S/N$).   A
detailed quantitative comparison is beyond the scope of this paper due to the many
intrinsic differences between these simulations.

\subsection{Limitations of these results}\label{SSS:limitations}

The shear calibration bias estimates in this section were intended 
primarily as a demonstration of one possible application (out of many)
of the {\sc shera}
pipeline.  Here we summarize why the specific numbers presented here should not be used
as a precise estimate of the shear systematics in
science papers using the SDSS re-Gaussianization shape catalogue:
\begin{enumerate}
\item In this section, we have averaged over all sources that are
  detected in the simulations.  In practice, the source population
  that is used depends on the lens redshift, since photometric
  redshifts are used to (roughly) select those sources that are behind
  the lenses.
\item When calculating signals for the real data, there is an
  additional weight factor $\Sigma_\mathrm{c}^{-2}$ to achieve
  optimal weight in the estimate of $\Delta\Sigma$.  This weight
  correlates with galaxy properties and, consequently, will change the
  mean calibration bias for a given sample population.
\item The range of observing conditions in the SDSS is quite broad.
  In practice these results might depend in detail on simulating a
  range of observing conditions.
  %For example, the COSMOS region has slightly
  %better seeing than average.  As a result, the resolution
  %distribution is not quite the same in the COSMOS region as in the full
  %survey.  The results in Sec.~\ref{SSS:shearcalibtrends-nonoise}
  %suggest a modest change in the average calibration bias as a result,
  %since the dependence on $R_2$ is not strong.
\item The real data use $r$ and $i$ band averaged shapes.  In practice
  these might have different shear calibration biases than the $i$
  band shapes used here, because of colour gradients changing the
  observed morphology slightly, and because of the different $S/N$ of
  detection in the two bands.
\item As demonstrated in Sec.~\ref{SSS:shearcalib-noisy}, the
  estimate of shear calibration depends on how the sample is cut in
  $S/N$.  Thus, our crude $\sigma_\gamma$ cut must be replaced with a
  more realistic approximation of the sample selection used in the
  SDSS data, based on the model flux from {\sc Photo}.
%\item As explained in more detail in Sec.~\ref{SS:shapecatalog} and
%  Reyes et al. 2011 (in prep.), the
%  old and new shape catalogs are nearly the same in terms of shear
%  calibration, but there have been slight updates that tend to change
%  it by $\sim 1.5$ per cent.  All results here correspond to the new shape
%  catalog, and those with the old shape catalog must differ as stated
%  here.
\item Using a realistic simulation, we should be able to understand
  the impact of selection biases in determining the shear
  calibration. \newtext{Because the simulations described here have
    several atypical features (particularly the use of 90 degree
    rotated galaxy pairs), the selection biases do not operate in the
    same way here as in real data.}
\end{enumerate}

A precise calculation of the shear calibration bias for the real
data would have to account for these effects.

%\subsection{Deblend failures}%
%
%\textbf{look for size, mag discrepancies}
%
%\subsection{Selection biases}
%
%\textbf{Multiple noise maps: what can we conclude about ellipticity,
%  resolution correlation?  Or, detectability vis a vis alignment with
%  PSF vs. not?}

\section{Discussion}\label{S:discussion}

We have described new, publicly-available software, {\sc shera}, that can be used to
simulate (optionally sheared) ground-based images with realistic
morphologies and any PSF that has worse resolution compared to COSMOS.
This software is independent of modeling assumptions about what galaxies
or PSFs look like, properly handles the pixel response functions, and
has been tested to sub-per cent precision \reftext{(Sec.~\ref{S:techvalidation})}.  The code has been publicly
released, along with CTI-corrected, cleaned COSMOS galaxy postage
stamps for a flux-limited sample at $F814W<22.5$.  This code should be
highly useful for realistically assessing the systematics of lensing
analysis (or indeed any other detailed image analysis, such as galaxy 
profile determination) in ground-based data, including the effects of
a range of observing conditions.  It should also allow for assessment
of parameter uncertainties and degeneracies, and selection biases, via
the generation of many noise realizations for a single galaxy.  As a
basic demonstration, we have shown a crude estimation of shear
calibration bias for SDSS single-epoch lensing analysis.

In order for the code and data that are described in this paper to be
more broadly applicable, there are several improvements that would be
needed.  First, we would like the ability to simulate data in other
passbands besides $i$ \newtext{(for which there is precedent, e.g. \citealt{2008APh....30...65F}
used data from the Ultra Deep Field to do so)}.  Realistic
galaxies have colour gradients that will cause an intrinsically
different appearance in significantly different bands.  While this may
be unimportant for current surveys that have statistical errors that
are $>5$ per cent, future surveys such as LSST that aim for better
than $1$ per cent precision in the lensing signal have correspondingly
a need for very well-constrained shear systematics.  Thus, such
effects must be handled realistically, and it is possible that
sufficient well-sampled data\footnote{Data with some instruments may
  not have a Nyquist-sampled PSF, depending on the choice of dither
  pattern.} in other bands and in random fields exists in the {\it HST}
archive that could be used for this purpose.  We specify random fields
because those that were selected due to, e.g., presence of a galaxy
cluster, may not have a representative morphological mix.  There is
additional value to obtaining data in other, random fields, given that
even with the relatively large (for {\it HST}) size of the COSMOS field, it
still exhibits significant cosmic variance (e.g.,
\citealt{2010ApJ...708..505K}) in the redshift distribution, which may
manifest at some level as an atypical morphology distribution.

Additionally, we need a way to simulate a deeper sample.  Presently we
have stopped at $I<22.5$ because the method described here requires
modification if the input images have non-negligible noise, and the
COSMOS data have $S/N>50$ for that magnitude range.  Thus, we require
either a generalization of the method to lower $S/N$ in the input
images, or we require deeper input data than is available in the
COSMOS field. \newtext{In some cases, the former would be possible: since the
{\sc shera} algorithm is a linear operation on the COSMOS postage stamps,
one could propagate any noise covariance matrix through the pipeline and arrive
at an output covariance matrix $N^{\rm IN}_{ij}$ on the output postage stamps.
If the noise covariance in the data we wish to simulate is $N^{\rm OUT}_{ij}$,
then so long as ${\mathbfss N}^{\rm OUT}-{\mathbfss N}^{\rm IN}$ is semi-positive
definite, one could add in the appropriate additional noise and thereby extend
the methodology of this paper to noisy input data. The software implementation of such a method, exploration of
its range of applicability, and investigation of complications such as
non-Gaussian noise is left to a future paper. In the opposite case, namely
that where the input data has more noise in some mode than the data we wish to
simulate, it seems likely that the problem is hopeless and deeper input data
would be required.}

Despite the need for future work to make the code and/or input dataset
as useful as possible for lensing surveys that are coming up on the
timescale of $\sim 1$ decade, we anticipate that {\sc shera v1.0} has
numerous applications for better understanding of current data and
those surveys that are starting in the next year, such as KIDS, HSC,
and DES.

\section*{Acknowledgments}

We thank the referee for many constructive comments about
the organization and content of this paper.  The authors would also like to thank Jim Gunn, Robert Lupton, Dustin Lang,
David Hogg, Michael Blanton, Barney Rowe, Peter Capak,
Chiaki Hikage, Uros Seljak, and Gary Bernstein for useful
conversations about this project; and Eric Huff both for discussing it
and giving the
software a name.     
C.H. is supported by the US National Science Foundation (AST-0807337),
the US Department of Energy (DE-FG03-02-ER40701), the Alfred P. Sloan
Foundation, and the David and Lucile Packard Foundation.
A.L. acknowledges support from the Chamberlain Fellowship at LBNL and
from the Berkeley Center for Cosmological Physics.  RJM is supported
by STFC Advanced Fellowship \#PP/E006450/1 and ERC grant
MIRG-CT-208994.  This work was done in part at JPL, run under a
contract for NASA by Caltech. 

\newtext{The HST
COSMOS Treasury program was supported through NASA grant
HST-GO-09822. We wish to thank Tony Roman, Denise Taylor, and David
Soderblom for their assistance in planning and scheduling of the
extensive COSMOS observations.  We gratefully acknowledge the
contributions of the entire COSMOS collaboration consisting of more
than 70 scientists.  More information on the COSMOS survey is
available at {\texttt http://cosmos.astro.caltech.edu/}. It is a pleasure to
acknowledge the excellent services provided by the NASA IPAC/IRSA
staff (Justin Howell, Anastasia Laity, Anastasia Alexov, Bruce
Berriman and John Good) in providing online archive and server
capabilities for the COSMOS data-sets.}

\appendix

\section{SDSS imaging properties used for simulations}\label{S:sdssprops}

In order to mimic SDSS data, we have determined several
properties of the SDSS data at the position of each galaxy.  While not
all
users will want to mimic specifically SDSS data, we include this
information with the data release as well.

\subsection{SDSS observations of the COSMOS field}\label{SS:sdssobscosmos}

It is worth noting that there are several atypical aspects to
the SDSS imaging in that region.  First, the median seeing is slightly
better than typical for the SDSS survey as a whole (by $\sim 10$ per
cent) although in fact the range of seeing values is rather broad.
Second, the sky level and therefore the photometric noise at fixed
magnitude is higher than usual for SDSS.  As a consequence, the object
detection and star/galaxy separation are somewhat less efficient than
in most of the survey area for $r\gtrsim 21$ or
$i\gtrsim 20.6$ (for more details, see \citealt{2011arXiv1107.1395N}).

\subsection{SDSS PSF}

The SDSS PSF is determined for all galaxies in an SDSS field by the
{\sc psp} (postage stamp pipeline) using a
procedure described in \cite{2001ASPC..238..269L}.  In brief, it
involves modeling the temporally and spatially varying PSF using a
Karhunen-Lo\'eve (KL) transform, which uses a set of bright stars to
determine basis functions and then to fit their coordinates to
spatially varying (quadratic) functions.  The information about the basis
functions and how their coefficients vary across a field is included
in psField files. 

Thus, for each galaxy in our COSMOS sample for which we have postage
stamp images, we find its position in SDSS imaging.  Some of these
galaxies are too faint to be detected; however, given their position
on the sky we can still determine precisely the CCD position at which
they would have been detected in SDSS.  Given this information, 
we obtain a postage stamp image of the galaxy $i$-band PSF using the
publicly available {\sc read\_psf} C code\footnote{\tt
  http://www.astro.princeton.edu/\~{}rhl/readAtlasImages.tar.gz} that
reconstructs the basis functions and the variation of the
coefficients across the field from the SDSS psField files.

\subsection{Photometricity}\label{SSS:photometricity}

As stated previously, roughly 33 per cent of the \newtext{SDSS imaging
  in the} COSMOS field is
classified as non-photometric according to the ubercalibration
\citep{2008ApJ...674.1217P} procedure on the rerun 301 (DR8)
reductions.  There are four SDSS fields overlapping the COSMOS region:
1462 and 1907 include most of the galaxies, and 1458 and 2125 each
cover a small fraction of the area.  All of 1462 and 2125 in the
COSMOS region are classified as photometric, but only part of 1907 and
none of 1458 in that area are photometric.  When determining the
photometric offset between COSMOS and SDSS photometry we must be
careful to exclude the regions that are non-photometric, and the data
release includes information about photometricity.  

\subsection{Photometric calibration}

We start with the total galaxy magnitudes from COSMOS with an offset
to convert from $F814W$ to SDSS $i$ (Section~\ref{SS:catalogue}).  To
determine the total number of $i$-band counts to assign to a galaxy of
this magnitude, we first determine the relevant number of nanomaggies
using
\beq
\mathrm{mag} = 22.5 - 2.5\,\log_{10}{[\mathrm{flux}\,\, \mathrm{(nanomaggies)}]}.
\eeq
We then use the SDSS photometric calibration (in units of nanomaggies
per count) from ubercal in the relevant SDSS run, camcol, and field
for each position.  This will allow us to generate images with units
of counts per pixel.

\subsection{Noise level}

When determining the level of noise to put into the fake data, 
we ignore the noise in the COSMOS observations, which is very small
relative to that in SDSS (as stated in Sec.~\ref{SS:shear}, the faintest
magnitude that we use in COSMOS corresponds to a minimum $S/N \sim
50$, and most are $>100$).  Accounting for it in detail would be quite
challenging given that it exhibits non-negligible pixel-to-pixel
correlations after we convolve it with the SDSS PSF.   We have confirmed that
for these $S/N$ levels, the noise fields that result from adding the
desired level of uncorrelated Gaussian noise are statistically
consistent with the noise fields we hope to introduce; that is, the KS
test shows no deviations due to the noise in the
original COSMOS postage stamps.  

For these simulations, we approximate the noise in SDSS as being a
random, \newtext{uncorrelated} Gaussian noise field with variance given by 
\beq\label{E:skyvar}
\sigma_\mathrm{sky}^2 = \frac{\mathrm{sky}}{\mathrm{gain}} +
\sigma_\mathrm{dark}^2, 
\eeq 
where the first term results from the
Poisson noise due to the photons in the sky, and the second is due to
the dark current (current that builds up due to heat even in the
absence of photons).  %Thus we are making two approximations: first,
%that the sky level is high enough that its Poisson noise can be
%approximated as a Gaussian, and second, that the Poisson noise due to
%the galaxy flux is negligible.  In \cite{2005MNRAS.361.1287M}, it was
%shown that this approximation is acceptable for $i\gtrsim 19.3$, which
%includes the large majority of our sample.  To ensure maximum
%flexibility in noise models, the simulation code is able to input both
%Gaussian and Poisson noise, and to include noise due to the object
%flux if desired; however, we do not use these options when simulating
%SDSS data.  %In order to implement these simulations, we include with
%the data release the SDSS $i$-band sky level, gain, and dark variance
%at the position of each galaxy.


\begin{thebibliography}{}

\bibitem[\protect\citeauthoryear{{Abazajian} et~al.}{2009}]{2009ApJS..182..543A}
 {Abazajian} K. et~al., 2009, ApJS, 182, 543

\bibitem[\protect\citeauthoryear{{Aihara} et~al.}{2011}]{2011ApJS..193...29A}
 {Aihara} H. et~al., 2011, ApJS, 193, 29

\bibitem[\protect\citeauthoryear{{Bartelmann} \& {Schneider}}{2001}]{2001PhR...340..291B}
 {Bartelmann} M., {Schneider} P., 2001, Phys. Rep., 340, 291

\bibitem[\protect\citeauthoryear{{Bernstein}}{2010}]{2010MNRAS.406.2793B}
 {Bernstein} G., 2010, MNRAS, 406, 2793

\bibitem[\protect\citeauthoryear{{Bernstein} \& {Jarvis}}{2002}]{2002AJ....123..583B}
 {Bernstein} G., {Jarvis} M., 2002, AJ, 123, 583

\bibitem[\protect\citeauthoryear{{Bertin} \& {Arnouts}}{1996}]{1996A&AS..117..393B}
 {Bertin} E., {Arnouts} S., 1996, A\&AS, 117, 393

\bibitem[\protect\citeauthoryear{{Bridle} et~al.}{2010}]{2010MNRAS.405.2044B}
 {Bridle} S. et~al., 2010, MNRAS, 405, 2044

\bibitem[\protect\citeauthoryear{{Bridle} et~al.}{2009}]{2009AnApS...3....6B}
 {Bridle} S. et~al., 2009, Ann. App. Stat., 3, 6

\bibitem[\protect\citeauthoryear{{Dark Energy Survey Collaboration}}{2005}]{2005astro.ph.10346T}
 {Dark Energy Survey Collaboration}, 2005, preprint (arXiv:astro-ph/0510346)

\bibitem[\protect\citeauthoryear{{Dobke} et~al.}{2010}]{2010PASP..122..947D}
 {Dobke} B., {Johnston} D., {Massey} R., {High} F., {Ferry} M., {Rhodes} J., {Vanderveld} R., 2010, PASP, 122, 947

\bibitem[\protect\citeauthoryear{{Eisenstein} et~al.}{2001}]{2001AJ....122.2267E}
 {Eisenstein} D. et~al., 2001, AJ, 122, 2267

\bibitem[\protect\citeauthoryear{{Ferry}
    et~al.}{2008}]{2008APh....30...65F} 
{Ferry} M.,  {Rhodes} J.,  {Massey} R.,  {White} M.,  
	{Coe} D.,  {Mobasher} B., 2008, Astroparticle Phys., 30, 65

\bibitem[\protect\citeauthoryear{{Fukugita} et~al.}{1996}]{1996AJ....111.1748F}
 {Fukugita} M., {Ichikawa} T., {Gunn} J., {Doi} M., {Shimasaku} K., {Schneider} D., 1996, AJ, 111, 1748

\bibitem[\protect\citeauthoryear{{Fu} et~al.}{2008}]{2008A&A...479....9F}
 {Fu} L. et~al., 2008, A\&A, 479, 9

\bibitem[\protect\citeauthoryear{{Gunn} et~al.}{1998}]{1998AJ....116.3040G}
 {Gunn} J. et~al., 1998, AJ, 116, 3040

\bibitem[\protect\citeauthoryear{{Hamana} et~al.}{2003}]{2003ApJ...597...98H}
 {Hamana} T. et~al., 2003, ApJ, 597, 98

\bibitem[\protect\citeauthoryear{{Hao} et~al.}{2006}]{2006MNRAS.370.1339H}
{Hao} C.~N., {Mao} S., {Deng} Z.~G., {Xia} X.~Y., {Wu} H., 2006,
MNRAS, 370, 1339

\bibitem[\protect\citeauthoryear{{Heymans} et~al.}{2006}]{2006MNRAS.368.1323H}
 {Heymans} C. et~al., 2006, MNRAS, 368, 1323

\bibitem[\protect\citeauthoryear{{High} et~al.}{2007}]{2007PASP..119.1295H}
 {High} F., {Rhodes} J., {Massey} R., {Ellis} R., 2007, PASP, 119, 1295

\bibitem[\protect\citeauthoryear{{Hirata} \& {Seljak}}{2003}]{2003MNRAS.343..459H}
 {Hirata} C., {Seljak} U., 2003, MNRAS, 343, 459

\bibitem[\protect\citeauthoryear{{Hirata} et~al.}{2004}]{2004MNRAS.353..529H}
 {Hirata} C. et~al., 2004, MNRAS, 353, 529

\bibitem[\protect\citeauthoryear{{Hoekstra}}{2007}]{2007MNRAS.379..317H}
 {Hoekstra} H., 2007, MNRAS, 379, 317

\bibitem[\protect\citeauthoryear{{Hoekstra} \&
    {Jain}}{2008}]{2008ARNPS..58...99H} 
{Hoekstra} H., {Jain} B., 2008, Ann. Rev. Nuclear and Particle
Science, 58, 99

\bibitem[\protect\citeauthoryear{{Hogg} et~al.}{2001}]{2001AJ....122.2129H}
 {Hogg} D., {Finkbeiner} D., {Schlegel} D., {Gunn} J., 2001, AJ, 122, 2129

\bibitem[\protect\citeauthoryear{{Ilbert} et~al.}{2009}]{2009ApJ...690.1236I}
 {Ilbert} O. et~al., 2009, ApJ, 690, 1236

\bibitem[\protect\citeauthoryear{{Ivezi{\'c}} et~al.}{2004}]{2004AN....325..583I}
 {Ivezi{\'c}} {\v Z}. et~al., 2004, Astron. Nachr., 325, 583

\bibitem[\protect\citeauthoryear{{Kaiser}}{2000}]{2000ApJ...537..555K}
 {Kaiser} N., 2000, ApJ, 537, 555

\bibitem[\protect\citeauthoryear{{Kaiser} et~al.}{2010}]{2010SPIE.7733E..12K}
 {Kaiser} N. et~al., 2010, Proc. SPIE, 7733, 12

\bibitem[\protect\citeauthoryear{{Kaiser} et~al.}{1995}]{1995ApJ...449..460K}
 {Kaiser} N., {Squires} G., {Broadhurst} T., 1995, ApJ, 449, 460

\bibitem[\protect\citeauthoryear{{Kasliwal} et~al.}{2008}]{2008ApJ...684...34K}
 {Kasliwal} M., {Massey} R., {Ellis} R., {Miyazaki} S., {Rhodes} J., 2008, ApJ, 684, 34

\bibitem[\protect\citeauthoryear{{Kitching} et~al.}{2010}]{2010arXiv1009.0779K}
 {Kitching} T. et~al., 2010, preprint (arXiv:1009.0779)

\bibitem[\protect\citeauthoryear{{Koekemoer} et~al.}{2007}]{2007ApJS..172..196K}
 {Koekemoer} A. et~al., 2007, ApJS, 172, 196

\bibitem[\protect\citeauthoryear{{Koekemoer} et~al.}{2002}]{2002hstc.conf..337K}
 {Koekemoer} A., {Fruchter} A., {Hook} R., {Hack} W., 2002, {The 2002 HST Calibration Workshop: Hubble after the Installation of the ACS and the NICMOS Cooling System}, eds. Arribas, Koekemoer, \& Whitmore, p. 337

\bibitem[\protect\citeauthoryear{{Kormendy} et~al.}{2009}]{2009ApJS..182..216K}
 {Kormendy} J., {Fisher} D.~B., {Cornell} M.~E., {Bender} R., 2009,
 ApJS, 182, 216

\bibitem[\protect\citeauthoryear{{Kova{\v c}} et~al.}{2010}]{2010ApJ...708..505K}
 {Kova{\v c}} K. et~al., 2010, ApJ, 708, 505

\bibitem[\protect\citeauthoryear{{Kron}}{1980}]{1980ApJS...43..305K}
 {Kron} R., 1980, ApJS, 43, 305

\bibitem[\protect\citeauthoryear{{LSST Science Collaborations}}{2009}]{2009arXiv0912.0201L}
 {LSST Science Collaborations}, 2009, preprint (arXiv:0912.0201)

\bibitem[\protect\citeauthoryear{{Lauer}}{1985}]{1985MNRAS.216..429L}
{Lauer} T.~R., 1985, MNRAS, 216, 429

\bibitem[\protect\citeauthoryear{{Leauthaud} et~al.}{2007}]{2007ApJS..172..219L}
 {Leauthaud} A. et~al., 2007, ApJS, 172, 219

\bibitem[\protect\citeauthoryear{{Leauthaud} et~al.}{2011}]{2011arXiv1104.0928L}
 {Leauthaud} A. et~al., 2011, preprint (arXiv:1104.0928)

\bibitem[\protect\citeauthoryear{{Lupton} et~al.}{2001}]{2001ASPC..238..269L}
 {Lupton} R., {Gunn} J., {Ivezi{\'c}} Z., {Knapp} G., {Kent} S., 2001, ASP Conf. Ser., 238, 269

\bibitem[\protect\citeauthoryear{{Mandelbaum} et~al.}{2005}]{2005MNRAS.361.1287M}
 {Mandelbaum} R. et~al., 2005, MNRAS, 361, 1287

\bibitem[\protect\citeauthoryear{{Mantz} et~al.}{2008}]{2008MNRAS.387.1179M}
 {Mantz} A., {Allen} S., {Ebeling} H., {Rapetti} D., 2008, MNRAS, 387, 1179

\bibitem[\protect\citeauthoryear{{Marleau} \& {Simard}}{1998}]{1998ApJ...507..585M}
 {Marleau} F., {Simard} L., 1998, ApJ, 507, 585

\bibitem[\protect\citeauthoryear{{Massey} et~al.}{2007a}]{2007MNRAS.376...13M}
 {Massey} R. et~al., 2007a, MNRAS, 376, 13

\bibitem[\protect\citeauthoryear{{Massey}
    et~al.}{2007b}]{2007MNRAS.380..229M}
{Massey} R., {Rowe} B., {Refregier} A., {Bacon} D.~J., Berg{\'e} J.,
2007b, MNRAS, 380, 229

\bibitem[\protect\citeauthoryear{{Massey}
    et~al.}{2010a}]{2010RPPh...73h6901M}
{Massey} R., {Kitching} T., {Richard} J., 2010a, Rep. Prog. Phys., 73, 086901

\bibitem[\protect\citeauthoryear{{Massey} et~al.}{2010b}]{2010MNRAS.401..371M}
 {Massey} R., {Stoughton} C., {Leauthaud} A., {Rhodes} J., {Koekemoer} A., {Ellis} R., {Shaghoulian} E., 2010b, MNRAS, 401, 371


\bibitem[\protect\citeauthoryear{{Melchior} et~al.}{2010}]{2010A&A...510A..75M}
 {Melchior} P., {B{\"o}hnert} A., {Lombardi} M., {Bartelmann} M., 2010, A\&A, 510, A75

\bibitem[\protect\citeauthoryear{{Miller} et~al.}{2007}]{2007MNRAS.382..315M}
 {Miller} L., {Kitching} T., {Heymans} C., {Heavens} A., {van Waerbeke} L., 2007, MNRAS, 382, 315

\bibitem[\protect\citeauthoryear{{Miyazaki} et~al.}{2002}]{2002PASJ...54..833M}
 {Miyazaki} S. et~al., 2002, PASJ, 54, 833

\bibitem[\protect\citeauthoryear{{Miyazaki} et~al.}{2006}]{2006SPIE.6269E...9M}
 {Miyazaki} S. et~al., 2006, Proc. SPIE, 6269, 9

\bibitem[\protect\citeauthoryear{{Nakajima}
    et~al.}{2011}]{2011arXiv1107.1395N} 
{Nakajima} R.,  {Mandelbaum} R.,  {Seljak} U.,  {Cohn} J.~D.,  
	{Reyes} R.,  {Cool} R., 2011, preprint (arXiv:1107.1395)

\bibitem[\protect\citeauthoryear{{Okabe} et~al.}{2010}]{2010PASJ...62..811O}
 {Okabe} N., {Takada} M., {Umetsu} K., {Futamase} T., {Smith} G., 2010, \pasj, 62, 811

\bibitem[\protect\citeauthoryear{{Padmanabhan} et~al.}{2008}]{2008ApJ...674.1217P}
 {Padmanabhan} N. et~al., 2008, ApJ, 674, 1217

\bibitem[\protect\citeauthoryear{{Park} \& {Schowengerdt}}{1983}]{1983CGIP...23..258P}
 {Park} S., {Schowengerdt} R., 1983, Comp. Gr. Im. Proc., 23, 258

\bibitem[\protect\citeauthoryear{{Pier} et~al.}{2003}]{2003AJ....125.1559P}
 {Pier} J., {Munn} J., {Hindsley} R., {Hennessy} G., {Kent} S., {Lupton} R., {Ivezi{\'c}} {\v Z}., 2003, AJ, 125, 1559

\bibitem[\protect\citeauthoryear{{Refregier}}{2003a}]{2003ARA&A..41..645R}
 {Refregier} A., 2003a, ARA\&A, 41, 645

\bibitem[\protect\citeauthoryear{{Refregier}}{2003b}]{2003MNRAS.338...35R}
 {Refregier} A., 2003b, MNRAS, 338, 35

\bibitem[\protect\citeauthoryear{{Refregier} \& {Bacon}}{2003}]{2003MNRAS.338...48R}
 {Refregier} A., {Bacon} D., 2003, MNRAS, 338, 48

\bibitem[\protect\citeauthoryear{{Reyes} et~al.}{2010}]{2010Natur.464..256R}
 {Reyes} R., {Mandelbaum} R., {Seljak} U., {Baldauf} T., {Gunn} J., {Lombriser} L., {Smith} R., 2010, \nat, 464, 256

\bibitem[\protect\citeauthoryear{{Rhodes} et~al.}{2000}]{2000ApJ...536...79R}
 {Rhodes} J., {Refregier} A., {Groth} E., 2000, ApJ, 536, 79

\bibitem[\protect\citeauthoryear{{Rhodes} et~al.}{2007}]{2007ApJS..172..203R}
 {Rhodes} J. et~al., 2007, ApJS, 172, 203

\bibitem[\protect\citeauthoryear{{Richards} et~al.}{2002}]{2002AJ....123.2945R}
 {Richards} G. et~al., 2002, AJ, 123, 2945

\bibitem[\protect\citeauthoryear{{Rines} et~al.}{2007}]{2007ApJ...657..183R}
 {Rines} K., {Diaferio} A., {Natarajan} P., 2007, ApJ, 657, 183

\bibitem[\protect\citeauthoryear{{Rozo} et~al.}{2010}]{2010ApJ...708..645R}
 {Rozo} E. et~al., 2010, ApJ, 708, 645

\bibitem[\protect\citeauthoryear{{Sargent} et~al.}{2007}]{2007ApJS..172..434S}
 {Sargent} M. et~al., 2007, ApJS, 172, 434

\bibitem[\protect\citeauthoryear{{Schlegel} et~al.}{1998}]{1998ApJ...500..525S}
 {Schlegel} D., {Finkbeiner} D., {Davis} M., 1998, ApJ, 500, 525

\bibitem[\protect\citeauthoryear{{Schrabback} et~al.}{2007}]{2007A&A...468..823S}
 {Schrabback} T. et~al., 2007, A\&A, 468, 823

\bibitem[\protect\citeauthoryear{{Schrabback} et~al.}{2010}]{2010A&A...516A..63S}
 {Schrabback} T. et~al., 2010, A\&A, 516, A63

\bibitem[\protect\citeauthoryear{{Schulz} et~al.}{2010}]{2010MNRAS.408.1463S}
 {Schulz} A., {Mandelbaum} R., {Padmanabhan} N., 2010, MNRAS, 408, 1463

\bibitem[\protect\citeauthoryear{{Scoville} et~al.}{2007a}]{2007ApJS..172....1S}
 {Scoville} N. et~al., 2007a, ApJS, 172, 1

\bibitem[\protect\citeauthoryear{{Scoville} et~al.}{2007b}]{2007ApJS..172...38S}
 {Scoville} N. et~al., 2007b, ApJS, 172, 38

\bibitem[\protect\citeauthoryear{{S\'ersic}}{1968}]{1968adga.book.....S}
 {S\'ersic} J., 1968, {Atlas de galaxias australes}, Cordoba, Argentina: Observatorio Astronomico

\bibitem[\protect\citeauthoryear{{Sirianni}
    et~al.}{1998}]{1998SPIE.3355..608S}
{Sirianni} M et~al., 1998, {Society of Photo-Optical Instrumentation
  Engineers (SPIE) Conference Series}, Ed. D'Odorico, S., 3355, 608


\bibitem[\protect\citeauthoryear{{Smith} et~al.}{2002}]{2002AJ....123.2121S}
 {Smith} J. et~al., 2002, AJ, 123, 2121

\bibitem[\protect\citeauthoryear{{Stoughton} et~al.}{2002}]{2002AJ....123..485S}
 {Stoughton} C. et~al., 2002, AJ, 123, 485

\bibitem[\protect\citeauthoryear{{Strauss} et~al.}{2002}]{2002AJ....124.1810S}
 {Strauss} M. et~al., 2002, AJ, 124, 1810

\bibitem[\protect\citeauthoryear{{Tucker} et~al.}{2006}]{2006AN....327..821T}
 {Tucker} D. et~al., 2006, Astron. Nachr., 327, 821

\bibitem[\protect\citeauthoryear{{Vikhlinin} et~al.}{2009}]{2009ApJ...692.1060V}
 {Vikhlinin} A. et~al., 2009, ApJ, 692, 1060

\bibitem[\protect\citeauthoryear{{Voigt}
    et~al.}{2011}]{2011arXiv1105.5595V} 
{Voigt} L.~M.,  {Bridle} S.~L.,  {Amara} A.,  {Cropper} M.,  
	{Kitching} T.~D.,  {Massey} R.,  {Rhodes} J.,  {Schrabback}
        T., 2011, preprint (arXiv:1105.5595)

\bibitem[\protect\citeauthoryear{{York} et~al.}{2000}]{2000AJ....120.1579Y}
 {York} D. et~al., 2000, AJ, 120, 1579

\bibitem[\protect\citeauthoryear{{Zhang} \& {Komatsu}}{2011}]{2011MNRAS.414.1047Z}
 {Zhang} J., {Komatsu} E., 2011, MNRAS, 414, 1047

\end{thebibliography}
\end{document}